\documentclass[a4paper]{article}

\title{Zero Crossings on the Accuracy-Smoothness Frontier: Maximal Monotone and Minimal Inflection Non Stationary Predictors}
\author{Marc Wildi\footnote{Zurich University of Applied Sciences (ZHAW). 
Technikumstrasse 71, 8400 Winterthur, Switzerland.  Website: https://marcwildi.com.
Email: marc.wildi@zhaw.ch.}.}

\usepackage{a4wide}
\usepackage{amsmath}
\usepackage{amsfonts} 
\usepackage{amsthm}
\usepackage{amssymb}
\usepackage{appendix}
\usepackage[utf8]{inputenc}
\usepackage{hyperref}
\usepackage{float}
\usepackage{booktabs}
\newtheorem{Proposition}{Proposition}
\newtheorem{Corollary}{Corollary}
\newtheorem{Theorem}{Theorem}

\usepackage{Sweave}
\begin{document}

\maketitle

\begin{abstract}
\noindent

The Smooth Sign Accuracy (SSA) framework resolves the smoothness--accuracy dilemma for stationary predictors by balancing mean squared error (MSE) against the frequency of sign changes. We generalize it to non-stationary integrated processes (I-SSA), deriving MSE-optimal predictors constrained by the expected duration between turning points (I(1)) or inflection points (I(2)). The resulting maximal-monotone and minimal-inflection predictors, filters, and smoothers customize classic benchmarks along a new efficient Accuracy--Smoothness (AS) frontier. Tracking the identity (the original data) along this frontier defines the \emph{I-SSA trend}, a novel smoothing concept that, unlike classical filters, imposes no artificial structure on the data-generating process and is directly interpretable through explicit control of turning- or inflection-point frequency. The optimization is a quadratically constrained quadratic program (QCQP) that reduces to a Generalized Trust Region Subproblem (GTRS) with hidden convexity. On its feasible, invertible region the problem is convex with a well-defined global optimum, and an exact closed-form expression delivers the efficient AS frontier. Under fairly general assumptions, finite-sample solutions converge to an asymptotic I-SSA predictor, devoid of integration boundary perturbations. The framework unifies mean squared error on non-stationary levels, and sign accuracy, zero-crossing frequency, and target correlation on stationary differences.

\end{abstract}

~
\newpage


\section{Introduction}

Forecasting requires balancing conflicting objectives. The SSA framework (Wildi, 2024, 2026a) targets {accuracy} (estimating future levels) and {smoothness} (suppressing spurious noise). We propose {I-SSA}, extending SSA to non-stationary integrated series and letting users balance accuracy (A) and smoothness (S) along a novel, efficient AS-frontier. We focus on univariate, homoscedastic series (the multivariate stationary case is treated in Wildi, 2026b). I-SSA overcomes the limitations of the frequency-domain forecast trilemma (McElroy and Wildi, 2019), offering better interpretability, numerical efficiency, and robustness against overfitting.

Traditional smoothing methods include curvature-penalized approaches (Whittaker, 1923; Henderson, 1924; Hodrick and Prescott, 1997)---despite recent critiques (Hamilton, 2018; Phillips and Shi, 2021)---splines (Eilers, Marx, and Durban, 2015; Wood, 2017), trend extraction (Harvey, 1989; Durbin and Koopman, 2012), and spectral filtering (Baxter and King, 1999; Pollock, 2000; Percival and Walden, 2000). Since our approach emphasizes turning and inflection points (TP/IP), a natural benchmark is the HP filter, a special case of Whittaker--Henderson smoothing, still widely used in business-cycle analysis and crisis prediction (Drehmann and Yetman, 2021); we complement the analysis with the boosted HP filter (Phillips and Shi, 2021, 2024). Broader comparisons are available in our R package.

While the HP filter minimizes curvature, SSA defines smoothness via the {expected duration between sign changes} (zero crossings, ZC) of a stationary, zero-mean predictor, building on Rice (1944). I-SSA extends this to integrated processes by targeting ZCs in {stationary differences} while minimizing MSE at the {non-stationary levels}. Economically, sign changes in growth rates mark transitions between expansions and contractions, so capturing business cycles requires sufficiently long intervals between crossings. This mean duration, the {holding time} (HT) (Wildi, 2024), is explicitly regulated in I-SSA for the growth (I(1)) or acceleration (I(2)) of the predictor via the rate of TPs or IPs.

I-SSA defines a novel Accuracy-Smoothness (AS) frontier: no other linear predictor of the same, or shorter length can improve both MSE on non-stationary levels (accuracy) and ZC rate in stationary differences (smoothness). Within this family, I-SSA is characterized by the {fewest growth sign-changes}, which justifies the terms {maximal monotone} (I(1)) and {minimal inflection} (I(2)).\footnote{The method also yields minimal monotone and maximal inflection predictors for a given MSE, though such extreme `un-smoothing' is typically irrelevant in prediction; see Wildi (2026a,b).} The classic MSE predictor is one point on this frontier, whereas I-SSA covers it in full. We also give a closed-form link between HT and MSE along the frontier, extending the stationary AS dilemma (Wildi, 2024, 2026a,b) to integrated processes. 

Extending SSA to the non-stationary setting raises technical caveats. The cointegration constraint between the acausal target and the causal predictor, which ensures finite MSE on levels, introduces asymmetries in the Toeplitz structure of the system matrices. Permanent `unit-root' memory prevents convergence to circulants, so the system matrices do not commute and a fixed Fourier eigenbasis is unavailable. We therefore use a generalized eigenvalue decomposition of a matrix pencil, as in cointegration (Johansen, 1988, 1991) and canonical correlation analysis (Box and Tiao, 1977), but with a structure originating in the optimization literature on the generalized trust-region subproblem (GTRS; Mor\'e, 1993; Adachi et al., 2017) because the Hessian is generally indefinite. The pencil separately encodes the non-stationary level and stationary autocorrelation operators, which the generalized decomposition diagonalizes simultaneously. This yields a closed-form, MSE-optimal predictor constrained by the expected duration between turning points, to our knowledge not previously explored for integrated processes. Although the formulation matches the stationary case, the geometry differs fundamentally: the generalized eigenvectors are non-orthogonal and depend on the HT, and the acausal target and causal predictor rely on different spectral bases, leading to a counterfactual spectral decomposition (details below). 

Formally, the optimization is a generalized trust-region subproblem (GTRS) with hidden convexity. If the classic MSE predictor is known, I-SSA reduces to a convolution with it that modifies its ZC rate while preserving MSE optimality; we call this transformation {customization}. For this problem, the region of invertibility and strict monotonicity reduce prediction to a one-dimensional search with certification of the global optimum. Under fairly general assumptions, finite-length solutions converge to a well-structured infinite-length I-SSA problem, free of finite-length boundary perturbations.

I-SSA customization means the projection of existing benchmark predictors onto the non-stationary AS frontier. In the applications, we customize the HP filter and its boosted variant, introducing the {maximal monotone} HP trend nowcast, obtained by targeting the acausal two-sided (boosted) HP filter in I-SSA. This reconciles growth sign analysis, in differences, with optimal {level} tracking. With the cursor on the frontier set to replicate the benchmark's TP rate, the I-SSA trend achieves lower MSE, notably at TPs, with faster response (smaller retardation) and no over- or undershooting. All examples are replicable with our open-source R package (\url{https://github.com/wiaidp/R-package-SSA-Predictor}), which also extends customization to Hamilton's regression filter (Hamilton, 2018), the Baxter--King filter (Baxter and King, 1999), and a refined Beveridge--Nelson decomposition (Kamber, Morley, and Wong, 2024).

Replacing the two-sided HP target by the identity, i.e., the original data, defines the \emph{I-SSA trend}: minimal MSE subject to maximal monotonicity (I(1)) or minimal inflection (I(2)). This novel smoothing principle imposes minimal structural assumptions on the data generating process (DGP), consistent with the typical spectral shape of economic time series (Granger, 1966), and prescribes no functional form (polynomial, spline, minimum curvature). TPs or IPs emerge from the interaction between the `neutral' (amorphous-memory) HT constraint and optimal tracking, which supports efficiency and interpretability. Relative to the classic Whittaker--Henderson smoother, the I-SSA trend yields a lower nowcast MSE at directional changes, confirming the findings for the HP customization. Moreover, the concurrent I-SSA trend distributes TPs more evenly and gives clearer signals, avoiding clusters of spurious TPs around directional changes.

The remainder of the paper is organized as follows. Section \ref{ssa} reviews the SSA criterion (Wildi, 2024, 2026a,b); Section \ref{i-ssa} extends it to integrated processes through I-SSA; Section \ref{applications} applies customization to business cycle nowcasting and the I-SSA trend to data smoothing; Section \ref{conclu} concludes.

\section{SSA} \label{ssa}

The following summarizes the stationary Smooth Sign Accuracy framework of Wildi (2024, 2026a,b), which provides the foundation for I-SSA.

Consider a zero-mean stationary series $x_t$ (non-zero-mean processes are centered) $x_t=\sum_{i=0}^{\infty}\xi_i\epsilon_{t-i}$, with $\xi_0=1$. The one-sided sequence $\boldsymbol{\xi}_{\infty}:=(\xi_0,\xi_1,...)'$ is square summable and corresponds to the weights in the (purely non-deterministic) Wold decomposition of $x_t$, as detailed by Brockwell and Davis (1993). Let $z_{t}=\sum_{k=-\infty}^{\infty}\gamma_k x_{t-k}$
be another stationary process and assume we want to predict $z_{t+\delta}$, the prediction target, with $\delta \in \mathbb{Z}$. We construct a length-$L$ causal predictor $y_t=\sum_{k=0}^{L-1}b_kx_{t-k}$ for $z_{t+\delta}$, termed forecasting, nowcasting, or backcasting as $\delta>0$, $=0$, or $<0$.\footnote{Our target specification also accommodates signal extraction, where $\gamma_k$ are coefficients of a two-sided (possibly bi-infinite) filter, see our examples.}

For example, with $x_t=\epsilon_t$, $\gamma_0=1$, $\gamma_1=0.5$, $z_t=\epsilon_t+0.5\epsilon_{t-1}$ is an MA(1); one-step forecasting ($\delta=1$) gives the MSE predictor $y_t=0.5\epsilon_t$, i.e., $b_0=0.5$, $b_k=0$ otherwise.

The MSE predictor tracks the level optimally but can be `noisy': here $y_t=0.5\epsilon_t$ is WN while $z_{t+1}$ is more persistent. Similarly, in real-time signal extraction the one-sided concurrent filter is markedly noisier than its acausal, two-sided counterpart (see examples below). To address this, Wildi (2024) proposes to control the HT, the mean duration between consecutive zero-crossings, of the predictor.

Specifically, let $\boldsymbol{\Xi}$ denote the $L\times L$ matrix with $i$-th row $\boldsymbol{\Xi}_{i\cdot}:=(\xi_{i-1},\xi_{i-2},...,\xi_0,\mathbf{0}_{L-i})$, $i=1,...,L$, where $\mathbf{0}_{L-i}$ is a zero vector of length $L-i$. We define $\mathbf{x}_t:=(x_t,...,x_{t-(L-1)})'$ and $\boldsymbol{\epsilon}_t:=(\epsilon_t,...,\epsilon_{t-(L-1)})'$. Additionally, we define
$\mathbf{b}_{\epsilon}:=\boldsymbol{\Xi}\mathbf{b}_x$. Consequently, we obtain:
$$
y_t=\mathbf{b}_x'\mathbf{x}_t\approx\left(\boldsymbol{\Xi}\mathbf{b}_x\right)'\boldsymbol{\epsilon}_{t}=\mathbf{b}_{\epsilon}'\boldsymbol{\epsilon}_t,
$$
where the approximation via the finite MA inversion of $x_t$ holds if the filter coefficients decay to zero sufficiently rapidly or, equivalently, if $L$ is sufficiently large (exact results can be derived but are omitted here). Wildi (2024,2026a) proposes the following SSA criterion
  \begin{eqnarray}\label{crit_gen}
&\min_{\mathbf{b}_{\epsilon}}E[(z_{t+\delta}-y_t)^2]\label{mse_f}\\
&\mathbf{b}_{\epsilon}'\mathbf{Mb_{\epsilon}}-\rho_1\mathbf{b}_{\epsilon}'\mathbf{b}_{\epsilon}=0\label{ht_f},
\end{eqnarray}
where the MA approximation holds at the stationary point of the Lagrangian (see details below). The $L\times L$ autocovariance generating matrix $\mathbf{M}$,
\[
\mathbf{M}=\left(\begin{array}{ccccccccc}0&0.5&0&0&0&...&0&0&0\\
0.5&0&0.5&0&0&...&0&0&0\\
...&&&&&&&&\\
0&0&0&0&0&...&0.5&0&0.5\\
0&0&0&0&0&...&0&0.5&0
\end{array}\right),
\]
satisfies $\mathbf{b_{\epsilon}}'\mathbf{Mb_{\epsilon}}=\sum_{k=1}^{L-1}b_{\epsilon,k-1}b_{\epsilon,k}$, the first-order autocovariance of the predictor, assuming standardized white noise. The constraint \eqref{ht_f} imposes a first-order autocorrelation (ACF(1)) of $\rho_1$ and is feasible if $|\rho_1|<\rho_{\max}(L)$, the largest eigenvalue of $\mathbf{M}$ determining the Rayleigh quotient. Since $|\rho_1|<\rho_{\max}(L)$, the matrix $\mathbf{M}-\rho_1\mathbf{I}$ is indefinite, with both positive and negative eigenvalues, which gives the SSA problem its GTRS structure.

Let $ht(y|\mathbf{b}_{\epsilon})$ denote the (stationarity-invariant) expected duration between consecutive zero-crossings of $y_t$, the HT.\footnote{Crossings satisfy $t_{i-1}<t_i$, $t_1\geq L$, $y_{t_i-1}y_{t_i}<0$, and $y_{t-1}y_t>0$ for $t_{i-1}<t<t_i$.} Assuming Gaussianity,
\begin{equation}\label{ht}
ht(y|\mathbf{b}_{\epsilon})=\displaystyle{\frac{\pi}{\arccos(\rho(y))}},
\end{equation}
linking HT and ACF(1) (Wildi, 2024); this relation is fairly robust to non-Gaussianity (central limit theorem, see Wildi, 2026b). Since $\arccos$ is monotonic on $]-1,1[$, the criterion  becomes
\begin{eqnarray*}
\left.\begin{array}{cc}
&\min_{\mathbf{b}_{\epsilon}}E[(z_{t+\delta}-y_t)^2]\\
&ht(y|\mathbf{b}_{\epsilon})=ht_1\\
\end{array}\right\}.
\end{eqnarray*}
Specifying $ht_1>ht_{MSE}$ ($ht_1<ht_{MSE}$), the HT of the MSE predictor, yields a smoother (unsmoother) predictor with fewer (more) ZCs. The resulting SSA predictor minimizes MSE subject to the ZC rate; dually, it minimizes the ZC rate for given MSE (efficient frontier). The MSE predictor is recovered by setting $ht_1=ht_{MSE}$. Wildi (2024,2026b) proposes variants of the SSA criterion in which the MSE objective \eqref{crit_gen} is replaced by either the target correlation, i.e., the correlation between the predictor $y_t$ and the target $z_{t+\delta}$, or the sign accuracy, i.e., the probability that predictor and target have the same sign (assuming Gaussianity). Since all variants are equivalent, SSA unifies MSE, target correlation, sign accuracy, and the rate of zero-crossings within a single interpretable framework.

\section{I-SSA}\label{i-ssa}

Here $\tilde{x}_t$, $\tilde{z}_t=\sum_{|k|<\infty}\gamma_k \tilde{x}_{t-k}$, and $y_t=\sum_{k=0}^{L-1}b_{\tilde{x}k}\tilde{x}_{t-k}$ are all assumed non-stationary I($d$), with $d>0$, so HT and zero-crossing rate, autocorrelation, target correlation or sign accuracy become ill-defined; we now show how to resolve this.

\subsection{Concept}\label{ext_i1_ssa}

Let $\Delta(B)\tilde{x}_t=(1-B)^d\tilde{x}_t=:x_t$ be stationary and invertible, with Wold weights $\xi_k$, $k\geq0$. The MSE predictor is $y_{t,MSE}=\hat{\boldsymbol{\gamma}}_{MSE}'\mathbf{\tilde{x}}_t$, with weights $(\hat{\gamma}_{\tilde{x}\delta k})_{k=0,...,L-1}$ and $\mathbf{\tilde{x}}_t:=(\tilde{x}_t,...,\tilde{x}_{t-(L-1)})'$, valid under general assumptions (McElroy and Wildi, 2016; McElroy and Livsey, 2022; McElroy, 2022). We henceforth assume the MSE predictor $(\hat{\gamma}_{\tilde{x}\delta k})_{k=0,...,L-1}$ to be known. 

The MSE principle guarantees stationarity of $e_t:=\tilde{z}_{t+\delta}-y_{t,MSE}$, so $\tilde{z}_{t+\delta}$ and $y_{t,MSE}$ are cointegrated with vector $(1,-1)$. To eliminate the unit root(s) of $\tilde{x}_t$, the error filter $\gamma_{k+\delta}-\hat{\gamma}_{\tilde{x}\delta k}$ must satisfy
\begin{eqnarray}\label{coint_eq}
\sum_{k=-\infty}^{\infty} (\gamma_{k+\delta}-\hat{\gamma}_{\tilde{x}\delta k})k^j=0,\quad j=0,...,d-1,
\end{eqnarray}
with $\hat{\gamma}_{\tilde{x}\delta k}=0$ for $k\notin\{0,...,L-1\}$, (McElroy and Wildi, 2016), see below for details. 

We focus on $d=1$, so $\tilde{x}_t,\tilde{z}_t\sim I(1)$; the extension to $d>1$ is analogous (Appendix \eqref{i2_ext}). We assume throughout that $\sum_{k=-\infty}^{\infty} \gamma_{k+\delta}\neq 0$, with continuous transfer function at zero frequency, so that the filter does not cancel the unit root, i.e., $\tilde{z}_t$ is non-stationary  I($1$) (if $\tilde{z}_t$ is stationary, the problem can be re-framed in SSA through suitable transformations). 
Define the $L\times L$ summation and differencing matrices
\[\boldsymbol{\Sigma}=\left(\begin{array}{ccccc}1&0&0&...&0\\
1&1&0&...&0\\
...&&&&\\
1&1&1&...&1\end{array}\right) ~\textrm{and} ~\boldsymbol{\Delta}:=\boldsymbol{\Sigma}^{-1}=\left(\begin{array}{cccccc}1&0&0&...&0&0\\-1&1&0&...&0&0\\
&&&&\\
0&0&0&...&-1&1\end{array}\right).
\]
For a predictor $y_t$ with weights $\mathbf{b}$, define $e_{y,t}:=y_{t,MSE}-y_t$, replacing target $\tilde{z}_{t+\delta}$ with the MSE benchmark $y_{t,MSE}$—justified by the law of iterated projections: the MSE predictor serves as an equivalent I-SSA target, and I-SSA is a constrained predictor of $\tilde{z}_{t+\delta}$ or a smoother of the MSE predictor. The I(1)-SSA filter $\mathbf{b}_{\tilde{x}}=(b_{\tilde{x}0},...,b_{\tilde{x}L-1})'$ minimizes $E[e_{y,t}^2]$ subject to the HT constraint (detailed below); without constraint it reduces to the MSE design ($\mathbf{b}_{\tilde{x}}=\hat{\boldsymbol{\gamma}}_{MSE}$ and $e_{y,t}=0$). The added cointegration constraint
\[
\sum_{k=0}^{L-1}b_{\tilde{x}k}=\sum_{k=0}^{L-1}\hat{\gamma}_{ \tilde{x}\delta k}
\]
ensures $e_{SSA,t}$ is stationary:
\begin{eqnarray}\label{coint_r}
e_{SSA,t}=(\hat{\boldsymbol{\gamma}}_{MSE}-\mathbf{b}_{\tilde{x}})'\mathbf{\tilde{x}}_t=  (\hat{\boldsymbol{\gamma}}_{MSE}-\mathbf{b}_{\tilde{x}})' \boldsymbol{\Sigma}'\boldsymbol{\Delta}'\mathbf{\tilde{x}}_t=
(\hat{\boldsymbol{\gamma}}_{MSE}-\mathbf{b}_{\tilde{x}})'\boldsymbol{\Sigma}'\mathbf{x}_t,
\end{eqnarray}
the last equality holding because the final entry of $(\hat{\boldsymbol{\gamma}}_{MSE}-\mathbf{b}_{\tilde{x}})'\boldsymbol{\Sigma}'$ vanishes by the cointegration constraint, allowing the non-stationary $\tilde{x}_{t-(L-1)}$ in $\boldsymbol{\Delta}'\mathbf{\tilde{x}}_t$ to be replaced by the stationary $x_{t-(L-1)}=\tilde{x}_{t-(L-1)}-\tilde{x}_{t-L}$ (or any number). Assuming $L$ large enough, the finite MA approximation\footnote{Theorem \ref{thm:decay_reformulated_uniform} in the main theoretical results, Section \ref{the_sec} below, shows that filter weights decay sufficiently fast for the MA approximation to be valid under fairly general assumptions on the target.}
\[
(\hat{\boldsymbol{\gamma}}_{MSE}-\mathbf{b}_{\tilde{x}})'\boldsymbol{\Sigma}'\mathbf{x}_t\approx
(\hat{\boldsymbol{\gamma}}_{MSE}-\mathbf{b}_{\tilde{x}})'\boldsymbol{\Sigma}'\boldsymbol{\Xi}'\boldsymbol{\epsilon}_t
\]
gives
\begin{eqnarray}\label{main_int_res}
e_{SSA,t}&=&y_{t,MSE}-y_t=(\hat{\boldsymbol{\gamma}}_{MSE}-\mathbf{b}_{\tilde{x}})'\boldsymbol{\Sigma}'\mathbf{x}_t\approx
(\hat{\boldsymbol{\gamma}}_{MSE}-\mathbf{b}_{\tilde{x}})'\boldsymbol{\Sigma}'\boldsymbol{\Xi}'\boldsymbol{\epsilon}_t\nonumber\\
&=&\hat{\boldsymbol{\gamma}}_{MSE}'\boldsymbol{\Sigma}'\boldsymbol{\Xi}'\boldsymbol{\epsilon}_t-\mathbf{b}_{\epsilon}'\boldsymbol{\Sigma}'\boldsymbol{{\epsilon}}_t,
\end{eqnarray}
where $\mathbf{b}_{{\epsilon}}:=\boldsymbol{\Xi}\mathbf{b}_{\tilde{x}}$ inherits the cointegration constraint from $\mathbf{b}_{\tilde{x}}$  (see below), and the last equality uses commutativity of $\boldsymbol{\Sigma}$ and $\boldsymbol{\Xi}$ under convolution (lower-triangular Toeplitz matrices). In \eqref{main_int_res}, $y_{t,MSE}$ and $y_t$ are non-stationary cointegrated predictors of $\tilde{z}_{t+\delta}$, while the right-hand processes $\hat{\boldsymbol{\gamma}}_{MSE}'\boldsymbol{\Sigma}'\boldsymbol{\Xi}'\boldsymbol{\epsilon}_t$ and $\mathbf{b}_{\epsilon}'\boldsymbol{\Sigma}'\boldsymbol{\epsilon}_t$ are stationary and no longer predictors—yet their cross-sectional differences remain (virtually) identical, which is key to optimization.\\

The stationary MA representations in \eqref{main_int_res} thus yield a valid target correlation for the objective, while an HT constraint can be based on $\mathbf{b}_{\epsilon}'\boldsymbol{\epsilon}_t=\mathbf{b}_{\tilde{x}}'\boldsymbol{\Xi}'\boldsymbol{\epsilon}_t\approx y_t-y_{t-1}$, controlling zero-crossings of the stationary first difference $y_t-y_{t-1}$; a length constraint on its variance is likewise feasible. This yields a conceptual I(1)-SSA criterion:
\begin{eqnarray}\label{crit_int}
\left.\begin{array}{cc}
&\min_{\mathbf{b}_{{\epsilon}}}(\boldsymbol{\gamma}_{\tilde{\Xi}\delta}-\boldsymbol{\Sigma}\mathbf{b}_{{\epsilon}})'(\boldsymbol{\gamma}_{\tilde{\Xi}\delta}-\boldsymbol{\Sigma}\mathbf{b}_{{\epsilon}})\\
&\mathbf{b}_{\epsilon}'\mathbf{M}\mathbf{b_{\epsilon}}=\rho_1\mathbf{b}_{\epsilon}'\mathbf{b}_{\epsilon}\\
\end{array}\right\},
\end{eqnarray}
where $\boldsymbol{\tilde{\Xi}}:=\boldsymbol{\Xi}\boldsymbol{\Sigma}=\boldsymbol{\Sigma}\boldsymbol{\Xi}$ and $\boldsymbol{\gamma}_{\tilde{\Xi}\delta}:=\boldsymbol{\tilde{\Xi}}\boldsymbol{\hat{\gamma}}_{MSE}$. The criterion presupposes the cointegration constraint `ex nihilo' and is thus incomplete, but remains useful for interpretability.
The target derivative $-2\boldsymbol{\Sigma}'\boldsymbol{\gamma}_{\tilde{\Xi}\delta}+2\boldsymbol{\Sigma}'\boldsymbol{\Sigma}\mathbf{b}_{{\epsilon}}$ gives the Lagrangian normal equation $\left(-\tilde{\lambda}\left(\mathbf{M}-\rho_1\mathbf{I}\right)
+\boldsymbol{\Sigma}'\boldsymbol{\Sigma}\right)\mathbf{b}_{{\epsilon}}=\boldsymbol{\Sigma}'\boldsymbol{\gamma}_{\tilde{\Xi}\delta}$, hence $\mathbf{b}_{\epsilon}=F\boldsymbol{\tilde{\mathcal{V}}}^{-1}\boldsymbol{\Sigma}'\boldsymbol{\gamma}_{\tilde{\Xi}\delta}$
with $\tilde{\boldsymbol{\mathcal{V}}}:=\mathbf{M}-\rho_1\mathbf{I}-1/\tilde{\lambda}\boldsymbol{\Sigma}'\boldsymbol{\Sigma}$ and $F:=-1/\tilde{\lambda}$.  Wildi (2026a) proposes a variant of I-SSA based on the target correlation of the stationary processes on the right of \eqref{main_int_res}; both criteria are equivalent.

This expression extends the SSA solution, in which the integration operator $\boldsymbol{\Sigma}$ is replaced by $\mathbf{I}$, so that the `reduced' Hessian $\tilde{\boldsymbol{\mathcal{V}}}:=\mathbf{M}-\rho_1\mathbf{I}-1/\tilde{\lambda}\mathbf{I}$ admits a simple decomposition into the Fourier eigenvectors of $\mathbf{M}$, i.e., a fixed geometry. Its presence in I-SSA is the main technical difficulty: $\tilde{\boldsymbol{\mathcal{V}}}$ `grows' unboundedly, and the classic eigenvalue decomposition now depends on both $\rho_1$ and $\tilde{\lambda}$, because $\mathbf{M}$ and $\boldsymbol{\Sigma}'\boldsymbol{\Sigma}$ do not commute and cannot be diagonalized simultaneously. The resulting classic geometry is parameter-dependent and unsuited to our purpose.

For a (nearly) Gaussian process $y_t$, $ht_1 := \pi/\arccos(\rho_1)$ defines the HT of the first differences $y_t - y_{t-1}$, assuming zero mean (if a linear drift is present, the constraint instead addresses crossover points around the trend). Dually, $y_t$ is {maximally monotone} if sign changes of $y_t - y_{t-1}$ are minimized for a given tracking accuracy (e.g., MSE, detailed below). We next formally integrate the cointegration constraint, imposing additional structure and asymmetry on the problem.

\subsection{Solution}\label{red_sol}

Let $\Gamma(0):=\sum_{k=0}^{L-1}\hat{\gamma}_{ \tilde{x}\delta k}\neq 0$ be the zero-frequency transfer function of the MSE filter, assuming continuity at zero (McElroy and Wildi, 2016). So the I(1)-SSA cointegration constraint reads $\sum_{k=0}^{L-1}b_{\tilde{x}k}=\Gamma(0)\neq 0$,
or in vector form $$\mathbf{b}_{\tilde{x}}=\Gamma(0)\mathbf{e}_1+\mathbf{B}\tilde{\mathbf{b}},$$ 
where $\mathbf{e}_1=(1,0,...,0)'$ ($L\times1$), $\tilde{\mathbf{b}}=(\tilde{b}_1,...,\tilde{b}_{L-1})'$ ($(L-1)\times1$), and $\mathbf{B}$ is the $L\times(L-1)$ matrix
\[
\mathbf{B} = \begin{pmatrix}
-1 & -1 & -1 & \cdots & -1 \\
1 & 0 & 0 & \cdots & 0 \\
0 & 1 & 0 & \cdots & 0 \\
\vdots & & & \ddots & \vdots \\
0 & 0 & 0 & \cdots & 1
\end{pmatrix}
\]
(a row of $-1$s atop the $(L-1)\times(L-1)$ identity).

We reframe the I-SSA problem under the cointegration constraint with simplified notation:
\begin{equation}\label{beps_btilde}
    \mathbf{b} = \Gamma(0) \boldsymbol{\Xi} \mathbf{e}_1 + \boldsymbol{\Xi} \mathbf{B} \mathbf{\tilde{b}} = \mathbf{v} + \mathbf{P} \mathbf{\tilde{b}},
\end{equation}
with $\mathbf{b}=\mathbf{b}_\epsilon$, $\mathbf{v} = \Gamma(0) \boldsymbol{\Xi} \mathbf{e}_1$ and the full-column-rank matrix $\mathbf{P} = \boldsymbol{\Xi} \mathbf{B} \in \mathbb{R}^{L \times (L-1)}$. The MSE objective in \eqref{crit_int} becomes:
$$
    J(\mathbf{\tilde{b}}) = \mathbf{\tilde{b}}' \mathbf{\tilde{A}} \mathbf{\tilde{b}} - 2 \mathbf{\tilde{c}}' \mathbf{\tilde{b}} + \text{const}
$$
where $\mathbf{\tilde{A}} = \mathbf{P}' \boldsymbol{\Sigma}' \boldsymbol{\Sigma} \mathbf{P}$, $\mathbf{\tilde{c}} = \mathbf{P}' \boldsymbol{\Sigma}' (\boldsymbol{\gamma} - \boldsymbol{\Sigma} \mathbf{v})$ where the constant depends on the MSE predictor $\boldsymbol{\gamma}_{\tilde{\Xi}\delta}$ which we rewrite as $\boldsymbol{\gamma}=\boldsymbol{\gamma}_{\tilde{\Xi}\delta}$.

The HT constraint can be rewritten as $\mathbf{b}' \mathbf{W} \mathbf{b} = 0$, where $\mathbf{W} = \mathbf{M} - \rho_1 \mathbf{I}$. Substituting the cointegration constraint yields:
$$
    \mathbf{\tilde{b}}' \tilde{\mathbf{W}} \mathbf{\tilde{b}} + 2 \mathbf{\tilde{u}}' \mathbf{\tilde{b}} + d = 0,
$$
where $\tilde{\mathbf{W}} = \mathbf{P}' \mathbf{W} \mathbf{P}$, $\mathbf{\tilde{u}} = \mathbf{P}' \mathbf{W} \mathbf{v}$, and $d = \mathbf{v}' \mathbf{W} \mathbf{v}$. In the stationary SSA, feasibility assumes $|\rho_1|\leq \rho_{\max}(L)$, the largest eigenvalue of $\mathbf{M}$ or its spectral radius (Wildi (2026a), Proposition 1). As shown below, this feasibility condition extends to I-SSA. Because the boundary case $|\rho_1| = \rho_{\max}(L)$ renders the optimization `trivial',\footnote{In this case, the solution defaults to $\mathbf{b} = \pm c \mathbf{v}_{\rho{\max}}$, the corresponding eigenvector of $\mathbf{M}$, leaving only sign and scaling $c>0$ for optimization (see Wildi, 2026a).} we restrict our focus to $|\rho_1| < \rho_{\max}(L)$. Under this assumption, $\mathbf{W}$ possesses both positive and negative eigenvalues, reducing the I-SSA problem to a GTRS with an indefinite equality constraint.

The Lagrangian is given by:
$$
    \mathcal{L}(\mathbf{\tilde{b}}, \lambda) = \mathbf{\tilde{b}}' \mathbf{\tilde{A}} \mathbf{\tilde{b}} - 2 \mathbf{\tilde{c}}' \mathbf{\tilde{b}} + \lambda (\mathbf{\tilde{b}}' \tilde{\mathbf{W}} \mathbf{\tilde{b}} + 2 \mathbf{\tilde{u}}' \mathbf{\tilde{b}} + d)
$$
Setting the gradient with respect to $\mathbf{\tilde{b}}$ to zero yields the normal equations:
$$
    \mathbf{\tilde{H}}(\lambda) \mathbf{\tilde{b}}(\lambda) = \mathbf{\tilde{c}} - \lambda \mathbf{\tilde{u}}
$$
where 
$$
\mathbf{\tilde{H}}(\lambda) = \mathbf{\tilde{A}} + \lambda \tilde{\mathbf{W}}
$$
is the  Hessian. Assuming $\mathbf{\tilde{H}}(\lambda)$ is invertible (invertibility is addressed below), the optimal $\mathbf{\tilde{b}}$ for a given $\lambda$ is:
\begin{equation}
    \mathbf{\tilde{b}}(\lambda) = \mathbf{\tilde{H}}(\lambda)^{-1} (\mathbf{\tilde{c}} - \lambda \mathbf{\tilde{u}}). \label{eq:b_lambda}
\end{equation}
The  I(1)-SSA solution to \eqref{crit_int}, under the cointegration constraint reparameterized via $\mathbf{\tilde{b}}(\lambda^*)$, follows from \eqref{eq:b_lambda} with $\lambda^*$ satisfying the secular equation
\begin{equation}\label{seceq}
\mathbf{b}(\lambda^*)'\mathbf{W}\mathbf{b}(\lambda^*)=0,
\end{equation}
where $\mathbf{b}(\lambda^*)=\mathbf{v}+\mathbf{P}\tilde{\mathbf{b}}(\lambda^*)$. Below, we derive strict monotonicity---implying uniqueness---and a simple numerical solution to \eqref{seceq}  via bisection.

Extensions to $d>1$ follow by using $\boldsymbol{\Sigma}^d$ throughout, with additional cointegration constraints
\[
\sum_{k=0}^{L-1}b_{\tilde{x}k}k^j=\sum_{k=0}^{L-1}\hat{\gamma}_{ \tilde{x}\delta k}k^j,\quad j=0,...,d-1,
\]
eliminating the higher-order unit root of $\tilde{x}_t$ in $e_t=y_{t,MSE}-y_t$ for I(d)-SSA; see Appendix \eqref{i2_ext}. For $\lambda=0$, I-SSA recovers the MSE predictor, noting that the cointegration constraint could be dropped since the MSE is already cointegrated.

\subsection{Main Theoretical Results}\label{the_sec}

The cointegration matrix $\mathbf{B}$ introduces asymmetries that render direct derivations feasible but needlessly complex. We therefore impose the cointegration constraint not a priori, but through an additional penalty term in the Lagrangian, maintaining $\Gamma(0)\neq 0$ and, for simplicity, setting $d=1$ throughout (the proofs readily extend to $d>1$). The problem formulation of the previous section nonetheless remains useful in deriving these results.

The first proposition delimits the region of strict positive definiteness of the Hessian $\mathbf{H}(\lambda)$ in the reformulated problem (Lagrangian), which does not depend on $\mathbf{B}$ anymore.

\begin{Proposition}\label{propo_pos_def}
Assume $|\rho_1|<\rho_{\max}(L)$ (feasibility) in the HT constraint and define $\mathbf{A} = \boldsymbol{\Sigma} \boldsymbol{\Xi}$ and $\mathbf{Q} = \boldsymbol{\Xi}' (\mathbf{M} - \rho_1 \mathbf{I}) \boldsymbol{\Xi}$.
Consider the Hessian 
\[
\mathbf{H}(\lambda) = \mathbf{A}' \mathbf{A} + \lambda \mathbf{Q},
\]
and let $\nu_{\max}$ and $\nu_{\min}$ denote the maximum and minimum eigenvalues of the matrix $(\mathbf{A}' \mathbf{A})^{-1} \mathbf{Q}$. Then $\nu_{\max} > 0$ and $\nu_{\min} < 0$, and the matrix $\mathbf{H}(\lambda)$ is strictly positive definite if and only if $\lambda$ lies in the open interval:
\[
\lambda \in \left] -\frac{1}{\nu_{\max}}, -\frac{1}{\nu_{\min}} \right[=:J_{++}.
\]
\end{Proposition}

The proof is given in the Appendix. Strict convexity of $\mathbf{H}(\lambda)$ in $J_{++}$---termed `hidden convexity'---allows us to formulate the main result characterizing the finite-length, $L<\infty$, I-SSA solution.

\begin{Theorem}[Algebraic Solution to I-SSA GTRS Problem]\label{prop_alt}
 Consider the constrained optimization problem:
\begin{align*}
    \min_{\mathbf{\tilde{b}} \in \mathbb{R}^{L-1}} \quad & J(\mathbf{\tilde{b}}) = (\boldsymbol{\gamma} - \boldsymbol{\Sigma} \mathbf{b}_{\epsilon})' (\boldsymbol{\gamma} - \boldsymbol{\Sigma} \mathbf{b}_{\epsilon}) \\
    \text{subject to} \quad & \mathbf{b_{\epsilon}' \mathbf{M} b_{\epsilon}} = \rho_1 \mathbf{b_{\epsilon}' b_{\epsilon}}\\
    \text{and} \quad& \mathbf{b}_{\epsilon} = \boldsymbol{\Xi} (\Gamma(0) \mathbf{e}_1 + \mathbf{B} \mathbf{\tilde{b}}).
\end{align*}
\begin{enumerate}
\item \textbf{Stationary Point}: For $\lambda\in J_{++}$, a stationary point $\mathbf{b}_{\tilde{x}}(\lambda)$ of the Lagrangian is given by
\[
\mathbf{b}_{\tilde{x}}(\lambda) = \mathbf{H}(\lambda)^{-1} (\mathbf{A}' \boldsymbol{\gamma} - \mu(\lambda) \mathbf{1}),
\]
with the scalar $\mu(\lambda)$ defined as
\[
\mu(\lambda) = \frac{\mathbf{1}' \mathbf{H}(\lambda)^{-1} \mathbf{A}' \boldsymbol{\gamma} - \Gamma(0)}{\mathbf{1}' \mathbf{H}(\lambda)^{-1} \mathbf{1}},
\]
where $\mathbf{1} \in \mathbb{R}^L$ is a vector of ones, and $\mathbf{\tilde{b}}(\lambda)=\mathbf{b}_{\tilde{x}}(\lambda)_{-0}$ (excluding the first lag-zero entry).

\item \textbf{Feasibility}: For $|\rho_1|<\rho_{\max}(L)$ there exists $\lambda^*\in J_{++}$ such that 
\[
\rho_{\mathbf{b}_{\epsilon}}(\lambda^*)
=\frac{\mathbf{b}_{\epsilon}(\lambda^*)'\mathbf{M}\mathbf{b}_{\epsilon}(\lambda^*)}{\mathbf{b}_{\epsilon}(\lambda^*)'\mathbf{b}_{\epsilon}(\lambda^*)}
=\rho_1,
\]
with $\mathbf{b}_{\epsilon}=\boldsymbol{\Xi}\mathbf{b}_{\tilde{x}}$.

\item \textbf{Uniqueness}: $\lambda^*\in  J_{++}$ is uniquely determined by the HT constraint $\rho_{\mathbf{b}_{\epsilon}}(\lambda)=\rho_1$.

\item \textbf{Global Minimum}:\label{gl_min} the I-SSA solution $\mathbf{b}_{\tilde{x}}(\lambda^*)$, $\lambda^*\in J_{++}$, is the global minimum of the MSE objective under HT and cointegration constraints.

\item \textbf{Generalized Spectral Decomposition}: Let $\boldsymbol{\nu}=(\nu_1,...,\nu_L)$ denote the generalized eigenvalues of the pencil $(\mathbf{A'A},\mathbf{Q})$ with corresponding generalized eigenvectors $\mathbf{v}_i$, $i=1,...,L$. Then
\begin{equation}\label{frequency-domain}
\mathbf{b}_{\tilde{x}}(\lambda^*)=\sum_{i=1}^L\frac{w_i}{1+\lambda^*\nu_i} \mathbf{v}_i,
\end{equation}
where $\mathbf{w}=(w_1,...,w_L)'=\mathbf{V'}\left(\mathbf{A}'\boldsymbol{\gamma}-\mu(\lambda^*)\mathbf{1}\right)$ is the `counterfactual' spectral factorization of the target.

\item \textbf{AS-Dilemma}: \label{as_dil}
The derivative of the MSE objective function $J(\lambda)$ with respect to $\lambda$, $\lambda \in J_{++}$, is
\begin{equation}\label{secu_as}
\frac{d J(\lambda)}{d \lambda} = - \lambda \frac{d F(\lambda)}{d \lambda},
\end{equation}
where $F(\lambda)=\mathbf{x(\lambda)'Qx(\lambda)}$ is the secular function with $\frac{d F(\lambda)}{d \lambda}<0$ (strict monotonicity).
This can be decomposed into an autocorrelation-driven term and a scale-driven term:
\begin{equation}
    \frac{d J(\lambda)}{d \lambda} = \underbrace{- \lambda \| \mathbf{b}_{\epsilon}(\lambda) \|^2 \frac{d \rho_{\mathbf{b}_{\epsilon}}(\lambda)}{d \lambda}}_{\text{Autocorrelation effect}} - \underbrace{\lambda \big( \rho_{\mathbf{b}_{\epsilon}}(\lambda) - \rho_1 \big) \frac{d \| \mathbf{b}_{\epsilon}(\lambda) \|^2}{d \lambda}}_{\text{Scale effect}}. \label{eq:decomposition_e}
\end{equation}
At the root $\lambda=\lambda^*(\rho_1)$ of the secular equation $F(\lambda^*)=0 $ we have  
\begin{equation}
    \left.\frac{d J(\lambda)}{d \lambda}\right|_{\lambda^*(\rho_1)} = \left.\underbrace{- \lambda^*(\rho_1) \| \mathbf{b}_{\epsilon}(\lambda^*(\rho_1)) \|^2 \frac{d \rho_{\mathbf{b}_{\epsilon}}(\lambda)}{d \lambda}}_{\text{Autocorrelation effect}}\right|_{\lambda^*(\rho_1)}  \label{eq:decomposition_ee},
\end{equation}
and
\begin{equation}\label{dilemma}
\left.\frac{d J(\lambda)}{d \rho_1}\right|_{\lambda=\lambda^*(\rho_1)}=-\lambda^*(\rho_1)(\mathbf{b}_{\epsilon}^*)'\mathbf{b}_{\epsilon}^*=\left\{\begin{array}{cc}<0&\rho_1<\rho_{MSE}\\
>0&\rho_1>\rho_{MSE}\end{array}\right.,
\end{equation}
with $\mathbf{b}_{\epsilon}^*:=\mathbf{b}_{\epsilon}(\lambda^*(\rho_1))$.

\end{enumerate}
\end{Theorem}
A proof is given in the Appendix. 

The strict monotonicity of $F(\lambda)$ (Assertion \ref{as_dil}) ensures the root $F(\lambda^*)=0$ is obtainable via efficient numerical optimization, e.g., bisection of $J_{++}$. Note that the theorem is essentially assumption-free since its proof relies only on generally valid algebraic identities, assuming $\boldsymbol{\Xi}$ to be invertible, which is ensured by $\xi_0=1$ on its main diagonal (even the standing assumption $\Gamma(0)\neq 0$ could be dropped). However, an important caveat concerns the interpretation of the HT constraint and the objective function. Both rely on finite MA approximations, justifying, e.g., \eqref{main_int_res}, which links the objective function (right-hand side) to the effective MSE (left-hand side). For these approximations to be valid, the predictor weights must decay sufficiently fast. This gap is closed by Theorem \ref{thm:decay_reformulated_uniform}, which provides the missing restrictions and assumptions. For now, we assume that the objective and the constraint admit their intended interpretations. A more thorough discussion is provided in Appendix \ref{disc}, contrasting I-SSA with SSA.

The following corollary `explains' the typical tipped or nose-shaped profile of the I-SSA predictor at the left boundary (early lags), as illustrated in Section \ref{applications}, as well as the decay of filter coefficients with increasing lag in typical applications\footnote{Excluding, e.g., excessive smoothing (or un-smoothing) by an extremal HT constraint $|\rho_1|\to\rho_{\max}(L)$.}, signifying stability of the difference equation despite the presence of non-stationary operators.

\begin{Corollary}
\label{prop:dirichlet_b0}
Assume $\rho_1\neq\rho_{MSE}$ (non-degenerate case). Then the solution of Theorem~\ref{prop_alt} can be written as a `time domain' discrete difference equation
\begin{equation}
    \big[ \boldsymbol{\Sigma}' \boldsymbol{\Sigma} + \lambda (\mathbf{M} - \rho_1 \mathbf{I}) \big] \mathbf{v} = \mathbf{d},
    \label{eq:foc_system_v_e}
\end{equation}
where $\mathbf{v} = \boldsymbol{\Xi} \mathbf{b}_{\tilde{x}}$ and $\mathbf{d} = \boldsymbol{\Sigma}' \boldsymbol{\gamma} - \mu(\lambda) (\boldsymbol{\Xi}')^{-1} \mathbf{1}$. This difference
equation has implicit Dirichlet and `extended' Dirichlet boundary constraints at the left and right boundaries. Specifically, the `ghost' values $b_{\tilde{x},-1},b_{\tilde{x},L}$  implied by these constraints satisfy
$$b_{\tilde{x},-1}=0\quad\textrm{and}\quad b_{\tilde{x},L}=-\sum_{j=0}^{L-1}\xi_{j+1}b_{\tilde{x},L-(j+1)},$$
where $\xi_k$ are the AR-inversion weights defined by $\boldsymbol{\Xi}^{-1}$. 
\end{Corollary}

The proof is in the Appendix \ref{disc} with a background discussion. Note that $\boldsymbol{\Xi}^{-1}$ always exists since $\xi_0=1$ on the main diagonal.\\ 

The next theorem establishes asymptotic convergence of the I-SSA solution as $L\to\infty$ and shows that the I-SSA coefficients decay fast enough to justify finite MA approximations under fairly general assumptions on target and Wold weights. For this purpose, let $\ell^2$ denote the Hilbert space of square-summable real sequences
$\mathbf{x}=(x_0,x_1,\dots)'$ with $\|\mathbf{x}\|_2^2=\sum_{k\ge 0}x_k^2<\infty$.
Finite-length vectors $\mathbf{x}\in\mathbb{R}^L$ are identified with elements of
$\ell^2$ by zero-padding, and an infinite matrix (operator) is called bounded on
$\ell^2$ if $\|\mathbf{A}\|:=\sup_{\mathbf{x}\neq 0}\|\mathbf{A}\mathbf{x}\|_2/\|\mathbf{x}\|_2<\infty$.

\begin{Theorem}[Asymptotic I-SSA and Interpretability]\label{thm:decay_reformulated_uniform} 
Under the standing assumptions $\Gamma(0)\neq 0$ (preservation of integration order, together with invertibility of the Wold decomposition, see below) and $\rho_1\neq\rho_{MSE}$ (non-degeneracy), and for a given target $\hat{\boldsymbol{\gamma}}_{MSE}$ and Wold decomposition $\boldsymbol{\xi}$, assume there exists $L_0>0$ such that $|\rho_1|<\rho_{\max}(L_0)$, the largest eigenvalue of $\mathbf{M}(L_0)$. For $L>L_0$, consider the I-SSA predictor $\mathbf{b}_{\tilde{x}}^* :=\mathbf{b}_{\tilde{x}}(\lambda^*)$ from Theorem \ref{prop_alt}, with weights $b^*_k$, $k=0,\dots,L-1$.
Suppose there exist $s>1$ and constants $C_\xi, C_\gamma > 0$, independent of $L$, such that:
\begin{enumerate}
\item[(A1)] $|\xi_j| \le C_\xi j^{-s}$ for $j=0,\dots,L-1$ (polynomial decay of MA weights), and $\xi(z)\neq 0$ for $|z|\leq 1$ (invertibility);
\item[(A2)] $|\gamma_{MSE,k}| \le C_\gamma k^{-s}$ for $k=0,\dots,L-1$ (polynomial decay of the target);
\item[(A3)] The sequence $\lambda^*(L)$ of optimal Lagrange parameters in Theorem \ref{prop_alt} is uniformly bounded away from the boundary of $J_{++}(L)$, i.e., $\inf_L \text{dist}(\lambda^*(L), \partial J_{++}(L)) > 0$.
\end{enumerate}
Then, as $L \to \infty$:
\begin{enumerate}

\item[(i)] The boundaries of the strict positivity domain, $\lambda_{\min}(L)$ and $\lambda_{\max}(L)$, converge to finite limits 
\begin{eqnarray*}
\lambda_{\min}(L) &\to& -\frac{2}{(1-\rho_1)^2}<0\\
\lambda_{\max}(L) &\to& \frac{1}{4(1 + \rho_1)}>0.
\end{eqnarray*}

\item[(ii)] There is a constant $C^*$, depending only on $(s, C_\xi, C_\gamma, \rho_1)$ and the uniform separation bound, but not on $L$, such that for all $L$ and all $k=0,\dots,L-1$:
\[
|b^*_k - \gamma_{MSE,k}| \le C^* k^{-s}
\qquad\text{and hence}\qquad
|b^*_k| \le (C_\gamma + C^*)k^{-s}.
\]
In particular, the weights $\mathbf{b}_{\tilde{x}}^* = (b^*_0,\dots,b^*_{L-1})'$ decay polynomially at rate $s$, uniformly in $L$.

\item[(iii)] The sequence of optimal Lagrange multipliers $\lambda^*(L)$ converges to a unique limit $\lambda^*_\infty$ strictly inside $J_{++}(\infty)$, satisfying the asymptotic secular equation 
$$
\mathbf{b}_\infty(\lambda_\infty^*)'\mathbf Q_{\infty}\mathbf{b}_\infty(\lambda_\infty^*)=0,~~\textrm{where}~~\mathbf Q_{\infty}=\boldsymbol\Xi_{\infty}'(\mathbf M_{\infty}-\rho_1\mathbf I_{\infty})\boldsymbol\Xi_{\infty},
$$
and
\begin{eqnarray}
\mathbf{b}_\infty(\lambda_\infty^*)&=&\mathbf d_\infty(\lambda_\infty^*)+\hat{\boldsymbol{\gamma}}_{MSE\infty}\label{exp_dev_mse}\\
\mathbf d_\infty(\lambda_\infty^*)&=&-\lambda_\infty^*\boldsymbol\Xi_\infty^{-1}\boldsymbol\Delta_\infty\,\mathbf K_\infty(\lambda_\infty^*)^{-1}\mathbf f_\infty \nonumber\\
\mathbf{f}_\infty&=&\boldsymbol\Delta_\infty'(\mathbf M_\infty-\rho_1\mathbf I_\infty)\boldsymbol\Xi_\infty\hat{\boldsymbol{\gamma}}_{MSE\infty}\nonumber\\
\mathbf{K}_\infty(\lambda_\infty^*)&=&\mathbf{I}_\infty+\lambda_\infty^*\boldsymbol\Delta_\infty'(\mathbf M_\infty-\rho_1\mathbf I_\infty)\boldsymbol{\Delta}_\infty\label{Kinf},
\end{eqnarray}
where $\boldsymbol\Xi_\infty$, $\boldsymbol\Delta_\infty$, $\mathbf I_\infty$, $\mathbf M_{\infty}$ are the one-sided infinite versions of $\boldsymbol\Xi_L$, $\boldsymbol\Delta_L$, $\mathbf I_L$, $\mathbf M_{L}$, and all operators are bounded on $\ell^2$; in particular, the modified Hessian $K_\infty(\lambda_\infty^*)$ is also invertible. Finite-length solutions, embedded in $\ell^2$ by zero-padding, converge in $\ell^2$-norm to the infinite-length solution.
\end{enumerate}
\end{Theorem}

The proof is in the appendix, together with a background discussion that covers, among other things, the modified and bounded Hessian $\mathbf{K}_L(\lambda)$, the assumptions, and their verification in applications.
 The last Corollary specifies the efficient AS frontier of I-SSA.

\begin{Corollary}\label{eff-as-front}
Let the assumptions of Theorem \ref{thm:decay_reformulated_uniform} hold. Then, for $\rho_1>\rho_{MSE}$, the optimized MSE objective of the I-SSA is strictly increasing in $\rho_1$, and no linear predictor of the same (or smaller) length $L$ can both reduce MSE (in levels) and increase $\rho_1$ in differences: for Gaussian processes with vanishing drift, I-SSA is maximally monotone, in the sense that for given MSE tracking accuracy (on non-stationary levels), no other linear predictor of that (or smaller) length achieves a smaller TP rate—defining the `right branch' of the efficient frontier ($\rho_1>\rho_{MSE}$). The reverse holds for $\rho_1<\rho_{MSE}$: no linear predictor of the same (or smaller) length can reduce both MSE and $\rho_1$, and I-SSA is minimally monotone, in the sense that for given MSE tracking accuracy (on non-stationary levels), no other linear predictor of that (or smaller) length achieves a larger TP rate—defining the `left branch' of the efficient frontier ($\rho_1<\rho_{MSE}$).
\end{Corollary}

The proof is in the Appendix. Conditioning on Gaussianity is generally not a limitation, since Eq.~\eqref{ht} linking HT and first-order ACF is fairly robust to departures from Gaussianity—particularly under accentuated smoothness, where smoothing tips `naturally' toward Gaussianity via the central limit theorem (Wildi 2026b). In case of a non-vanishing drift, the Corollary addresses the crossover rate about the linear trend. In applications, however, the drift, i.e., the mean of stationary first differences, is often negligible relative to the stationary growth-dynamics above and below the drift, so that the (mean) distance between consecutive zero-crossings (TPs) is identical or similar to that between consecutive mean-crossings (crossover points), effectively addressed by the HT. We do not treat the case of asymmetry, with crossings of a line markedly above or below the mean, since this would break the symmetries induced by the first-order ACF. In any case, I-SSA maximizes smoothness in terms of the first-order ACF of stationary differences, retaining interpretability and affecting the crossing rate effectively at any  level.

\section{Applications: Real-Time Business-Cycle Analysis and I-SSA Trend}\label{applications}

We first apply I-SSA to HP, customizing the filter on the novel AS frontier. The data is the US INDPRO (industrial production) series, and the target is the two-sided HP filter; we compute an optimal MSE predictor on levels replicating the TP rate of the one-sided HP in differences. We expect the I-SSA nowcast to outperform the classic concurrent HP in tracking the two-sided HP while matching the latter's smoothness. In the second example, the target is the original data instead of the two-sided HP, illustrating the novel I-SSA trend (and smoothing) concept. We apply the design to a random-walk process and to INDPRO, benchmarking against HP and the {boosted} HP (Mei, Phillips and Shi, 2024), imposing the respective TP rates on I-SSA. For identical smoothness, I-SSA outperforms in terms of MSE, specifically at peaks and dips of the series, where the nowcast is faster (advanced) without over- or undershooting, which makes it well suited to real-time assessment.

\subsection{HP Customization: Maximal Monotone  HP Trend-Nowcast}\label{max_mon_issa}

\begin{figure}[H]\begin{center}\includegraphics[height=3in, width=4in]{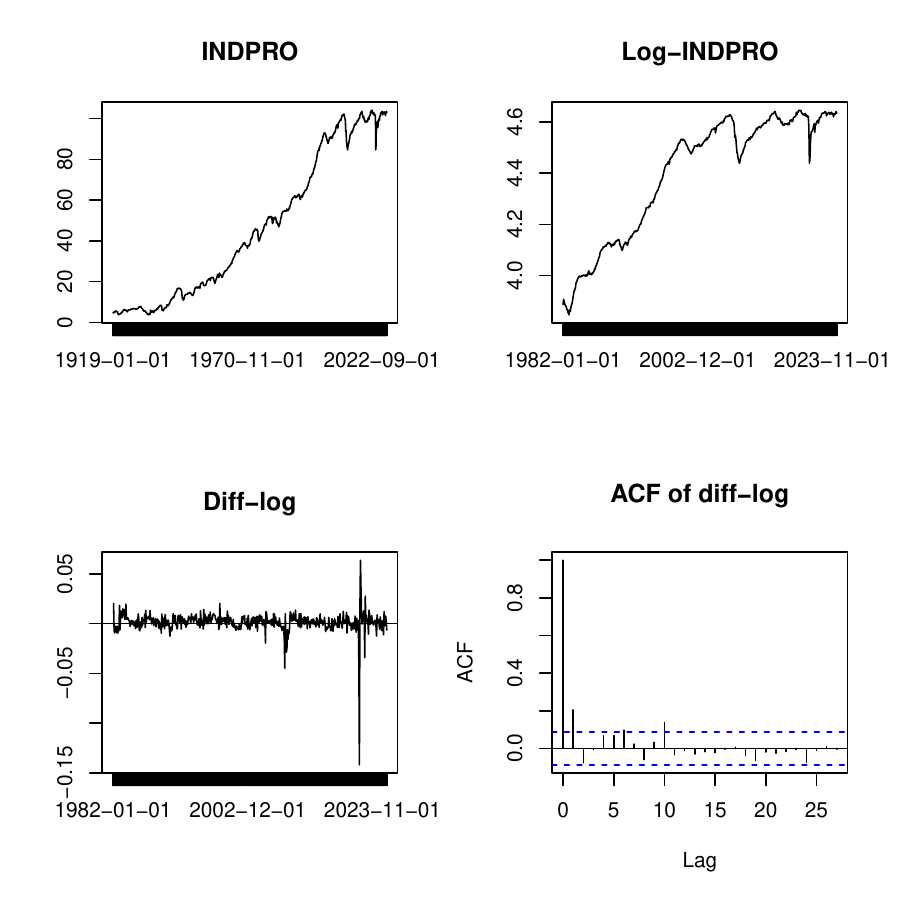}\caption{INDPRO original entire sample (top left), log-transformed INDPRO from 1982 onwards (top right), log-differenced data (bottom left) and ACF of log-differenced series (bottom right).\label{indpro}}\end{center}\end{figure}

\begin{figure}[H]\begin{center}\includegraphics[height=5in, width=5in]{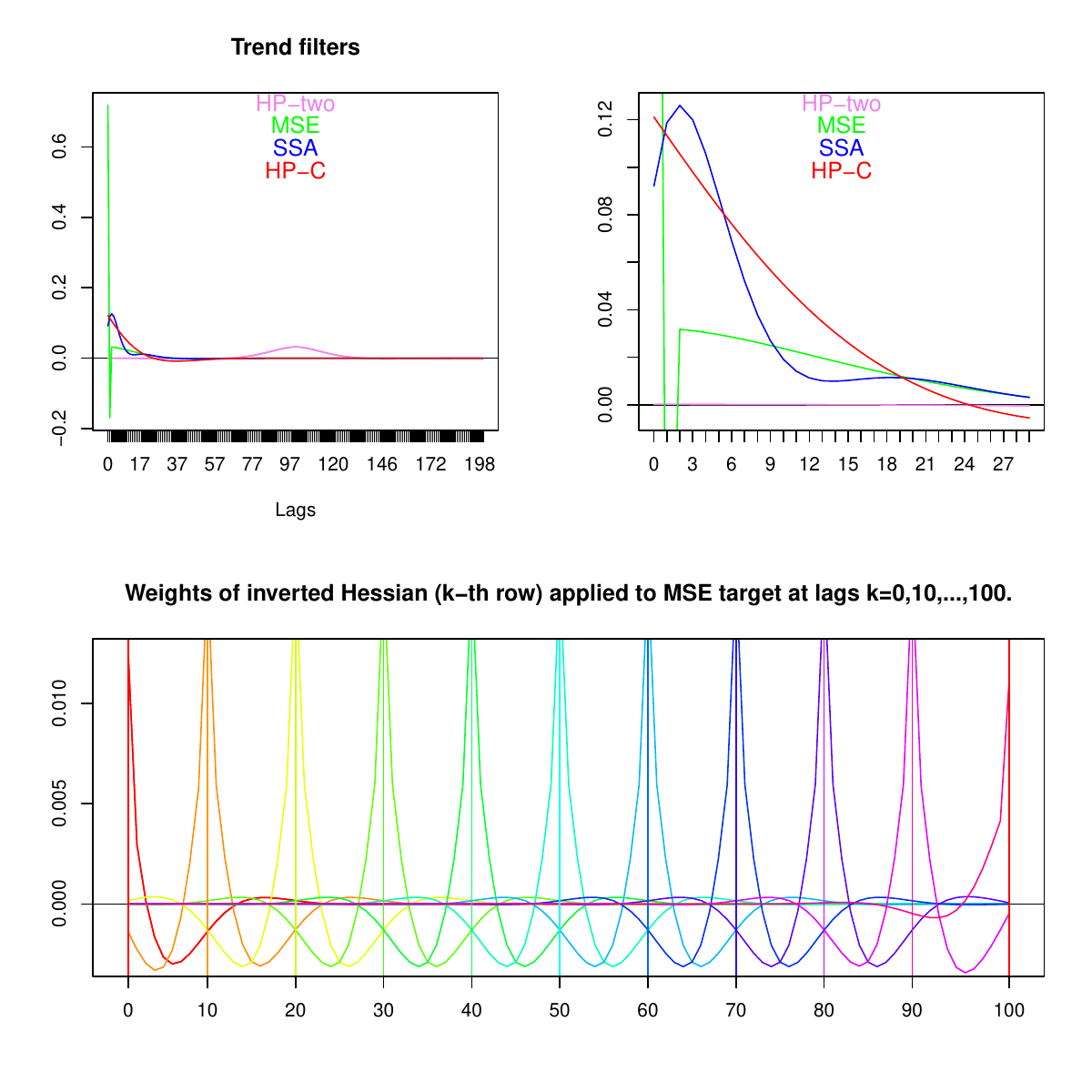}\caption{Top panels: two-sided HP target (violet), HP-C (red), MSE (green), and I(1)-SSA (blue); the latter two are based on an ARIMA(1,1,0) specification. All filter coefficients sum to one (cointegration constraint). The left panel shows all lags; the right panel shows the first thirty, with the vertical axis truncated. I-SSA exhibits the typical nose shape, peaking at lag $k=2$ in this example, consistent with Dirichlet boundary constraints.\
Bottom panel: weights assigned by $\mathbf{H}(\lambda^*)^{-1}$, with $\lambda^*=-102.57$, to the MSE target at lags $k=0,10,\ldots,L-1$. The $k+1$-th row of the inverted Hessian is concentrated at lag $k$, with symmetric decay in the middle ($k=50$) and asymmetric decay towards the boundaries $k=0, k=L-1=100$. \label{indpro_filt}}\end{center}\end{figure}

\begin{figure}[H]\begin{center}\includegraphics[height=3in, width=4in]{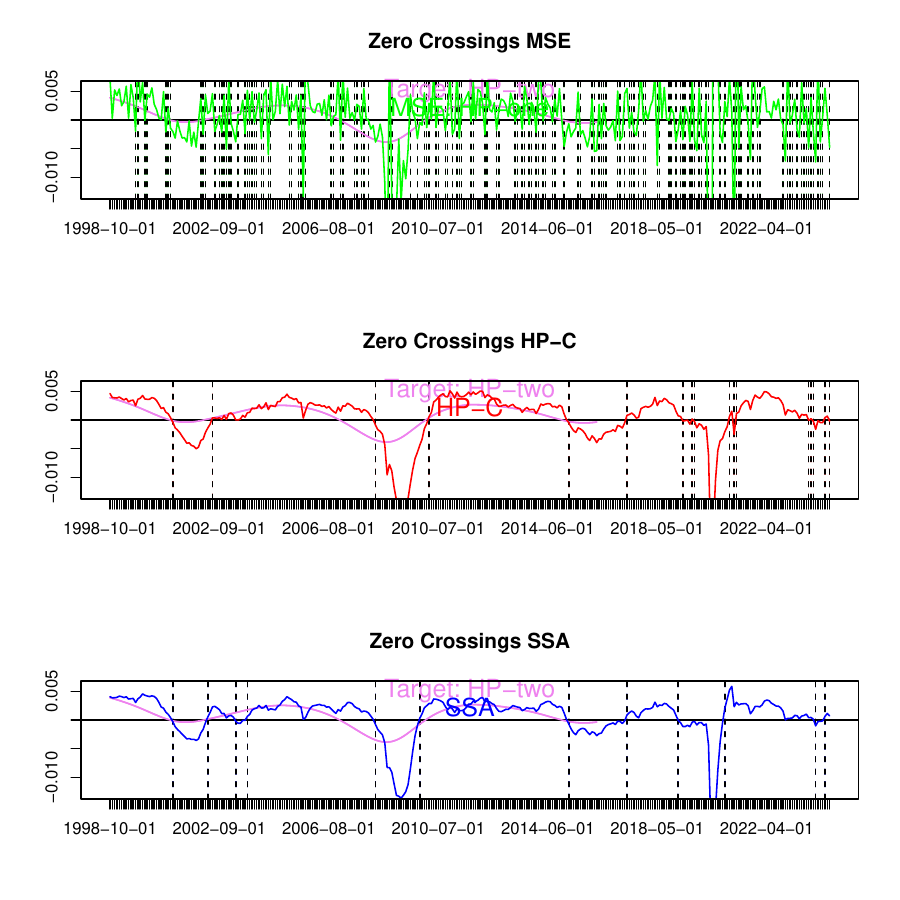}\caption{Differenced filter outputs: two-sided target (violet), MSE (green), HP-C (red) and I(1)-SSA (blue) with zero-crossings of the designs marked by vertical lines. MSE (top), HP-C (middle) and I(1)-SSA (bottom).\label{indpro_diff_filt_out}}\end{center}\end{figure}

We apply the I(1)-SSA predictor $\mathbf{b}_{\tilde{x}}=\Gamma(0)\mathbf{e}_1+\mathbf{B}\tilde{\mathbf{b}}$, where $\tilde{\mathbf{b}}$ is determined by Equation \eqref{eq:b_lambda}, to the {monthly} US-industrial production index INDPRO\footnote{Board of Governors of the Federal Reserve System (US), Industrial Production: Total Index [INDPRO], retrieved from FRED, Federal Reserve Bank of St. Louis; https://fred.stlouisfed.org/series/INDPRO, October 31, 2024.},  as illustrated in Fig.\eqref{indpro}. The target is specified using a two-sided HP filter with parameter $\lambda=$14400 for monthly data, assuming a nowcast (i.e., $\delta=0$) and $L=201$. We benchmark the I(1)-SSA approach against the MSE design, positing that the differenced data adheres to an AR(1)-model (results are robust against alternative MA or ARMA specifications). This assumption is supported by the weak, somewhat unsystematic, yet slightly persistent ACF pattern observed (see the bottom right panel of the figure). The AR(1) model is utilized to derive the weights $\boldsymbol{\xi}$ in the Wold decomposition of the (differenced) time series. For a comprehensive analysis, we also incorporate the classic one-sided HP concurrent filter, referred to as HP-C, as an additional benchmark, see McElroy (2008) and Cornea-Madeira (2017) for background. 
All (trend-) filters are displayed in Fig.~\eqref{indpro_filt}. The two-sided (truncated) HP filter, with coefficients $\gamma_k$ for $|k| \leq (L-1)/2$, is symmetrically centered at lag $k = (L-1)/2$ (violet line, left panel). The nowcast presumes that the differenced data follow the AR(1) specification above, with performance improving relative to the WN hypothesis. The classic MSE predictor (green) allocates most weight to the most recent observations, owing to the smooth trajectory of the associated ARIMA(1,1,0) process, which requires little additional noise suppression by the nowcast filter. This illustrates that a unilateral focus on MSE may yield `noisy' predictors, as shown in Fig.~\eqref{indpro_diff_filt_out}. 

The first-order ACF of the differenced MSE nowcast is $\rho_{MSE}=0.105$, corresponding to $HT=2.143$, barely above white noise. In contrast, the first-order ACF of the HP-C filter, $\rho_{HP-C}=0.954$, is substantially higher, with $HT_{HP-C}=10.264$, and we therefore set $ht_1=HT_{HP-C}$ in the HT constraint. The resulting optimal Lagrange multiplier is $\lambda^*=-102.573$; its negative sign and magnitude indicate a markedly stronger smoothing effect than the MSE benchmark, as desired. The effective first-order ACF $\rho^*=0.954$ of $\mathbf{b}_{\epsilon}(\lambda^*)=\boldsymbol{\Xi}\left(\Gamma(0)\mathbf{e}_1+\mathbf{B}\tilde{\mathbf{b}}(\lambda^*)\right)$ matches the HT constraint, confirming convergence of the numerical optimization. In summary, I(1)-SSA replicates the smoothness (TP rate) of HP-C while improving tracking through a lower MSE on the AS-frontier.

The bottom panel of Fig.~\ref{indpro_filt} shows the weights of $\mathbf{H}(\lambda^*)^{-1}$ assigned to the MSE target for $\lambda^*=-102.573$ and at lags $k=0,10,20,\ldots,L-1=100$ (multiples in ten) noting that $b_{\tilde{x},k-1}^*=\mathbf{H}(\lambda^*)_{k\cdot}^{-1}\left(\mathbf{A}'\boldsymbol{\gamma}-\mu(\lambda^*)\mathbf{1}\right)$. The figure shows that the $k$-th row $\mathbf{H}(\lambda^*)_{k\cdot}^{-1}$ of the inverted Hessian is concentrated at lag $k$, with symmetric decay in the middle ($k=50$) and asymmetric decay towards the boundaries $k=0$ and $k=L-1$.

Assumption (A3) of Theorem~\ref{thm:decay_reformulated_uniform} concerns the sequence $\{\lambda^*(L)\}_L$. In our application it holds numerically: $\lambda^*(L)\simeq -102.573$ is essentially constant in $L$ for $L>100$ and well away from the finite-sample boundaries $(-944.87,0.13)$ of $J_{++}$ as well as from the asymptotic boundaries $(-925.92, 0.128)$ given by the first assertion of Theorem \ref{thm:decay_reformulated_uniform}. The proximity of finite and asymptotic boundaries confirms near convergence. Furthermore, the extreme eigenvalues of the `stationary' Hessian $\mathbf K(L)$ stabilize rapidly at $\lambda_{\min}=1.685>0$ and $\lambda_{\max}=474.7$, confirming invertibility and boundedness, respectively (see the accompanying I-SSA package). This is consistent with the fast decay of the I-SSA predictor weights ---implied by the theorem--- in the top two panels of the figure, and supports the validity of the finite MA approximation underlying \eqref{main_int_res}.
\\

\begin{figure}[H]\begin{center}\includegraphics[height=2.5in, width=6in]{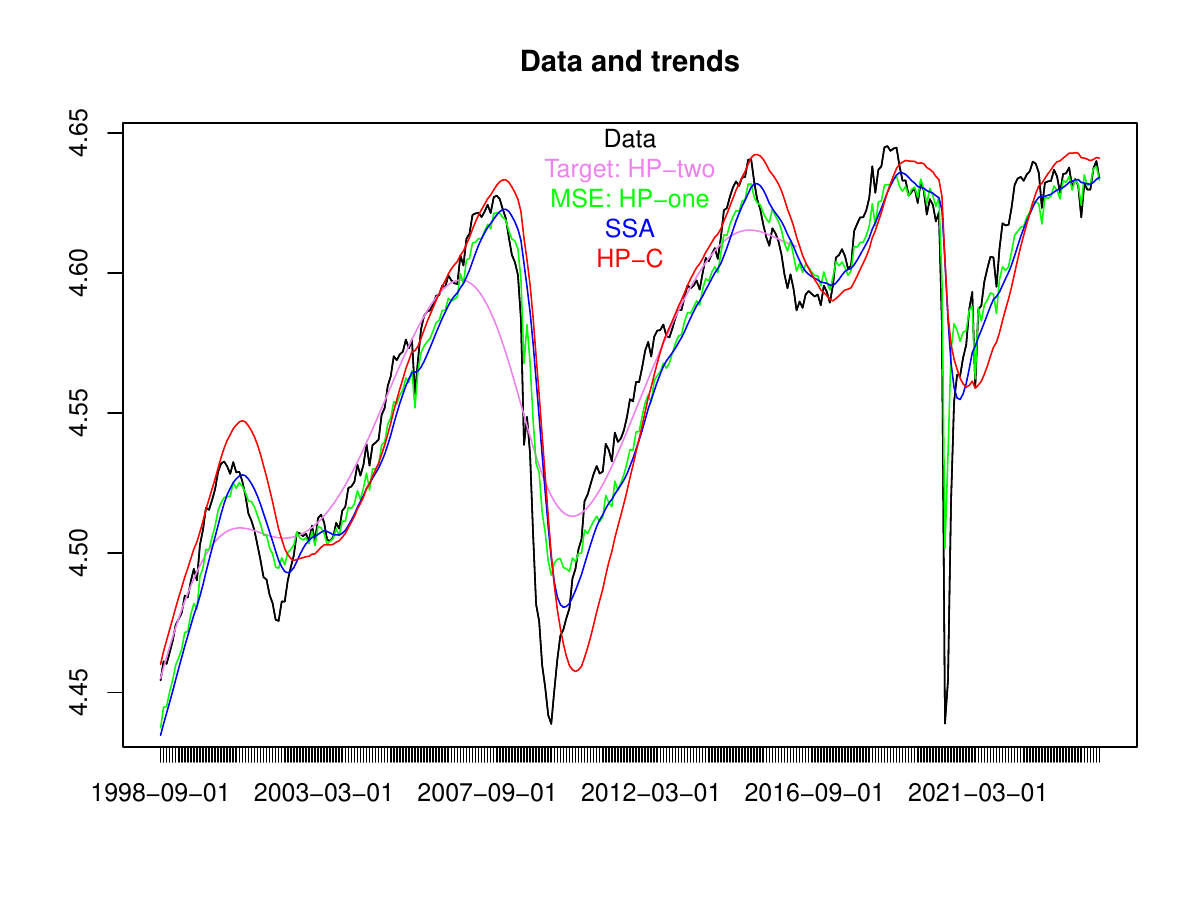}\caption{(Log-) INDPRO (black), two-sided HP target (violet), MSE (green), HP-C (red), and I(1)-SSA (blue). HP-C tracks the symmetric HP better in the middle of longer up- or downturns but is prone to over- and undershooting relative to the HP target at peaks and troughs, where it also lags I-SSA, yielding a net MSE loss.
.\label{indpro_filt_out}}\end{center}\end{figure}

The non-stationary (logarithmic) INDPRO series (black), along with the filter outputs, is illustrated in Fig.\eqref{indpro_filt_out}, covering a historical timeframe spanning the last three recessions, during which the index lost previous momentum. The acausal two-sided HP filter (violet) cannot extend to the sample's actual endpoint. I(1)-SSA (blue) is characterized as maximal monotone for given MSE tracking, or as maximal tracking for a given TP rate. The figure shows that HP-C (red) tracks the symmetric HP better in the middle of longer up- or downturns but is prone to over- and undershooting at peaks and troughs, where it also lags I-SSA, yielding a net MSE loss quantified in Table \eqref{table_indpro}. If the series were effectively I(2), with up- or down-swings of unbounded length asymptotically, HP-C would outperform I-SSA in MSE; but increasing $d=1$ to $d=2$ would then re-establish I-SSA's outperformance, sitting on the adjusted frontier.
 
Table \eqref{table_indpro} summarizes sample performances in MSE, referenced against the two-sided HP filter, and empirical HTs of the differenced predictors. HP-C exhibits the lowest accuracy, with empirical MSE exceeding twice the optimal benchmark. I(1)-SSA, in between, confirms the anticipated MSE hierarchy. Exchanging MSE for TP rate along the efficient AS frontier incurs minimal tracking loss, with I(1)-SSA's TP rate seemingly surpassing HP-C's\footnote{The discrepancy in sample HTs between I(1)-SSA and HP-C, despite identical first-order ACFs, may stem from the limited number of zero-crossings, affecting the reliability of sample estimates. Both empirical holding times can be matched with arbitrary fidelity for sufficiently long simulated samples within our I-SSA package.}. This trade-off is justified by the analysis's specific objectives: given its prevalence in applications, the smoothness offered by HP-C appears more desirable than pure MSE performance—otherwise a rational analyst would resort to the latter. We contend that I(1)-SSA can enhance accuracy while making up in smoothness. For this target and series, a stringent focus on MSE-accuracy, as in classic signal extraction, cannot be achieved without substantial TP-rate smoothness losses, as evidenced by the MSE predictor's performance.

\begin{table}[ht]
\centering
\begin{tabular}{rrrr}
  \hline
 & MSE & SSA & HP-C \\ 
  \hline
Sample mean square error & 0.00024 & 0.00037 & 0.00050 \\ 
  Sample holding time & 2.30597 & 14.04545 & 12.87500 \\ 
   \hline
\end{tabular}
\caption{Sample performances of MSE, HP-C and I(1)-SSA predictors: mean-square error relative to the acausal two-sided HP (first row) and empirical holding times of the stationary differenced predictors (second row).} 
\label{table_indpro}
\end{table}To underscore this point, Fig.\eqref{indpro_diff_filt_out} visualizes the HT discrepancy by juxtaposing the outputs of the {differenced} filters with the zero-crossings of each design, marked by corresponding vertical lines.

Our findings suggest that smoothness can be substantially improved over the MSE benchmark without overtly sacrificing accuracy. More precisely, the marginal cost of increased TP rate, in terms of increased MSE, is characterized by the AS-dilemma \eqref{dilemma} (using \ref{ht} to link ACF and HT). Fig.\eqref{indpro_diff_filt_out} further shows that the last three recession episodes are effectively tracked through the first differences of the smooth HP-C and I-SSA nowcasts, with minimal `false alarms' during extended expansions, noting that TP dating by I-SSA is either coincident or advanced. In short, the proposed I(1)-SSA framework reconciles first-difference analysis (zero-crossings associated with TPs) and optimal level tracking within a single, logically consistent, statistically efficient, and interpretable design. It addresses MSE and HT simultaneously, as well as target correlation and sign accuracy through the stationary proxy predictor on the right of \eqref{main_int_res}.
 \\


\subsection{I-SSA Trend}

In contrast to the previous section, the target here is the original series (identity filter $\mathbf{I}$ instead of HP): I-SSA tracks the level of the data as closely as possible subject to an HT constraint on first differences. This resembles classical Whittaker-Henderson smoothing, though the penalty term is the TP rate rather than curvature or mean-squared second-order differences.

For illustration, we rely on a random-walk process---$d=1$ and $\boldsymbol{\Xi}=\mathbf{I}$---covering both the mid-sample acausal {symmetric} trend smoother and the real-time causal {one-sided} trend nowcast (similar findings apply to $d>1$ or $\boldsymbol{\Xi}\neq \mathbf{I}$, not shown). The design is then applied `as is' to the INDPRO series used previously. I-SSA can accommodate a formulated model of the data through $d$ and $\boldsymbol{\Xi}$, but the random-walk setting conforms well with typical economic prediction problems (see the accompanying R-package for background), justifying a `model-free' I-SSA trend based on these default settings.\\

\textbf{1) Random-Walk}:

We  compare I-SSA to HP and I-SSA-B to the boosted variant, abbreviated, HP-B, as implemented in the R-package \emph{MacroFilters}. For the random walk,  the HP-B algorithm selected  100 iterations for $\lambda=129600$ (monthly setting), based on BIC. Empirical performances are based on a single very long sample, of length $10^7$, and we compare the I-SSA two-sided smoothers and one-sided nowcasts, each replicating the respective true HTs of HP and HP-B on white noise (differenced random-walk), namely $59.51$  (HP-two), $ 7.651$ (HP-one), $30.28$ (HP-B-two) and $ 5.633$ (HP-B-one).

Performance is reported in Tables \ref{table_issa_acausal_bhp} (two-sided acausal smoothers) and \ref{table_issa_causal_bhp} (one-sided causal nowcasts). We verify throughout that the sample HT converge to the expected HT. HP-B designs are more adaptive (smaller HT and MSE, larger curvature) than the original HP, as expected. I-SSA-two reduces MSE by roughly $50\%$ relative to HP-two for identical HT and nearly identical curvature; I-SSA-B-two reduces MSE by roughly $7\%$ for identical HT and virtually identical curvature---in both cases a worthwhile trade-off. I-SSA-two has slightly smaller ZC-distance variance than HP (marginally more regular), while I-SSA-B-two is marginally more irregular. HP's twice-as-large HT corresponds to roughly a four-times-larger variance than HP-B's (similarly for I-SSA).

Table \ref{table_issa_causal_bhp} shows that two-sided designs exert much stronger smoothing, with HP's HT nearly 8 times larger than its nowcast's, yielding an MSE that even exceeds the nowcast's---hurting a priori expectations, since the two-sided design `knows' the future. In contrast, HP-B-two has smaller MSE than its nowcast, in agreement with expectations. However, I-SSA-B-one nowcast has substantially smaller MSE still. In general, I-SSA nowcasts reduce MSE by roughly $50\%$ for identical HT but at a cost of roughly $25\%$ larger curvature---a worthwhile trade-off, particularly when tracking matters. Marginal costs, derived from Eq.\eqref{dilemma}, quantify a 8.796$\%$ increase in MSE for a $1\%$ increase in $\rho_1$, equivalently a 1.569$\%$ MSE increase per $1\%$ increase in HT, for I-SSA-one; for I-SSA-B-one the corresponding losses are 5.46$\%$ and 1.899$\%$, allowing for more informed trade-offs on the frontier than just replicating the benchmarks' HTs (not pursued here). Interestingly, I-SSA-one outperforms HP-B-one on all three measures. 

\begin{figure}[H]\begin{center}\includegraphics[height=5in, width=5in]{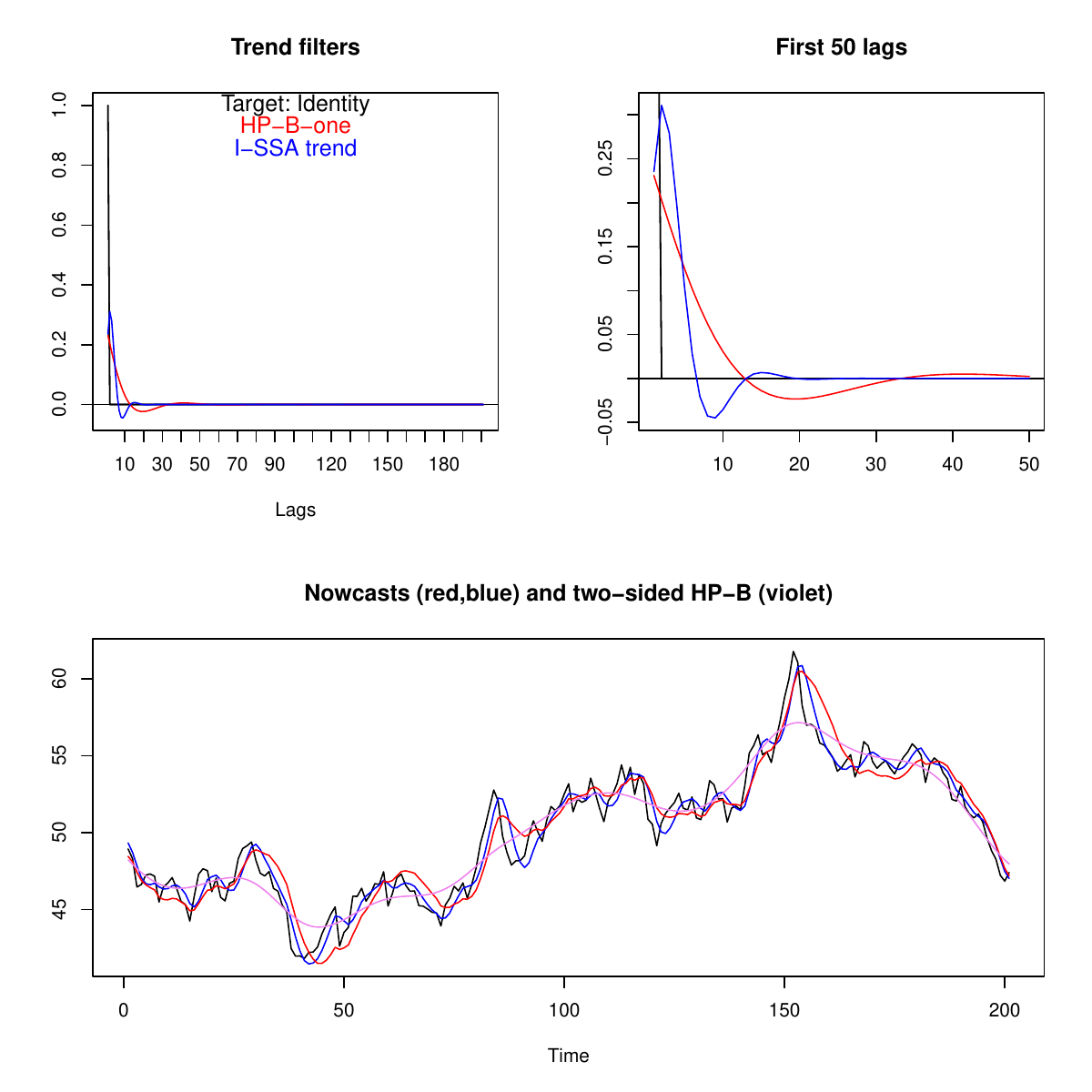}\caption{Causal one-sided HP-B and I-SSA trend: filter coefficients (top) and smoothed series (bottom). Top left: all coefficients; top right: magnified view. One-sided I-SSA (blue), two-sided HP-B (violet, bottom panel), one-sided HP-B (red), and identity target filter (black). I-SSA replicates the HT of one-sided HP-B, so it must outperform HP in MSE (efficient frontier). I-SSA weights decay faster to zero (higher adaptivity, smaller retardation), with more pronounced side lobes to preserve the HT constraint. The filtered series (bottom) show that I-SSA (blue) tracks the timing and level of peaks and dips better than one-sided HP-B (red), at a strictly identical TP rate.\label{issa_nowcast_trend_bhp}}\end{center}\end{figure}

In general, the ZC-distance variance of I-SSA nowcasts is substantially smaller than that of HP and HP-B, indicating more evenly spaced TPs. This can be explained by the filter profiles in Fig.~\ref{issa_nowcast_trend_bhp} (we highlight HP-B; HP is similar): faster decay (greater adaptivity, less phase-lag and retardation, i.e., systematic left-shift) and a stronger side lobe (needed to preserve equal HT) let one-sided I-SSA track peaks and dips of the series faster and more tightly, at an identical TP rate (bottom panel), with less hesitation, thereby avoiding spurious (`noisy') TPs around them; all are useful features in a real-time dating exercise. In contrast, the curvature concept underlying HP, and thus HP-B, imposes less inflection, possibly conflicting with the irregular dynamics of random-walk-like series, i.e., series lacking overt smoothness. This places `true' TPs farther apart while clustering spurious ones around them (larger variance, more irregular dating), especially since HP's implicit ARIMA(0,2,2) representation (McElroy, 2008) has a TP rate that vanishes asymptotically. Given this structural unit-root misspecification implied by HP's minimal-curvature concept, together with the measured MSE loss, we may prefer the more regular and evenly spaced I-SSA TP dating for series lacking the strong intrinsic smoothness entailed by I(2) processes. For the latter, we recommend increasing the integration order to $d=2$ in I-SSA.

Evidently, our comparisons are biased against HP,HP-B, since I-SSA relies on knowledge of the true model. The results above thus indicate what is left on the table by ignoring the true DGP, assuming an identical TP rate across indicators. \\

\begin{table}[ht]
\centering
\begin{tabular}{rrrrr}
  \hline
 & HP-two & I-SSA-two & HP-B-two & I-SSA-B-two \\ 
  \hline
MSE & 5.94693 & 2.83754 & 1.52707 & 1.43873 \\ 
  Empirical HT & 59.51446 & 59.51729 & 30.30022 & 30.28829 \\ 
  Curvature & 0.00820 & 0.00868 & 0.02376 & 0.02392 \\ 
  Variance of ZC distance & 1870.07671 & 1708.17217 & 435.67544 & 439.23862 \\ 
   \hline
\end{tabular}
\caption{Acausal two-sided smoothers: sample performance of HP-two, I-SSA-two (replicating HP's HT), and HP-B-two, I-SSA-B-two (replicating HP-B's HT) for a single, very long realization of the random walk. The sample HT replicates the expected HT up to negligible sample variation. HP-B is more adaptive than HP, with smaller MSE and HT but larger curvature. I-SSA-two reduces MSE by roughly $50\%$ for identical HT and nearly identical curvature; I-SSA-B-two reduces MSE by roughly $7\%$ for identical HT and virtually identical curvature. I-SSA-two has slightly smaller ZC-distance variance than HP (marginally more regular); I-SSA-B-two is marginally more irregular. HP's twice-as-large HT corresponds to roughly a four-times-larger variance.
} 
\label{table_issa_acausal_bhp}
\end{table}
\begin{table}[ht]
\centering
\begin{tabular}{rrrrrrr}
  \hline
 & HP-two & HP-one & I-SSA-one & HP-B-two & HP-B-one & I-SSA-B-one \\ 
  \hline
MSE & 5.94693 & 3.39404 & 1.50722 & 1.44740 & 1.62800 & 0.88949 \\ 
  Empirical HT & 59.51482 & 7.65129 & 7.65620 & 30.30282 & 5.62714 & 5.63246 \\ 
  Curvature & 0.00820 & 0.12404 & 0.18577 & 0.02376 & 0.24184 & 0.29457 \\ 
  Variance & 1870.12055 & 96.15100 & 31.87054 & 435.72189 & 34.90346 & 16.98246 \\ 
   \hline
\end{tabular}
\caption{Causal one-sided smoothers (two-sided HP, HP-B included for reference, first and fourth columns): sample performance of HP-one, I-SSA-one (replicating HP's HT), and HP-B-one, I-SSA-B-one (replicating HP-B's HT) for a single, very long realization of the random walk. The sample HT replicates the expected HT up to negligible sample variation. Two-sided designs exert much stronger smoothing, with HP's HT nearly 8 times larger than its nowcast's, resulting in an MSE exceeding the nowcast's. In contrast, HP-B-two has smaller MSE than its nowcast, while I-SSA-B-one has substantially smaller MSE still. In general, I-SSA reduces MSE by roughly $50\%$ for identical HTs but larger curvature. Interestingly, I-SSA-one outperforms HP-B-one on all three measures. The ZC-distance variance of the I-SSA nowcasts is substantially smaller than that of the HP and HP-B nowcasts, indicating markedly greater regularity of real-time TP dating (discussed in the main text).
} 
\label{table_issa_causal_bhp}
\end{table}

\textbf{2) INDPRO}:

We apply the above filters `as is,' in their default settings, to INDPRO, assessing the effects of potential model misspecification on I-SSA. We focus on HP-B, a more challenging benchmark for I-SSA, and on nowcasts (additional comparisons are offered in the accompanying R-package). When applied to white noise, the HT of the one-sided HP-B is $HT\approx 5.63$ (see Table \ref{table_issa_causal_bhp}). However, differences of log-INDPRO are weakly autocorrelated (Fig.~\ref{indpro}), so we also provide a corrected, slightly larger value, $HT\approx 7.43$, obtained by applying HP-B to an AR(1) with $a_1=0.3$ from the previous section. This correction affects the optimization only through $ht_1$ in the constraint; we still assume $\boldsymbol{\Xi}=\mathbf{I}$ (white noise).

Table \ref{table_hpb_indpro} reports empirical MSE, HT, and Curvature statistics for the data in Section \ref{max_mon_issa}. The HT of HP-B lies between the two I-SSA values, confirming weak positive dependence in the series, though the AR(1) correction seems slightly conservative (excessive smoothness). In any case, both I-SSA filters outperform HP-B, reducing MSE by roughly $25\%$ and $50\%$, respectively. Interestingly, the smoother I-SSA also outperforms in terms of Curvature. Table \ref{table_hpb_indpro_true} reports the true (expected) MSE (based on optimized objective functions) and HT for I-SSA, under the assumption of the RW model. The slight positive bias in the empirical numbers confirms the misspecification: if the differenced data were truly white noise, the smoothed series would exhibit slightly more zero crossings and a slightly smaller MSE.

\begin{figure}[H]\begin{center}\includegraphics[height=6in, width=6in]{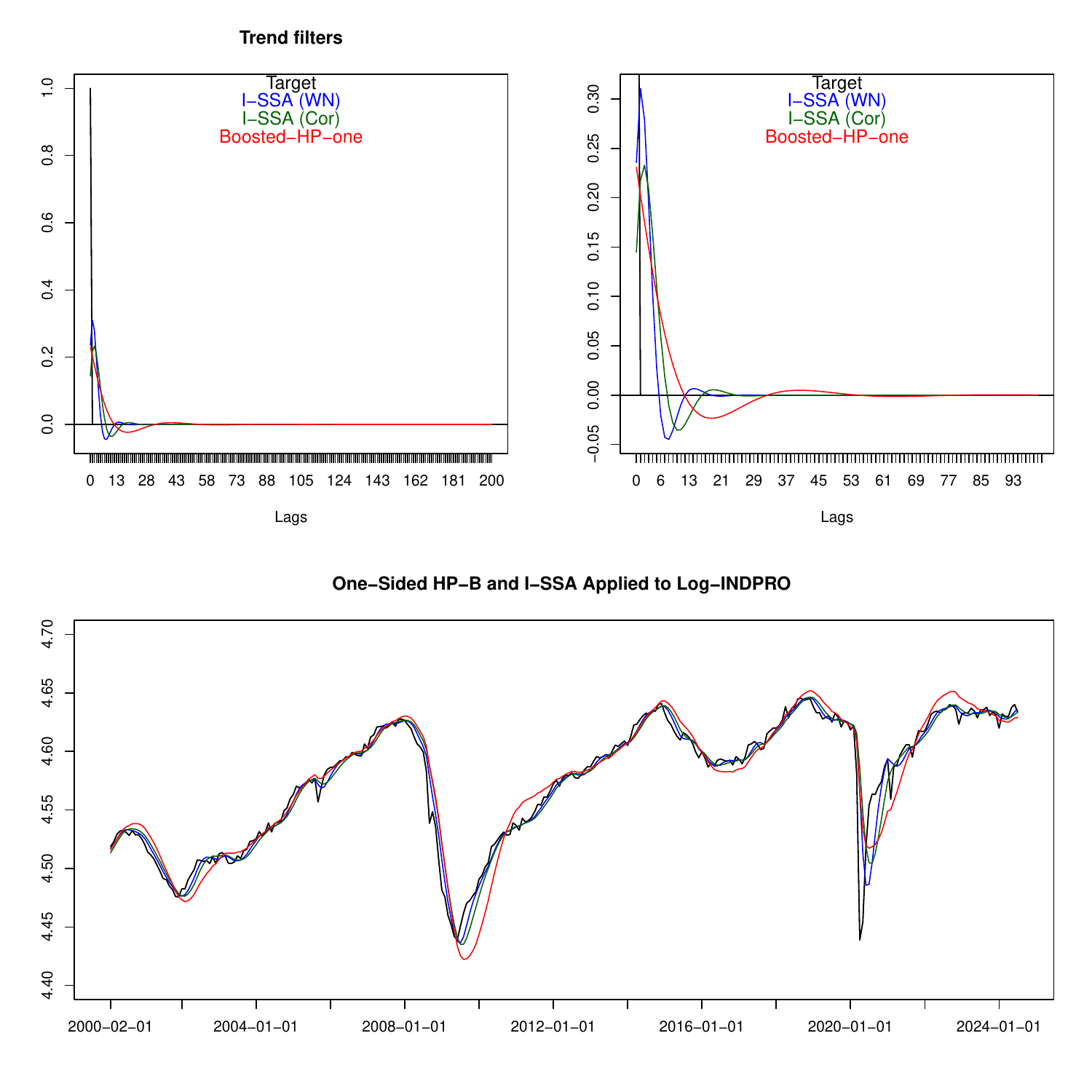}\caption{Log-INDPRO (black), HP-one (red), I-SSA assuming white noise (blue), and corrected I-SSA assuming an AR(1) (green)  from Jan:2000-Jan:2024, covering the last three recession episodes. For similar ---slightly smaller and slightly larger--- TP rates, I-SSA is more adaptive and faster, with smaller retardation, tracking maxima and minima at TPs better.  \label{issa__bhp_indpro}}\end{center}\end{figure}

Fig. \ref{issa__bhp_indpro} compares filter coefficients (top) and smoothed series (bottom); (blue and red filter coefficients are the same as in Fig.\ref{issa_nowcast_trend_bhp}). 
Increasing the HT in I-SSA (green vs. blue) shifts the weights slightly to the right, with slower decay and a right-shifted side lobe positioned between the faster I-SSA and HP-B. The smoothed series in the bottom panel suggest a tighter tracking of the data at peaks and dips, where both I-SSA generally lead HP-B, without over- or under-shooting. \\

This example illustrates that I-SSA based on the default random-walk specification can perform well on economic time series, provided first differences are not too far away from white noise. In the presence of weak autocorrelation, we recommend a slight adjustment of the HT constraint to correct for the resulting bias. In the case of stronger dependence, we additionally suggest inserting $\boldsymbol{\Xi}\neq \mathbf{I}$, reflecting the MA structure of the differenced data, to adjust the optimization beyond the constraint alone; in extreme cases, $d=2$ may be warranted.
 
 To conclude, the discrepancy between expected and empirical performance (table \ref{table_hpb_indpro} vs. \ref{table_hpb_indpro_true}) could be exploited as a diagnostic tool to infer the pertinence of the  model underlying $\boldsymbol{\Xi}$; see McElroy and Wildi (2010). We do not pursue this idea here.

\begin{table}[ht]
\centering
\begin{tabular}{rrrr}
  \hline
 & I-SSA (RW) & I-SSA (Cor) & HP-B-one \\ 
  \hline
MSE & 0.00010 & 0.00016 & 0.00022 \\ 
  HT & 7.03846 & 10.16667 & 8.56250 \\ 
  Curvature & 0.00305 & 0.00200 & 0.00235 \\ 
   \hline
\end{tabular}
\caption{Empirical MSE, HT and Curvature for one-sided HP-B and I-SSA applied to log-INDPRO. The two I-SSA are based on the 
default RW-model but the imposed HTs differ: the first (RW) replicates the HT of HP-B when applied to white noise; the second (Cor) replicates the HT of HP-B when applied to an AR(1) with $a_1=0.3$. This correction affects only the specification of the HT in the constraint and does not affect optimization which is based on the plain random-walk model. 
} 
\label{table_hpb_indpro}
\end{table}
\begin{table}[ht]
\centering
\begin{tabular}{rrr}
  \hline
 & I-SSA (RW) & I-SSA (Cor) \\ 
  \hline
MSE & 0.00008 & 0.00013 \\ 
  HT & 5.63323 & 7.43516 \\ 
   \hline
\end{tabular}
\caption{True (expected) MSE and HT of I-SSA based on the default RW model, with the HT in the constraint based on white noise (left) or AR(1) (right). MSE is the minimized value of the criterion. The positive bias of the empirical estimates in each HT setting suggests positive dependence in the differenced data: were the differenced INDPRO white noise, the table suggests the empirical MSE and HT would be smaller than observed.
} 
\label{table_hpb_indpro_true}
\end{table}

\section{Conclusion}\label{conclu}

I-SSA extends the SSA framework, and its underlying AS dilemma and efficient AS frontier, to non-stationary integrated time series. Its criterion defines I-SSA as the MSE-optimal predictor for a given TP/IP rate or, dually, as the maximally monotonous (for $d=1$) or minimally flexed (for $d=2$) predictor for a given MSE, with straightforward extensions to $d>2$. The dilemma between the conflicting A- and S-components is thereby expressed in closed form along an efficient AS frontier.

I-SSA generalizes the classic MSE approach, a single point on this frontier, to the entire frontier. It is simple and interpretable, incorporating sign accuracy, MSE, and smoothing requirements. It identifies TPs or IPs of non-stationary series, key determinants for decision making, by optimally tracing the changing levels of a dynamically evolving phenomenon, under a neutral (amorphous-memory) smoothness constraint that is unbiased toward any specific pattern or structure.

The I-SSA solution is the global minimum of the optimization GTRS problem with hidden convexity: once the classic MSE or benchmark predictor is known, the task reduces to a one-dimensional strictly monotonic problem, solved by fast algorithms such as bisection over the strict-positivity interval with known boundaries. The problem admits an asymptotic solution, devoid of the boundary perturbation induced by the cointegration constraint. The finite sample I-SSA converges in suitable norm to the limiting I-SSA.

Under correct model specification (integration order and Wold decomposition of the stationary differences), empirical HT and MSE asymptotically match their expected values; under misspecification, they are biased. In economic applications, the default random-walk model often performs well, owing to the typical spectral shape of series accumulating noise or news. Otherwise, the approach extends readily by correcting the HT bias or by incorporating the (estimated) Wold decomposition into the problem structure, e.g., the Hessian.

A powerful use case is the customization of generic linear benchmark predictors, filters, or smoothers onto the non-stationary AS frontier, as illustrated for the HP and boosted HP filters, with further benchmarks in the I-SSA R-package. Tracking the identity, the simplest benchmark, defines the {I-SSA trend}, which minimizes MSE relative to the original data subject to maximal monotonicity (or minimal inflection). For matched HTs, I-SSA filters decay faster than HP or boosted HP, emphasizing adaptivity and advancement of nowcasts, with stronger side lobes accommodating the imposed HT. Tracking thus improves, particularly at key TPs (or IPs). 

Looking forward, we aim to generalize I-SSA to multivariate prediction and to investigate timeliness, i.e., predictor retardation and advancement, thereby establishing a foundational prediction trilemma.

\appendix

\section{Discussion and Interpretation of Theoretical Results}\label{disc}

\textbf{Discussion Theorem 1}:

Equation \eqref{frequency-domain} generalizes the spectral decomposition of the Hessian from stationary SSA (Theorem 1, Wildi, 2024 and 2026a). Unlike SSA, where $\mathbf{V}$ is the fixed Fourier eigenvector matrix of $\mathbf{M}$, here $\mathbf{V}=\mathbf{V}(\rho_1)$ is the {generalized} eigenvector basis of the pencil $(\mathbf{A'A},\mathbf{Q}(\rho_1))$: generally non-orthogonal and $\rho_1$-dependent, coinciding with the classic Fourier eigenvectors only at the boundary $\rho_1\to\pm\rho_{\max}(L)$, $|\lambda|\to\infty$ (extremal smoothing or un-smoothing, see Appendix). The classic eigendecomposition of $\mathbf{H}(\lambda)$, by contrast, is orthogonal but depends on both $\rho_1$ {and} $\lambda$, and cannot achieve the decluttered form of \eqref{frequency-domain}: this requires the eigenvectors to simultaneously diagonalize the non-stationary unit-root operator $\mathbf{A'A}$ (level tracking) and the stationary autocorrelation operator $\mathbf{Q}$ (smoothness)—impossible for the ordinary basis, since long memory renders these matrices non-commutable in I-SSA (see proof). The generalized decomposition is thus key to extending SSA to I-SSA subject to this structural difference tied to long memory.

Since $\mathbf{V}$ is generally non-orthogonal ($\mathbf{V}'\neq \mathbf{V}^{-1}$), the generalized spectral factorization lacks the usual traits of a classic frequency-domain representation, yet it extends stationary SSA, where the interpretation is clear. There, the Hessian takes the simpler form $\mathbf{H}(\lambda)=\lambda\mathbf{M}+\mathbf{I}$\footnote{Here, the Hessian does not explicitly depend on $\rho_1$ but the dependence is implicit, since the optimal $\lambda^*$ must comply with the HT constraint determined by $\rho_1$.}, with $|\lambda|<1/\rho_{\max}(L)$ corresponding to $J_{++}$ in I-SSA (Wildi 2024, 2026a). Then $\mathbf{H}(\lambda)^{-1}=\mathbf{V}\mathbf{D}^{-1}\mathbf{V}'$, with Fourier eigenvectors $\mathbf{V}$ and diagonal elements $1/(1+\lambda\nu_k)=1/(1+\lambda\cos(k/(L+1)))$, where $\nu_k=\cos(k/(L+1))$ is the $k$-th eigenvalue of $\mathbf{M}$, independent of $\rho_1$. In the time domain, the Lagrangian normal equations of SSA, $\mathbf{H}(\lambda)\mathbf{b}=D\boldsymbol{\gamma}$, read
\begin{equation}\label{tdo}
0.5 \lambda \Big(b_{k-1}+b_k/(0.5 \lambda)+b_{k+1}\Big)=D\gamma_k,
\end{equation}
with $\boldsymbol{\gamma}$ the MSE target and $D$ a scaling. Its inversion $\mathbf{b}=D\mathbf{H}(\lambda)^{-1}\boldsymbol{\gamma}$ admits the frequency-domain form
\begin{equation}\label{ftdo}
\mathbf{b}=D\sum_{k=0}^{L}\frac{w_k}{1+\lambda\nu_k}\mathbf{v_k},
\end{equation}
where $w_k=\mathbf{v}_k'\boldsymbol{\gamma}$ — the special case of \eqref{frequency-domain} when the generalized eigenvectors reduce to classic Fourier eigenvectors (Wildi 2024, 2026a). Rewriting
\begin{eqnarray}\label{ar2_op}
\frac{1}{1+\lambda\nu_k}=\frac{1}{0.5\lambda}\frac{1}{ \Big(\exp(-ik/(L+1))+1/(0.5\lambda)+\exp(ik/(L+1))\Big)}
\end{eqnarray}
links both domains explicitly: $\exp(-ik/(L+1))$ and $\exp(ik/(L+1))$ are the forward and backward operators matching $b_{k-1}$ and $b_{k+1}$ in \eqref{tdo}. Thus $1/(1+\lambda\nu_k)$ is the transfer function of a non-stationary, acausal, time-reversible AR(2) operator 
(Wildi, 2026a); $w_k/(1+\lambda\nu_k)$ is its convolution with the MSE target. This operator is lowpass for $0>\lambda>-1$ ($\rho_1>\rho_{MSE}$: stronger smoothing, fewer crossings), highpass for $0<\lambda<1$ ($\rho_1<\rho_{MSE}$: more crossings), and allpass at $\lambda=0$, $\rho_1=\rho_{MSE}$ (identity). At $\lambda=\pm 1$, the stable and unstable real roots merge into a real I(2) double unit root at frequency $0$ or $\pi$, respectively; beyond this, $|\lambda|>1$,  the double root bifurcates into  a pair of complex conjugate unit roots at $\pm\omega\neq 0,\pi$, with the operator turning bandpass, peaking at $\pm\omega$. At $\lambda=-1/\rho_{\max}(L)$ the Hessian is semi-definite and the unit-root frequency matches the Fourier eigenvector $\mathbf{v}_{\rho_{\max}}(L)$ of $\mathbf{M}$, and similarly for $\lambda=1/\rho_{\max}(L)$. Typically, only the unstable lowpass case (smoothing), $0>\lambda>-1$, is relevant in applications. Instability is contained by the Fourier eigenvectors, formalized as implicit Dirichlet boundary constraints on \eqref{tdo} (Theorem 1, Wildi, 2024, 2026a).

With this background, and the necessary caveats, we interpret the I-SSA decomposition \eqref{frequency-domain} similarly: $\frac{1}{1+\lambda\nu_i}$ extends the non-stationary SSA AR(2) operator, and its product with $w_i$ extends the convolution with the MSE target. Note, however, that both the Hessian and $\nu_i$ now depend on $\rho_1$, and that the generalized spectral decompositions of I-SSA and target generally rest on different, non-orthogonal bases---what we term the `counterfactual' spectral decomposition. The immanent instability of the AR(2) is likewise contained by the generalized eigenvectors $\mathbf{V}$, formalized by {identical} Dirichlet boundary constraints (Corollary \ref{prop:dirichlet_b0} below), which give rise to the characteristic tipped, nose-shaped predictor profiles (see next section) observed in applications, shared by both SSA and I-SSA.


To conclude our discussion of the Theorem, a subtle but decisive difference separates SSA and I-SSA: in the stationary case, the geometry of the spectral decomposition is independent of the HT hyperparameter $\rho_1$, since the Fourier eigenvectors are fixed. Thus, a partially optimized solution $\mathbf{b}_{\epsilon}(\tilde{\lambda})$—i.e., a solution to the Lagrangian normal equations—at $\tilde{\lambda}\neq \lambda^*(\rho_1)$ is the SSA solution for some alternative $\tilde{\rho}_1\neq \rho_1$: the Lagrange multiplier $\lambda$ and the hyperparameter $\rho_1$ represent equivalent parametrizations of the problem and are exchangeable. This no longer holds in I-SSA, since the generalized eigenvectors $\mathbf{V}=\mathbf{V}(\rho_1)$ depend on $\rho_1$, affecting the entire geometry of the spectral representation: the partially optimized $\mathbf{b}_{\epsilon}(\tilde{\lambda})$ at $\tilde{\lambda}\neq \lambda^*(\rho_1)$ is generally not the I-SSA solution for a $\tilde{\rho}_1\neq \rho_1$, since $\mathbf{V}(\rho_1)\neq\mathbf{V}(\tilde{\rho}_1)$. As $\lambda$ and $\rho_1$ are no longer exchangeable parametrizations of a fixed geometric space, I-SSA solutions are obtained exclusively at $\lambda=\lambda^*(\rho_1)$. Equations \eqref{eq:decomposition_ee} and \eqref{dilemma} therefore single out this point—a distinction unnecessary in SSA. The I-SSA dilemma in \eqref{dilemma} thus offers a cleaner, more fundamental formulation of the direct dependence between $J(\lambda^*(\rho_1))$ (MSE tracking) and $\rho_1$ (smoothness), rather than an indirect link through $\lambda$—as in Wildi (2026a,b)—which is generally inappropriate in I-SSA. \\

\textbf{Discussion Corollary 1}:

If $\boldsymbol{\Xi}=\mathbf{I}$ (random walk), $b_{\tilde{x},L}=0$, falling back to the classic Dirichlet boundary constraint. Otherwise, $b_{\tilde{x},L}\to 0$ as $L\to\infty$, provided the (fairly general) assumptions of Theorem \ref{thm:decay_reformulated_uniform} below are satisfied. Equation \eqref{eq:foc_system_v_e} is the time-domain equivalent of the generalized spectral decomposition \eqref{frequency-domain}, which disentangles the combined permanent and transitory effects of the summation and linearized first-order ACF operators—the former absent in the SSA difference equation \eqref{tdo}. By diagonalizing simultaneously both operators through the generalized eigenvectors (see proof), it yields a simple, interpretable extension of the SSA-AR(2) operator in the spectral domain, as formalized by \eqref{frequency-domain}.

The symmetry of the matrices in  \eqref{eq:foc_system_v_e} suggests time reversibility of the I-SSA difference equation. Indeed, the first-order ACF in the HT constraint is indifferent to time orientation, so the I-SSA solution merges invertible (positive orientation: stable root) and non-invertible (negative orientation: unstable root) components of the homogeneous difference equation.\footnote{Invertibility does not refer to the causality of the predictor, which is determined solely by $\delta$ in $\tilde{z}_{t+\delta}$: for $\delta<0$ (backcast) the predictor is acausal, otherwise (nowcast/forecast) it is causal.} The implicit Dirichlet constraints at both ends control the aggregate solution, which combines forward and backward homogeneous solutions with the forward non-homogeneous solution---whose orientation is dictated by $d_k$ and hence $\gamma_k$---thus guaranteeing boundary stability irrespective of the MSE target. The latter shapes the I-SSA predictor on the inner lags $k=0,\ldots,L-1$, conditional on these immutable ghost boundaries. The degenerate case $\rho_1=\rho_{MSE}$, specifically excluded by the corollary, is nevertheless obtained in the limit $\rho_1\to\rho_{MSE}$; see the discussion ensuing the proof. \\

\textbf{Discussion Theorem 2}:

\textbf{Remarks}: Under fairly general assumptions, the theorem establishes convergence of the I-SSA problem structure and of its solution, together with polynomial decay  of the predictor weights and polynomial convergence for large lags.

 The polynomial decay justifies the finite MA approximation leading to \eqref{main_int_res}. This is non-trivial: an extreme smoothness constraint $\rho_1\to\rho_{\max}(L)$ forces $\mathbf{b}^*(L)$ towards $\mathbf{v}_{\max}(L)$, the Fourier eigenvector of the largest eigenvalue of $\mathbf{M}(L)$, whose coefficients do not decay polynomially. If instead $\rho_1$ is held fixed with $|\rho_1|<1$, polynomial decay at sufficiently large lags is preserved as $L>L_0(\rho_1)$ increases. Therefore, HT constraint and objective function receive their intended meaning, at least up to negligible deviations, valid uniformly in $L$ at the stationary points, i.e., under optimization.

The asymptotic form of I-SSA remains interpretable because it involves only `stationary' (bounded on $\ell^2$) operators, avoiding the infinite-length integration operator. In particular, the modified Hessian $K_\infty(\lambda_\infty^*)$ \eqref{Kinf} replaces the original Hessian $\mathbf{H}$, and $\mathbf{b}_\infty^*:=\mathbf{b}_\infty(\lambda_\infty^*)$ in \eqref{exp_dev_mse} is expressed through its deviation $\mathbf d^*_\infty$ from the infinite-length MSE predictor. This form allows the cointegration term of I-SSA to cancel asymptotically, suppressing the delicate rank-one boundary perturbation of finite-sample predictors induced by the cointegration constraint. In other words, $\mathbf{b}_\infty^*$ satisfies the cointegration constraint intrinsically.

The theorem assumes that the MA and target weights are known. We do not address their finite-sample estimation, which is covered by the existing literature, nor its interaction with the I-SSA equation system, and leave these topics for future research. In our applications, we typically assume a known target filter, e.g., the Hodrick--Prescott (HP) filter, the boosted HP filter, or the identity filter--- defining the I-SSA trend---, together with a generic random-walk specification for the underlying non-stationary series, so that $\boldsymbol{\Xi}=\mathbf{I}$. This choice complies with the theorem's assumption of known weights and, although not optimal, generally captures the dominant low-frequency dynamics of economic data reasonably well (see the examples in the next section).

Assumptions 1--2 are typically satisfied in practice, and our examples even exhibit exponential decay. Assumption 3 can be checked heuristically by computing the quantities involved at several lengths $L$; these typically converge rapidly to finite limits satisfying the constraints (cf.\ Section~\ref{applications}). In applications, where $L$ is fixed, this requires only that $\rho_1$ stay away from the extreme smoothing and un-smoothing boundaries, i.e.\ $|\rho_1|<\rho_{\max}(L)$ with a margin, a condition nearly always met in practice. 

Convergence or polynomial decay can alternatively be derived by constraining the target $\hat{\gamma}_{MSE}$, yielding a so-called `spectral balance' condition that extends the spectral completeness condition of Wildi (2024,2026a)--- not shown. However, Assumption 3, a simple condition on the scalar $\lambda^*(L)$, is generally easier to verify experimentally.\\

\section{I(2)-SSA: Lowest Curvature}\label{i2_ext}

This section provides an overview of the application of the SSA to integrated processes of order two, denoted as I(2), emphasizing that our methodology is generic and can be extended to processes with higher integration orders $d > 2$ through straightforward modifications. We denote by $y_{t,MSE}$, with coefficients $\boldsymbol{\hat{\gamma}}_{MSE}=(\hat{\gamma}_{ \tilde{x}\delta k})_{k=0,...,L-1}$, the  MSE filter of length $L$. For the case where $d = 2$, the cointegration constraints, as delineated in Equation \eqref{coint_eq}, can be divided into a `level'  constraint given by $\sum_{k=0}^{L-1} b_{\tilde{x}k} = \sum_{k=0}^{L-1}\hat{\gamma}_{ \tilde{x}\delta k} =: \Gamma(0)$, analogous to the I(1) case, and an additional `slope'  constraint defined as $\sum_{k=1}^{L-1} k b_{\tilde{x}k} = \sum_{k=1}^{L-1}k\hat{\gamma}_{ \tilde{x}\delta k} =: \dot{\Gamma}(0)$. 
The cancellation of the double unit-root by the error filter $\hat{\boldsymbol{\gamma}}_{MSE} - \mathbf{b}_{\tilde{x}}$ results in a stationary process represented by $e_t = y_{t,MSE} - y_t$. This is formalized as follows:
\begin{eqnarray}\label{i2canc}
y_{t,MSE}-y_t&=&(\hat{\boldsymbol{\gamma}}_{MSE}-\mathbf{b}_{\tilde{x}})'\mathbf{\tilde{x}}_t=  (\hat{\boldsymbol{\gamma}}_{MSE}-\mathbf{b}_{\tilde{x}})' \boldsymbol{\Sigma}^2~'\boldsymbol{\Delta}^2~'\mathbf{\tilde{x}}_t\nonumber\\
&=&
(\hat{\boldsymbol{\gamma}}_{MSE}-\mathbf{b}_{\tilde{x}})'\boldsymbol{\Sigma}^2~'\mathbf{x}_t.
\end{eqnarray}
Here, the first $L-2$ entries of $\boldsymbol{\Delta}^2~'\mathbf{\tilde{x}}_t$ are the stationary second order differences $x_t,x_{t-1},...,x_{t-(L-3)}$. We also  substituted the last two entries $x_{t-(L-2)},x_{t-(L-1)}$ of $\mathbf{x}_t$ to the right of the last equation for the last two non-stationary entries of $\boldsymbol{\Delta}^2~'\tilde{\mathbf{x}}_t$ to the left due to the structure of  $\boldsymbol{\Sigma}^2~'$, whose last two columns are $(L-1-k)_{k=0,...,L-1}$ and $(L-k)_{k=0,...,L-1}$. This implies that the last two entries of $(\hat{\boldsymbol{\gamma}}_{MSE}-\mathbf{b}_{\tilde{x}})'\boldsymbol{\Sigma}^2~'$ must vanish by the two cointegration constraints, a condition that similarly holds for $d>2$. The underlying problem structure mirrors that of the I(1) case as discussed in Section \eqref{ext_i1_ssa}, with the exception that $\boldsymbol{\Sigma}$ is replaced by $\boldsymbol{\Sigma}^2$ in Equation \eqref{i2canc}. Specifically, the processes situated to the right of Equation \eqref{i2canc} exhibit stationarity, thereby enabling the formulation of a meaningful objective function (target correlation) along with HT- and length-constraints. Consequently, the optimization criteria for I(2)-SSA are derived from Equation \eqref{crit_int}, where $\boldsymbol{\Sigma}^2~'$ replaces $\boldsymbol{\Sigma}'$:
\begin{eqnarray*}
\begin{array}{cc}
\left.\begin{array}{cc}
&\max_{\mathbf{b}_{\epsilon}}\mathbf{b}_{\epsilon}'\boldsymbol{\Sigma}^2~'\boldsymbol{\gamma}_{\tilde{\Xi}\delta}\\
&\mathbf{b}_{\epsilon}'\mathbf{M}\mathbf{b_{\epsilon}}=\rho_1\mathbf{b}_{\epsilon}'\mathbf{b}_{\epsilon}\\
&\mathbf{b}_{\epsilon}'\boldsymbol{\Sigma}^2~'\boldsymbol{\Sigma}^2\mathbf{b}_{\epsilon}=l
\end{array}\right\}
&\textrm{and~}
\left.\begin{array}{cc}
&\min_{\mathbf{b}_{{\epsilon}}}(\boldsymbol{\gamma}_{\tilde{\Xi}\delta}-\boldsymbol{\Sigma}^2\mathbf{b}_{{\epsilon}})'(\boldsymbol{\gamma}_{\tilde{\Xi}\delta}-\boldsymbol{\Sigma}^2\mathbf{b}_{{\epsilon}})\\
&\mathbf{b}_{\epsilon}'\mathbf{M}\mathbf{b_{\epsilon}}=\rho_1\mathbf{b}_{\epsilon}'\mathbf{b}_{\epsilon}\\
\end{array}\right\}
\end{array}.
\end{eqnarray*}
Here, $\boldsymbol{\gamma}_{\tilde{\Xi}\delta}=\boldsymbol{\Sigma}^2\boldsymbol{\Xi}\boldsymbol{\hat{\gamma}}_{MSE}$. The HT constraint regulates the frequency of zero-crossings in the stationary second order differences $y_t-2y_{t-1}+y_{t-2}$ of the predictor.  Strict monotonicity extends straightforwardly to the I($d$)-case, $d>1$, giving raise to the efficient AS-frontier for $\rho_1>\rho_{MSE}$ (right branch) and $\rho_1<\rho_{MSE}$ (left branch). 
As a result, $y_t - 2y_{t-1} + y_{t-2}$ minimizes the number of zero-crossings subject to a specified tracking accuracy of the target variable, thus ensuring that $y_t - y_{t-1}$ exhibits maximal monotonicity. Consequently, this results in $y_t$ having the fewest inflection points and the lowest curvature. In the final stage, it is necessary to impose level and slope (cointegration) constraints on  $\mathbf{b}_{\tilde{x}}$. Solving for $b_{\tilde{x}0}$ and $b_{\tilde{x}1}$ yields $\mathbf{b}_{\tilde{x}}=(\Gamma(0)-\dot{\Gamma}(0))\mathbf{e}_1+\dot{\Gamma}(0)\mathbf{e}_2+\mathbf{B}\tilde{\mathbf{b}}$, where $\tilde{\mathbf{b}}$ is a vector of length $L-2$, and $\mathbf{B}$ is defined as:
\[
\mathbf{B}=\left(\begin{array}{ccccc}1&2&3&...&L-2\\-2&-3&-4&...&-(L-1)\\
1&0&0&...&0\\
0&1&0&...&0\\
...&&&&\\
0&0&0&...&1\end{array}\right).
\] 
This matrix is of dimensions $L\times (L-2)$, while $\mathbf{e}_1$ and $\mathbf{e}_2$ are the first two unit-vectors of length $L$. An extension of Equation \eqref{eq:b_lambda} is  then straightforward.

\section{Proofs and Additional Results}

\subsection{Proofs}

\textbf{Proposition \ref{propo_pos_def}}\

\begin{proof}: 

By definition of $\boldsymbol{\Sigma}$ and $\boldsymbol{\Xi}$, the symmetric matrix $\mathbf{A}' \mathbf{A}$ is strictly positive definite (SPD). The matrix $\mathbf{M}$ has a decomposition with eigenvalues $\cos(j\pi/(L+1)),~j=1,...,L$ ---with $\rho_{\max}(L)=\cos(\pi/(L+1))$ and $\rho_{\min}(L)=-\rho_{\max}(L)$--- and (Fourier) eigenvectors $\mathbf{v}_j=(\sin(jk\pi))_{k=1,...,L}$.   
 Since $|\rho_1| < \rho_{\max}(L)$, the shifted matrix $\mathbf{M} - \rho_1 \mathbf{I}$ has both strictly positive and strictly negative eigenvalues, making it indefinite, so that $\mathbf{Q} = \boldsymbol{\Xi}' (\mathbf{M} - \rho_1 \mathbf{I}) \boldsymbol{\Xi}$ is also symmetric and indefinite.

To determine the positive definiteness of $\mathbf{H}(\lambda)$, we analyze the inequality:
\[
\mathbf{A}' \mathbf{A} + \lambda \mathbf{Q} \succ 0.
\]
Since $\mathbf{A}' \mathbf{A}$ is SPD, it possesses a unique SPD square root $(\mathbf{A}' \mathbf{A})^{1/2}$. Left- and right-multiplying the inequality by $(\mathbf{A}' \mathbf{A})^{-1/2}$ preserves the positive definiteness, yielding the equivalent condition:
\[
\mathbf{I} + \lambda (\mathbf{A}' \mathbf{A})^{-1/2} \mathbf{Q} (\mathbf{A}' \mathbf{A})^{-1/2} \succ 0.
\]
The matrix $(\mathbf{A}' \mathbf{A})^{-1/2} \mathbf{Q} (\mathbf{A}' \mathbf{A})^{-1/2}$ is symmetric, and its eigenvalues are identical to those of $(\mathbf{A}' \mathbf{A})^{-1} \mathbf{Q}$. Let these eigenvalues be denoted by $\nu_i$ for $i = 1, \dots, L$. Because $\mathbf{Q}$ is indefinite and $\mathbf{A}' \mathbf{A}$ is SPD, Sylvester's law of inertia dictates that $(\mathbf{A}' \mathbf{A})^{-1} \mathbf{Q}$ has both strictly positive and strictly negative eigenvalues. Thus, $\nu_{\max} > 0$ and $\nu_{\min} < 0$.

The positive definite condition requires that all eigenvalues of the shifted matrix are strictly positive:
\[
1 + \lambda \nu_i > 0 \quad \text{for all } i = 1, \dots, L.
\]
This imposes two simultaneous conditions on $\lambda$:
1. For positive eigenvalues ($\nu_i > 0$), the most restrictive bound is given by $\nu_{\max}$:
   \[
   1 + \lambda \nu_{\max} > 0 \implies \lambda > -\frac{1}{\nu_{\max}}.
   \]
2. For negative eigenvalues ($\nu_i < 0$), the most restrictive bound is given by $\nu_{\min}$:
   \[
   1 + \lambda \nu_{\min} > 0 \implies \lambda < -\frac{1}{\nu_{\min}}.
   \]
Combining these inequalities gives the exact open interval for which $\mathbf{H}(\lambda) \succ 0$:
\[
\lambda \in \left] -\frac{1}{\nu_{\max}}, -\frac{1}{\nu_{\min}} \right[.
\]
Within this interval, $\mathbf{H}(\lambda)$ is strictly positive definite.
\end{proof}

\textbf{Remark:} The proof holds for arbitrary integration orders since $\mathbf{A'A}=\boldsymbol{\Xi}'(\boldsymbol{\Sigma}^d)'\boldsymbol{\Sigma}^d\boldsymbol{\Xi}\succ 0$.\\

\textbf{Theorem \ref{prop_alt}}

\begin{proof}
\textbf{Part 1: Stationary Point} 

First, we reparameterize the problem. Let $\mathbf{x} = \Gamma(0) \mathbf{e}_1 + \mathbf{B} \mathbf{\tilde{b}} \in \mathbb{R}^L$. By the construction of $\mathbf{B}$, the sum of its column elements is zero ($\mathbf{1}' \mathbf{B} = \mathbf{0}'$). Therefore, multiplying $\mathbf{x}$ by $\mathbf{1}'$ yields:
\[
\mathbf{1}' \mathbf{x} = \Gamma(0) \mathbf{1}' \mathbf{e}_1 + \mathbf{1}' \mathbf{B} \mathbf{\tilde{b}} = \Gamma(0).
\]
Optimizing over $\mathbf{\tilde{b}} \in \mathbb{R}^{L-1}$ is strictly equivalent to optimizing over $\mathbf{x} \in \mathbb{R}^L$ subject to the linear constraint $\mathbf{1}' \mathbf{x} = \Gamma(0)$. Once $\mathbf{x}$ is found, $\mathbf{\tilde{b}}$ is simply $(x_2, \dots, x_L)'$.

Substitute $\mathbf{b}_{\epsilon} = \boldsymbol{\Xi} \mathbf{x}$ into the objective function:
\[
J(\mathbf{x}) = (\boldsymbol{\gamma} - \boldsymbol{\Sigma} \boldsymbol{\Xi} \mathbf{x})' (\boldsymbol{\gamma} - \boldsymbol{\Sigma} \boldsymbol{\Xi} \mathbf{x}) = (\boldsymbol{\gamma} - \mathbf{A} \mathbf{x})' (\boldsymbol{\gamma} - \mathbf{A} \mathbf{x}) = \mathbf{x}' \mathbf{A}' \mathbf{A} \mathbf{x} - 2 \boldsymbol{\gamma}' \mathbf{A} \mathbf{x} + \boldsymbol{\gamma}' \boldsymbol{\gamma}.
\]
Substitute $\mathbf{b}_{\epsilon} = \boldsymbol{\Xi} \mathbf{x}$ into the quadratic constraint $\mathbf{b}_{\epsilon}' \mathbf{M} \mathbf{b}_{\epsilon} - \rho_1 \mathbf{b_{\epsilon}' b_{\epsilon}} = 0$:
\[
\mathbf{x}' \boldsymbol{\Xi}' (\mathbf{M} - \rho_1 \mathbf{I}) \boldsymbol{\Xi} \mathbf{x} = \mathbf{x}' \mathbf{Q} \mathbf{x} = 0.
\]
The problem is now a Quadratically Constrained Quadratic Program (QCQP) in $\mathbf{x}$:
\begin{align*}
    \min_{\mathbf{x} \in \mathbb{R}^L} \quad & \mathbf{x}' \mathbf{A}' \mathbf{A} \mathbf{x} - 2 \boldsymbol{\gamma}' \mathbf{A} \mathbf{x} \\
    \text{subject to} \quad & \mathbf{x}' \mathbf{Q} \mathbf{x} = 0, \quad \mathbf{1}' \mathbf{x} = \Gamma(0).
\end{align*}
We introduce Lagrange multipliers $\lambda \in \mathbb{R}$ for the quadratic constraint and $2\mu \in \mathbb{R}$ for the linear constraint. Ignoring the constant term, the Lagrangian is:
\[
\mathcal{L}(\mathbf{x}, \lambda, \mu) = \mathbf{x}' \mathbf{A}' \mathbf{A} \mathbf{x} - 2 \boldsymbol{\gamma}' \mathbf{A} \mathbf{x} + \lambda \mathbf{x}' \mathbf{Q} \mathbf{x} + 2\mu (\mathbf{1}' \mathbf{x} - \Gamma(0)).
\]
Taking the gradient with respect to $\mathbf{x}$ and setting it to zero yields the first-order stationarity condition:
\[
2 \mathbf{A}' \mathbf{A} \mathbf{x} - 2 \mathbf{A}' \boldsymbol{\gamma} + 2 \lambda \mathbf{Q} \mathbf{x} + 2 \mu \mathbf{1} = 0 \implies (\mathbf{A}' \mathbf{A} + \lambda \mathbf{Q}) \mathbf{x} = \mathbf{A}' \boldsymbol{\gamma} - \mu \mathbf{1}.
\]
For $\lambda \in \Lambda_{++}$, the matrix $\mathbf{H}(\lambda) = \mathbf{A}' \mathbf{A} + \lambda \mathbf{Q}$ is invertible and positive definite, ensuring a global minimum for the Lagrangian. Solving for $\mathbf{x}$ gives:
\[
\mathbf{x}(\lambda, \mu) = \mathbf{H}(\lambda)^{-1} (\mathbf{A}' \boldsymbol{\gamma} - \mu \mathbf{1}).
\]
To satisfy the linear constraint $\mathbf{1}' \mathbf{x} = \Gamma(0)$, we multiply by $\mathbf{1}'$:
\[
\mathbf{1}' \mathbf{H}(\lambda)^{-1} \mathbf{A}' \boldsymbol{\gamma} - \mu \mathbf{1}' \mathbf{H}(\lambda)^{-1} \mathbf{1} = \Gamma(0).
\]
Solving for $\mu$ provides $\mu(\lambda)$:
\begin{equation}\label{mu_eq}
\mu(\lambda) = \frac{\mathbf{1}' \mathbf{H}(\lambda)^{-1} \mathbf{A}' \boldsymbol{\gamma} - \Gamma(0)}{\mathbf{1}' \mathbf{H}(\lambda)^{-1} \mathbf{1}}.
\end{equation}
Substituting $\mu(\lambda)$ back into $\mathbf{x}(\lambda, \mu)$ yields $\mathbf{x}(\lambda)$ strictly as a function of the dual variable $\lambda$. \\

\textbf{Part 2: Feasibility}: 

We prove the assertion by deriving a vector $ \mathbf{v}^*$ that satisfies both the cointegration constraint $ \mathbf{c}'\mathbf{v}^* = \Gamma(0)$, with $\mathbf{c}' = \mathbf{1}' \boldsymbol{\Xi}^{-1}$, and the HT constraint $R(\mathbf{v}^*) = \rho_1$, where $R(\mathbf{v}^*)=(\mathbf{v}^*)'\mathbf{M}\mathbf{v}^*/(\mathbf{v}^*)'\mathbf{v}^*$ denotes the Rayleigh quotient (first order ACF). 
Maximum $\rho_{\max}(L)=\cos(\pi/(L+1))$ and minimum $-\rho_{\max}(L)=-\cos(\pi/(L+1))$ of $R$ are attained at $\mathbf{f}_{\max}$ and $\mathbf{f}_{\min}$, the (Fourier) eigenvectors belonging to the extremal eigenvalues $\cos(\pi/(L+1))>\rho_1>-\cos(\pi/(L+1))$ of $\mathbf{M}$, with the inequalities holding by assumption.

First, we find a vector $\mathbf{v}_-$ such that $R(\mathbf{v}_-) < \rho_1$ and $\mathbf{c}'\mathbf{v}_- \neq 0$. We can start with $\mathbf{f}_{\min}$. If it happens to be orthogonal to $\mathbf{c}$\footnote{If $\boldsymbol{\Xi}=\mathbf{I}$ (the non-stationary process is a random-walk) and $L$ is even, then $\mathbf{c}'\mathbf{f}_{\min}=\mathbf{1}'\mathbf{f}_{\min}=0$.}, we can perturb it infinitesimally with any vector not orthogonal to $\mathbf{c}$; by continuity of $R$, the perturbed vector will still have a Rayleigh quotient strictly greater than $\rho_1$. We do the same to find $\mathbf{v}_+$ such that $R(\mathbf{v}_+) > \rho_1$ and $\mathbf{c}'\mathbf{v}_+ \neq 0$.

Next, we normalize both vectors so they satisfy the linear constraint:
\[
\mathbf{x}_+ = \frac{\Gamma(0)}{\mathbf{c}'\mathbf{v}_+} \mathbf{v}_+, \quad \quad \mathbf{x}_- = \frac{\Gamma(0)}{\mathbf{c}'\mathbf{v}_-} \mathbf{v}_-,
\]
with $R(\mathbf{x}_+) > \rho_1$ and $R(\mathbf{x}_-) < \rho_1$, and $\mathbf{c}'\mathbf{x}_+ = \Gamma(0)$, $\mathbf{c}'\mathbf{x}_- = \Gamma(0)$. Consider the convex combination:
\[
\mathbf{v}(t) = t \mathbf{x}_+ + (1-t) \mathbf{x}_-, \quad t \in [0, 1].
\]
For any $t$, the linear cointegration constraint is preserved. Since $\Gamma(0) \neq 0$, the vector $\mathbf{v}(t)$ is never the zero vector for $t \in [0,1]$, so the function $f(t) = R(\mathbf{v}(t))$ is well-defined and continuous on $[0, 1]$. Since $f(1) = R(\mathbf{x}_+) > \rho_1$ and $f(0) = R(\mathbf{x}_-) < \rho_1$, the Intermediate Value Theorem yields some $t^* \in (0, 1)$ with $f(t^*) = \rho_1$. Setting $\mathbf{v}^* = \mathbf{v}(t^*)$ gives a vector satisfying both $\mathbf{c}'\mathbf{v}^* = \Gamma(0)$ and $R(\mathbf{v}^*) = \rho_1$. Hence either there exists a feasible $\mathbf{b}^*$ satisfying both constraints with minimum MSE strictly smaller  than that of $\mathbf{v}^*$, the I-SSA solution, or $\mathbf{b}^*=\mathbf{v}^*$ is the (feasible) I-SSA solution.
\\

\textbf{Part 3: Uniqueness}:

The parametric vector $\mathbf{x}(\lambda)$ and the Lagrange multiplier $\mu(\lambda)$ satisfy the first-order optimality condition for the Lagrangian:
\begin{equation}
\mathbf{H}(\lambda) \mathbf{x}(\lambda) + \mu(\lambda) \mathbf{1} = \mathbf{A}' \boldsymbol{\gamma}, \label{eq:kkt}
\end{equation}
subject to the linear constraint $\mathbf{1}' \mathbf{x}(\lambda) = \Gamma(0)$.

Differentiating \eqref{eq:kkt} with respect to $\lambda$ yields:
\begin{equation}
\mathbf{Q} \mathbf{x}(\lambda) + \mathbf{H}(\lambda) \mathbf{\dot{x}}(\lambda) + \dot{\mu}(\lambda) \mathbf{1} = \mathbf{0}, \label{eq:diff_kkt}.
\end{equation}
Consider the secular function $F(\lambda) = \mathbf{x}(\lambda)' \mathbf{Q} \mathbf{x}(\lambda)$, with root
\begin{equation}\label{root}
F(\lambda^*)=0.
\end{equation} 
The derivative of $F(\lambda)$ is:
\[
\dot{F}(\lambda) = 2 \mathbf{\dot{x}}(\lambda)' \mathbf{Q} \mathbf{x}(\lambda).
\]
Pre-multiplying equation \eqref{eq:diff_kkt} by $\mathbf{\dot{x}}(\lambda)'$, we obtain:
\[
\mathbf{\dot{x}}(\lambda)' \mathbf{Q} \mathbf{x}(\lambda) + \mathbf{\dot{x}}(\lambda)' \mathbf{H}(\lambda) \mathbf{\dot{x}}(\lambda) + \dot{\mu}(\lambda) \mathbf{\dot{x}}(\lambda)' \mathbf{1} = 0.
\]
Since the constraint $\mathbf{1}' \mathbf{x}(\lambda) = \Gamma(0)$ holds for all $\lambda \in J_{++}$, differentiating both sides with respect to $\lambda$ gives:
\[
\mathbf{1}' \mathbf{\dot{x}}(\lambda) = 0.
\]
Substituting this back into the previous equation eliminates the term involving $\dot{\mu}(\lambda)$ so that:
\[
\mathbf{\dot{x}}(\lambda)' \mathbf{Q} \mathbf{x}(\lambda) = - \mathbf{\dot{x}}(\lambda)' \mathbf{H}(\lambda) \mathbf{\dot{x}}(\lambda).
\]
Therefore, the derivative of the secular function is:
\begin{equation}\label{s_m_sf}
\dot{F}(\lambda) = -2 \mathbf{\dot{x}}(\lambda)' \mathbf{H}(\lambda) \mathbf{\dot{x}}(\lambda)<0,
\end{equation}
since the Hessian is strictly positive definite in $J_{++}$. By strict monotonicity, the root $\lambda^*$ in \eqref{root} must be unique, as claimed.\\

\textbf{Part 4: Global Minimum}:  A proof could be derived by relying on strong duality for the primal GTRS structure of I-SSA.\footnote{By the S-Lemma (Polik 2007) and GTRS duality theory (Pong \& Wolkowicz, 2014), the duality gap is zero, so  global maximum of the dual and global minimum of the primal match. Since the former is constrained to positive (semi-) definiteness,  $\lambda^*\in J_{++}$ must be the global optimum.} Under the posited assumptions, however, a more elementary proof is available. The Lagrangian is a quadratic function of $\mathbf{x}$, with vanishing gradient $\left.\nabla_{\mathbf{x}} \mathcal{L}(\mathbf{x},\lambda^*,\mu^*)\right|_{\mathbf{x}^*}=\mathbf{0}$, and strictly positive definite Hessian $\mathbf{H}(\lambda^*) \succ 0$ at the I-SSA solution $\mathbf{x}^*=\mathbf{x}^*(\lambda^*)$, where we assume $\mu^*=\mu(\lambda^*)$ is given by \eqref{mu_eq}. Therefore $\mathbf{x}^*(\lambda^*)$ is the global minimum of $\mathcal{L}(\mathbf{x},\lambda^*,\mu^*)$. Moreover, $\lambda^*$ and $\mu(\lambda^*)$ are such that $\mathbf{x}^*$ is feasible, implying
$$ 
\mathcal{L}(\mathbf{x}^*,\lambda^*,\mu^*)=J(\mathbf{x}^*).
$$
Let $\mathbf{x}^{**},\lambda^{**},\mu^{**}$ be another feasible solution, so that
$$
J(\mathbf{x}^{**}) = \mathcal{L}(\mathbf{x}^{**}, \lambda^{**},\mu^{**})=\mathcal{L}(\mathbf{x}^{**}, \lambda,\mu),
$$
where equality holds for arbitrary $\lambda,\mu$ since the constraints vanish at any feasible solution. Then
$$
J(\mathbf{x}^{**}) = \mathcal{L}(\mathbf{x}^{**}, \lambda^{**},\mu^{**})=\mathcal{L}(\mathbf{x}^{**}, \lambda^*,\mu^*)> \mathcal{L}(\mathbf{x}^*, \lambda^*,\mu^*)=J(\mathbf{x}^*),
$$
showing that $\mathbf{x}^*$ is the strict global minimizer of MSE under HT and cointegration constraints, as claimed.\\

\textbf{Part 5: Generalized Spectral Decomposition}

Consider the generalized eigenvalue decomposition of $\mathbf{Q}$ with respect to the positive definite matrix $\mathbf{A}'\mathbf{A}$. There exists an invertible matrix $\mathbf{V}$ such that:
\[
\mathbf{V}' \mathbf{A}'\mathbf{A} \mathbf{V} = \mathbf{I}, \quad \mathbf{V}' \mathbf{Q} \mathbf{V} = \mathbf{D},
\]
where $\mathbf{D} = \text{diag}(\nu_1, \nu_2, \dots, \nu_L)$ contains the generalized eigenvalues in decreasing order, so $\nu_1 = \nu_{\max}$ and $\nu_L = \nu_{\min}$. The columns of $\mathbf{V}$ are the corresponding eigenvectors $\mathbf{v}_i$.

Using this decomposition, we can rewrite $\mathbf{H}(\lambda)$ as:
\[
\mathbf{H}(\lambda) = (\mathbf{V}^{-1})' (\mathbf{I} + \lambda \mathbf{D}) \mathbf{V}^{-1}.
\]
Its inverse is:
\begin{equation}\label{spect_dec}
\mathbf{H}(\lambda)^{-1} = \mathbf{V} (\mathbf{I} + \lambda \mathbf{D})^{-1} \mathbf{V}'.
\end{equation}
Therefore, we obtain:
\begin{eqnarray}\label{spec_dec_bis}
\mathbf{b}_{\tilde{x}}=\mathbf{H}(\lambda)^{-1} \mathbf{g}(\lambda)  = \mathbf{V} (\mathbf{I} + \lambda \mathbf{D})^{-1} \mathbf{V}'(\mathbf{V'})^{-1}\mathbf{w}=
\sum_{i=1}^L\frac{w_i}{1+\lambda\nu_i} \mathbf{v}_i,
\end{eqnarray}
where $\mathbf{w}=(w_1,...,w_L)'=\mathbf{V'}\mathbf{g}(\lambda)$ is the `counterfactual' spectral factorization of the target. If $\mathbf{V}$ were an orthogonal eigenvector basis, $(\mathbf{V'})^{-1}=\mathbf{V}$, and the decompositions of $\mathbf{b}_{\tilde{x}}$ and $\mathbf{g}(\lambda)$ would rely on the same `spectral' basis, e.g., the Fourier-eigenvectors in the stationary SSA case. Here, however, the two decompositions use different (transposed) bases --$\mathbf{V}$ for $\mathbf{b}_{\tilde{x}}$ and $(\mathbf{V'})^{-1}$ for $\mathbf{g}(\lambda)$---,  which we term `counterfactual'. In any case, \eqref{spec_dec_bis} generalizes the stationary SSA framework by diagonalizing simultaneously the permanent two-sided integration operator and the (linearized) first-order ACF operator, thereby unearthing an identical expression of the stationary decomposition, albeit in a generalized basis.
\\

\textbf{Part 6: AS Dilemma}: 

Let $V(\lambda)$ denote the partially optimized Lagrangian substituting the optimal $\mathbf{x}(\lambda)$ and $\mu(\lambda)$:
$$
    V(\lambda) = J(\lambda) + \lambda \mathbf{x}(\lambda)' \mathbf{Q} \mathbf{x}(\lambda).
$$
Note that the linear constraint $\mathbf{1}' \mathbf{x}(\lambda) - \Gamma(0) = 0$ holds for all $\lambda$, hence the term involving $\mu$ vanishes. By the Envelope Theorem, the total derivative of $V(\lambda)$ with respect to $\lambda$ is equal to its partial derivative:
$$
   \frac{d J(\lambda)}{d \lambda} + \mathbf{x}(\lambda)' \mathbf{Q} \mathbf{x}(\lambda) + \lambda \frac{d}{d \lambda} \big( \mathbf{x}(\lambda)' \mathbf{Q} \mathbf{x}(\lambda) \big)= \frac{d V(\lambda)}{d \lambda} = \frac{\partial \mathcal{L}}{\partial \lambda} = \mathbf{x}(\lambda)' \mathbf{Q} \mathbf{x}(\lambda).
$$
Therefore, we obtain
\begin{equation}\label{secu_dev}
    \frac{d J(\lambda)}{d \lambda} = - \lambda \frac{d}{d \lambda} \big( \mathbf{x}(\lambda)' \mathbf{Q} \mathbf{x}(\lambda) \big)=- \lambda \frac{d F(\lambda)}{d \lambda} ,
\end{equation}
yielding \eqref{secu_as}.
Substituting $\mathbf{Q} = \boldsymbol{\Xi}' (\mathbf{M} - \rho_1 \mathbf{I}) \boldsymbol{\Xi}$ and $\mathbf{b}_{\epsilon}(\lambda) = \boldsymbol{\Xi} \mathbf{x}(\lambda)$, the quadratic form can be factored as:
\begin{align*}
    \mathbf{x}(\lambda)' \mathbf{Q} \mathbf{x}(\lambda) &= \mathbf{b}_{\epsilon}(\lambda)' (\mathbf{M} - \rho_1 \mathbf{I}) \mathbf{b}_{\epsilon}(\lambda) \\
    &= \mathbf{b}_{\epsilon}(\lambda)' \mathbf{b}_{\epsilon}(\lambda) \left( \frac{\mathbf{b}_{\epsilon}(\lambda)' \mathbf{M} \mathbf{b}_{\epsilon}(\lambda)}{\mathbf{b}_{\epsilon}(\lambda)' \mathbf{b}_{\epsilon}(\lambda)} - \rho_1 \right) \\
    &= \| \mathbf{b}_{\epsilon}(\lambda) \|^2 \big( \rho_{\mathbf{b}_{\epsilon}}(\lambda) - \rho_1 \big).
\end{align*}
Thus
\begin{equation}
\frac{d J(\lambda)}{d \lambda} = - \lambda \frac{d}{d \lambda} \big( \mathbf{x}(\lambda)' \mathbf{Q} \mathbf{x}(\lambda) \big)
     = \| \mathbf{b}_{\epsilon}(\lambda) \|^2 \frac{d \rho_{\mathbf{b}_{\epsilon}}(\lambda)}{d \lambda} + \big( \rho_{\mathbf{b}_{\epsilon}}(\lambda) - \rho_1 \big) \frac{d \| \mathbf{b}_{\epsilon}(\lambda) \|^2}{d \lambda},
\end{equation}
yielding the decomposition in \eqref{eq:decomposition_e}. Equation \eqref{eq:decomposition_ee} then follows from $\rho_{\mathbf{b}_{\epsilon}}(\lambda^*) - \rho_1=0$.

To verify Equation \ref{dilemma} we resort again to the Envelope Theorem stating 
$$ 
\frac{d J^*(\rho_1)}{d \rho_1} = \left. \frac{\partial \mathcal{L}}{\partial \rho_1} \right|_{ \lambda^*} = -\lambda^* (\mathbf{b}_{\epsilon}^*)' \mathbf{b}_{\epsilon}^*. 
$$
To determine the sign of $\lambda^*$, define the constraint function
$C(\mathbf{\tilde{b}})=\mathbf{b}_{\epsilon}'\mathbf{M}\mathbf{b}_{\epsilon}-\rho_1\mathbf{b}_{\epsilon}'\mathbf{b}_{\epsilon}$,
where $\mathbf{b}_{\epsilon}$ is linked to $\mathbf{\tilde{b}}$ via \eqref{beps_btilde}. Here $\mathbf{\tilde{b}}\in \mathbb{R}^{L-1}$ denotes the unconstrained lower-dimensional vector in the cointegration constraint.
 Let $\mathbf{\tilde{b}}_{MSE}$ denote the unconstrained minimum, with associated  $\mathbf{b}_{\epsilon MSE}=\boldsymbol{\Xi}\hat{\boldsymbol{\gamma}}_{MSE}$ and $J(\mathbf{\tilde{b}}_{MSE})=0$. 

\textbf{Case 1: $\rho_1 < \rho_{MSE}$}. Then $C(\mathbf{\tilde{b}}_{MSE}) > 0$, and we consider the {inequality} constrained problem: 
$$
\min J(\mathbf{\tilde{b}}),\quad \text{subject~ to:}\quad C(\mathbf{\tilde{b}}) \leq 0,
$$
where the cointegration constraint does not appear explicitly. 
Because the unconstrained minimum strictly violates this constraint, the constraint must be active (binding) at the optimum, meaning the solution exactly matches the equality problem $C(\mathbf{\tilde{b}}) = 0$.\footnote{Let $\mathbf{\tilde{b}}^*$ denote the optimal I-SSA solution and suppose $\mathbf{\tilde{b}}^*$ is an `interior' point $C(\mathbf{\tilde{b}}^*) < 0$.  Define $\mathbf{\tilde{b}}(t) = (1-t)\mathbf{\tilde{b}}^* + t \mathbf{\tilde{b}}_{MSE}$, $t \in [0, 1]$. By convexity, $J(\mathbf{\tilde{b}}(t)) < J(\mathbf{\tilde{b}}^*)$ and by continuity we can select $\tilde{t}>0$ such that $C(\mathbf{\tilde{b}}(\tilde{t})) < 0$; so $\mathbf{\tilde{b}}(\tilde{t})$ is  feasible and with strictly smaller objective, contradicting optimality of $\mathbf{\tilde{b}}^*$.}  
The KKT conditions for the constraint $C(\mathbf{\tilde{b}}) \leq 0$ dictate an `inequality' Lagrangian $\mathcal{L}_{ineq} = J(\mathbf{\tilde{b}}) + \mu C(\mathbf{\tilde{b}})$ with multiplier $\mu > 0$.\footnote{If $\mu<0$, following the zero level-line of the constraint gradient (ensuring the constraint remains matched) would decrease the objective function, implying a maximum instead of a minimum.} Matching this to our `equality' Lagrangian, we have $\lambda^* = \mu > 0$. 
Therefore, $\frac{d J^*(\rho_1)}{d \rho_1} = -\lambda^* (\mathbf{b}_{\epsilon}^*)' \mathbf{b}_{\epsilon}^* < 0$.

\textbf{Case 2: $\rho_1 > \rho_{MSE}$}. Then $C(\mathbf{\tilde{b}}_{MSE}) < 0$, and we consider the inequality constrained problem: $\min J(\mathbf{\tilde{b}})$ subject to $C(\mathbf{\tilde{b}}) \geq 0$, which is equivalent to $-C(\mathbf{\tilde{b}}) \leq 0$. The unconstrained minimum violates this, so the constraint is active at the optimum, matching the equality problem $C(\mathbf{\tilde{b}}) = 0$.
The KKT conditions for $-C(\mathbf{\tilde{b}}) \leq 0$ dictate a Lagrangian $\mathcal{L}_{ineq} = J(\mathbf{\tilde{b}}) + \mu (-C(\mathbf{\tilde{b}}))$ with $\mu > 0$. Matching this to our equality Lagrangian, we obtain $\lambda^* = -\mu < 0$.
Therefore, $\frac{d J^*(\rho_1)}{d \rho_1} = -\lambda^* (\mathbf{b}_{\epsilon}^*)' \mathbf{b}_{\epsilon}^* > 0$, verifying \eqref{dilemma}.
\end{proof}


\textbf{Remarks}: While derived for $d=1$, the structural elements of the proof are invariant to the integration order, since $\mathbf{A'A}=\boldsymbol{\Xi}'(\boldsymbol{\Sigma}^d)'\boldsymbol{\Sigma}^d\boldsymbol{\Xi}\succ 0$, so Proposition \ref{propo_pos_def} holds and the Hessian can be decomposed into $\mathbf{H}(\lambda) = (\mathbf{V}^{-1})' (\mathbf{I} + \lambda \mathbf{D}) \mathbf{V}^{-1}$ through the generalized eigenvalue decomposition based on the pencil $(\mathbf{A'A,Q})$. Only the cointegration constraint changes: none when $d=0$ (generalizing SSA), and the adaptation in Appendix \ref{i2_ext} when $d=2$ (with straightforward extensions for $d>2$, though largely irrelevant in applications).

The Theorem does not explicitly address the extreme HT constraint $\rho_1=\rho_{\max}(L)$, which forces $\mathbf{b}_{\epsilon}=\mathbf{x}(\lambda^*)\propto\mathbf{f}_{\max}$, the Fourier eigenvector of $\mathbf{M}$ at its largest eigenvalue (and similarly $\mathbf{b}_{\epsilon}\propto\mathbf{f}_{\min}$ for $\rho_1=-\rho_{\max}(L)$). Since $\mathbf{f}_{\max}$ is fixed, only sign and scaling remain free to MSE optimization. 

We now briefly discuss this boundary solution via the corresponding asymptotics: as $\rho_1\to\rho_{\max}(L)$, $\nu_{\max}\to 0^+$, $-1/\nu_{\max}\to-\infty$ (the lower bound of $J_{++}$), $\lambda\to-\infty$, and $\mathbf{H}(\lambda)\simeq \lambda\mathbf{Q}=\lambda\boldsymbol{\Xi}'(\mathbf{M}-\rho_{\max}(L)\mathbf{I})\boldsymbol{\Xi}$. The  Hessian thus reduces to the stationary SSA case, with $\mathbf{M}$'s Fourier eigenvectors determining the spectral decomposition (Wildi, 2026a), recovering $\mathbf{b}_{\epsilon}\propto\mathbf{f}_{\max}$ as a limiting case. If the MSE target $\boldsymbol{\gamma}_{\tilde{\Xi}\delta}=\boldsymbol{\Sigma}\boldsymbol{{\Xi}}\boldsymbol{\hat{\gamma}}_{MSE}$ and the extremal I-SSA solution $\boldsymbol{\Sigma}\boldsymbol{\Xi}\mathbf{b}_{\tilde{x}}=\boldsymbol{\Sigma}\mathbf{b}_{\epsilon}\propto\boldsymbol{\Sigma}\mathbf{f}_{\max}$ are not mutually orthogonal, the I-SSA and MSE predictors have positive target correlation (left criterion in \eqref{crit_int}), so the extreme HT constraint $\rho_1=\rho_{\max}(L)$ remains compatible with the prediction problem. Otherwise, target and I-SSA are orthogonal, enforced by excessive smoothness: admissible from an optimization standpoint, but at odds with the prediction intent.

Similarly, for $\rho_1=-\rho_{\max}(L)$, $\lim_{\lambda\to-\infty}\mathbf{b}_{\epsilon}(\lambda)\propto\mathbf{f}_{\min}$, the Fourier eigenvector at $-\rho_{\max}(L)$. In this extreme un-smoothing case, a further problem arises, illustrated by the random walk $\boldsymbol{\Xi}=\mathbf{I}$ with $L$ even: $\mathbf{f}_{\min}'\mathbf{1}=0$ implies $\mathbf{b}_{\tilde{x}}(\lambda)'\mathbf{1}=\mathbf{b}_{\epsilon}(\lambda)'\mathbf{1}=0$, so $\mathbf{b}_{\tilde{x}}(\lambda)$ cancels a unit root, contradicting the standing assumption.

We therefore caution against extreme smoothing (or un-smoothing) scenarios in which $|\rho_1|$ approaches $\rho_{\max}(L)$, although in typical prediction problems $\rho_1$ lies well within the spectral boundaries. Moreover, increasing $L$ to $L^+>L$ while maintaining $|\rho_1|=\rho_{\max}(L)$ removes the singularities and releases previously frozen degrees of freedom for optimization.

To conclude, the I-SSA problem exhibits an interesting singularity cancellation as $\lambda$ approaches the boundaries of $J_{++}$, linked to the so-called `hard case' of the underlying GTRS problem. This cancellation becomes increasingly intricate as $|\rho_1|$ also approaches its boundary $\rho_{\max}(L)$. In particular, when $\rho_1\to \rho_{\max}(L)$ {and} $\lambda\to -1/\nu_{\max}\to-\infty$, the approximation $\mathbf{H}(\lambda)\simeq \lambda\mathbf{Q}$ resolves the main (asymptotic) singularity of the exact Hessian $\mathbf{H}(\lambda)=\mathbf{A'A}+ \lambda\mathbf{Q}$, removing the need for an explicit cancellation in this special case. Otherwise, the required cancellation is performed `indirectly' by the cointegration constraint, which thereby addresses two distinct instability issues at once: optimal level tracking, its original purpose, and the prevention of an asymptotically infinite predictor as the Hessian turns singular. These non-trivial interactions are further affected by $L$, since the cointegration term vanishes asymptotically as $L\to\infty$, assuming $\rho_1<1$ is fixed (see the proof of Theorem 2 below). They underscore the richness of the I-SSA criterion --- its `hidden complexity'--- and are illustrated in the accompanying R package by a simple example representative of typical prediction problems.\\

\textbf{Corollary \ref{prop:dirichlet_b0}}
\begin{proof} 
To simplify notation set $\mathbf{b}:=\mathbf{b}_{\tilde{x}}$. By Theorem \ref{prop_alt}, the optimal vector $\mathbf{b}=(b_0,...,b_{L-1}) \in \mathbb{R}^L$ satisfies the first-order condition (FOC) of the constrained optimization problem:
\begin{equation}\label{time_domain}
    \big[ \boldsymbol{\Xi}' \boldsymbol{\Sigma}' \boldsymbol{\Sigma} \boldsymbol{\Xi} + \lambda \boldsymbol{\Xi}' (\mathbf{M} - \rho_1 \mathbf{I}) \boldsymbol{\Xi} \big] \mathbf{b} = \boldsymbol{\Xi}' \boldsymbol{\Sigma}' \boldsymbol{\gamma} - \mu(\lambda) \mathbf{1}.
\end{equation}
Pre-multiplying the FOC by $(\boldsymbol{\Xi}')^{-1}$ yields:
\begin{equation}
    \big[ \boldsymbol{\Sigma}' \boldsymbol{\Sigma} + \lambda (\mathbf{M} - \rho_1 \mathbf{I}) \big] \mathbf{v} = \mathbf{d},
    \label{eq:foc_system_v}
\end{equation}
where $\mathbf{v} = \boldsymbol{\Xi} \mathbf{b}$ and $\mathbf{d} = \boldsymbol{\Sigma}' \boldsymbol{\gamma} - \mu(\lambda) (\boldsymbol{\Xi}')^{-1} \mathbf{1}$. 

Equation \eqref{eq:foc_system_v} represents a linear system generating a difference equation for the elements $v_k$ ($k=0, \dots, L-1$). For an interior index $k$ ($0 < k < L-1$), the $k+1$-th row of the first-order autocovariance matrix $\mathbf{M}$ contains $0.5$ at columns $k$ and $k+2$, and $0$ elsewhere. Thus, the $k+1$-th equation of the system \eqref{eq:foc_system_v} takes the form:
\begin{equation}
    \sum_{j=1}^L (\boldsymbol{\Sigma}' \boldsymbol{\Sigma})_{k+1j} v_{j-1} + \lambda \big( 0.5 v_{k-1} - \rho_1 v_{k } + 0.5 v_{k+1} \big) = d_{k},
    \label{eq:interior_diff_eq}
\end{equation}
where $\lambda\neq 0$ (non-degenerate case: $\rho_1\neq \rho_{MSE}$). 
This establishes the homogeneous structure of the difference equation governing the predictor coefficients. 

To determine the left boundary behavior, we examine the actual first row of the system ($k=0$):
$$
    \sum_{j=1}^L (\boldsymbol{\Sigma}' \boldsymbol{\Sigma})_{1j} v_{j-1} + \lambda \big( - \rho_1 v_0 + 0.5 v_1 \big) = d_0.
$$
For the general interior difference equation \eqref{eq:interior_diff_eq} to be universally consistent across the entire domain, its theoretical evaluation at $k=0$ 
$$
    \sum_{j=1}^L (\boldsymbol{\Sigma}' \boldsymbol{\Sigma})_{1j} v_{j-1} + \lambda \big( 0.5 v_{-1} - \rho_1 v_0 + 0.5 v_1 \big) = d_0
    \label{eq:eval_first_row}
$$
must be algebraically identical to the actual truncated first row: equality is obtained by imposing $\lambda (0.5 v_{-1}) = 0$, implying the natural boundary condition:
\begin{equation}
    v_{-1} = 0.
\end{equation}
We now map this condition back to the original predictor coefficients $\mathbf{b}$. The relationship at the boundary is $v_k = b_k + \sum_{j=0}^{k-1} \xi_{j+1} b_{k-(j+1)}$. Extending this mapping to the ghost index $k=-1$ yields $v_{-1} = b_{-1}=0$
(`empty' sum). 

Similar reasoning holds at the right boundary. For $k=L-1$, the homogeneous structure of the difference equation is
$$\sum_{j=1}^L(\boldsymbol{\Sigma}'\boldsymbol{\Sigma})_{Ljv_{j-1}}+\lambda(0.5v_{L-2}-\rho_1v_{L-1}+0.5v_L)=d_{L-1}.$$
It corresponds to the boundary behaviour
$$\sum_{j=1}^L(\boldsymbol{\Sigma}'\boldsymbol{\Sigma})_{Ljv_{j-1}}+\lambda(0.5v_{L-2}-\rho_1v_{L-1})=d_{L-1},$$
implying $0.5v_L=0$, i.e.,
$$0=v_L=b_L+\sum_{j=0}^{L-2}\xi_{j+1}b_{L-(j+1)},$$
as claimed.\end{proof}

\textbf{Remarks:} The proof falls back to the stationary case, because left and right boundary constraints for the `ghosts' $b_{-1}$ and $b_L$ are determined entirely by $\mathbf{M}$ at $k=0$ and $k=L-1$. The (purely causal) summation operator $\boldsymbol{\Sigma}$ and its  reflection $\boldsymbol{\Sigma}'$ at the right boundary do not transgress the lag limits $k=0,...,L-1$. 

As $\rho_1\to\rho_{MSE}$, the Lagrange multiplier $\lambda\to 0$ and $\mathbf{b}\to\hat{\boldsymbol{\gamma}}_{MSE}$. While the two Dirichlet conditions remain active at the `ghost' lags, the boundary adjustment term (the homogeneous solution) loses memory, confining its effect to $b_{-1}$ and $b_L$. Thus, as I-SSA converges to the MSE predictor, the HT constraint loses weight through diminishing $\lambda$: the Dirichlet constraints still hold strictly but cease to influence the predictor coefficients $b_k$, $k=0,\dots,L-1$ (see Wildi 2026a and the accompanying R-package for illustration). The Corollary therefore implicitly covers the degenerate case $\lambda=0$, where $\mathbf{b}=\hat{\boldsymbol{\gamma}}_{MSE}$ is asymptotically compatible with any values of $\hat{\gamma}_{MSE,-1},\hat{\gamma}_{MSE,L}$, while $b_{-1}=b_L=0$ holds strictly.\\

\textbf{Theorem \ref{thm:decay_reformulated_uniform}}:

\begin{proof}
By assumption and Theorem \ref{prop_alt}, a feasible I-SSA solution  $\mathbf{x}(\lambda^*(L))$ exists for all $L>L_0$ since $\rho_1<\rho_{\max}(L_0)<\rho_{\max}(L)$.\\

\textbf{Part (i): Convergence of the boundaries of $J_{++}(L)$.} \\
\textbf{Step 1: Elimination of $\boldsymbol{\Xi}$ and Reduction to an Eigenvalue Problem.}
By Sylvester's law of inertia
\begin{equation}
\label{eq:congruence_delta}
{\mathbf{H}}(\lambda) \succ \mathbf{0} 
\iff 
\boldsymbol{\Delta}'(\boldsymbol{\Xi}^{-1})'{\mathbf{H}}(\lambda)\boldsymbol{\Xi}^{-1}\boldsymbol{\Delta} 
= \mathbf{I} + \lambda \boldsymbol{\Delta}'(\mathbf{M} - \rho_1 \mathbf{I})\boldsymbol{\Delta} \succ \mathbf{0}.
\end{equation}
Define the symmetric matrix $\mathbf{S} := \boldsymbol{\Delta}'(\mathbf{M} - \rho_1 \mathbf{I})\boldsymbol{\Delta} $, and let $\mathbf{J}$ denote the orthogonal `anti-diagonal' identity, which satisfies $\mathbf{J}' = \mathbf{J} = \mathbf{J}^{-1}$. Then  $\mathbf{J}\boldsymbol{\Delta}\mathbf{J} = \boldsymbol{\Delta}'$, $\mathbf{J}\boldsymbol{\Delta}'\mathbf{J} = \boldsymbol{\Delta}$, and $\mathbf{J}(\mathbf{M} - \rho_1\mathbf{I})\mathbf{J} = \mathbf{M} - \rho_1\mathbf{I}$, by symmetry. Consequently, $\mathbf{S}$ is orthogonally similar to
\[
\mathbf{O} := \mathbf{J} \mathbf{S} \mathbf{J} = \boldsymbol{\Delta}(\mathbf{M} - \rho_1 \mathbf{I})\boldsymbol{\Delta}', 
\]
i.e., eigenvalues are identical. For all $L \ge L_0$, the assumption $|\rho_1| < \rho_{\max}(L_0) \le \rho_{\max}(L)$ implies
\[
\lambda_{\min}(\mathbf{M} - \rho_1 \mathbf{I}) = -\rho_{\max}(L) - \rho_1 < 0 
\quad \text{and} \quad 
\lambda_{\max}(\mathbf{M} - \rho_1 \mathbf{I}) = \rho_{\max}(L) - \rho_1 > 0.
\]
Hence, $\mathbf{M} - \rho_1 \mathbf{I}$ (and consequently $\mathbf{S}$ and $\mathbf{O}$, by Sylvester's law) is strictly indefinite for all $L \ge L_0$, possessing both strictly positive and strictly negative real eigenvalues.

Let $\nu_{\max}(L) > 0$ and $\nu_{\min}(L) < 0$ denote the largest and smallest eigenvalues of $\mathbf{S}$ (equivalently of $\mathbf{O}$). Since $\mathbf{I} + \lambda \mathbf{S} \succ \mathbf{0}$ holds if and only if $1 + \lambda \nu > 0$ for all eigenvalues $\nu$ of $\mathbf{S}$, the boundaries of $J_{++}(L) = (\lambda_{\min}(L), \lambda_{\max}(L))$ are given by
\begin{equation}
\label{eq:lambda_extremal_eigenvalues}
\lambda_{\min}(L) = -\frac{1}{\nu_{\max}(L)} < 0, 
\qquad 
\lambda_{\max}(L) = -\frac{1}{\nu_{\min}(L)} > 0.
\end{equation}

\textbf{Step 2: Exact Diagonalization in the Fourier Basis and Cauchy Interlacing.}
Consider the identity:
\[
\boldsymbol{\Delta}'\boldsymbol{\Delta} = 
\begin{pmatrix}
2 & -1 & 0 & \cdots & 0 \\
-1 & 2 & -1 & \cdots & 0 \\
0 & -1 & 2 & \cdots & 0 \\
\vdots & \vdots & \vdots & \ddots & -1 \\
0 & 0 & 0 & -1 & 1
\end{pmatrix}
= 2(\mathbf{I} - \mathbf{M}) - \mathbf{e}_L \mathbf{e}_L',
\]
where $\mathbf{e}_L = (0,\dots,0,1)'$. Define the $(L+1)\times L$ matrix obtained by appending one row to $\boldsymbol{\Delta}$,
\[
\widetilde{\boldsymbol{\Delta}} := \begin{pmatrix} \boldsymbol{\Delta} \\ -\mathbf{e}_L' \end{pmatrix} \in \mathbb{R}^{(L+1)\times L},
\qquad \text{for which} \qquad
\widetilde{\boldsymbol{\Delta}}'\widetilde{\boldsymbol{\Delta}} = \boldsymbol{\Delta}'\boldsymbol{\Delta} + \mathbf{e}_L\mathbf{e}_L' = 2(\mathbf{I} - \mathbf{M}) =: \mathbf{L}_D.
\]
Now $\mathbf{L}_D = 2(\mathbf{I} - \mathbf{M})$ and $\mathbf{M} - \rho_1\mathbf{I}$ commute and are simultaneously diagonalized by the Fourier eigenvectors $\{\mathbf{v}^{(m)}\}_{m=1}^L$ of $\mathbf{M}$, with components
\[
v^{(m)}_k = \sqrt{\frac{2}{L+1}} \sin(k \omega_m), \qquad \omega_m = \frac{m \pi}{L+1}, \qquad k, m \in \{1, \dots, L\}.
\]
Since $\mathbf{M}\mathbf{v}^{(m)} = \cos(\omega_m)\mathbf{v}^{(m)}$, the eigenvalues of $\mathbf{L}_D$ and $\mathbf{M} - \rho_1\mathbf{I}$ associated with mode $m$ are $2(1 - \cos(\omega_m))$ and $(\cos(\omega_m) - \rho_1)$, respectively. Consequently, the eigenvalues of  $\widetilde{\boldsymbol{\Delta}}'\widetilde{\boldsymbol{\Delta}}(\mathbf{M} - \rho_1\mathbf{I}) = \mathbf{L}_D(\mathbf{M} - \rho_1\mathbf{I})=\mathbf{VD_LV'VD_{M-\rho_1I}V'}=\mathbf{VD_LD_{M-\rho_1I}V'}$ are given  by
\begin{equation}
\label{eq:mu_m_formula}
\mu_m(L) = 2\big(1 - \cos(\omega_m)\big)\big(\cos(\omega_m) - \rho_1\big), 
\qquad \omega_m = \frac{m \pi}{L+1}, \quad m = 1, \dots, L.
\end{equation}
Now consider the augmented $(L+1)\times(L+1)$ symmetric matrix $\widetilde{\mathbf{O}} := \widetilde{\boldsymbol{\Delta}}(\mathbf{M} - \rho_1\mathbf{I})\widetilde{\boldsymbol{\Delta}}'$. By Sylvester's determinant identity $\det(\mu\mathbf{I}_{L+1} - \mathbf{XY}) = \mu\det(\mu\mathbf{I} - \mathbf{YX})$, 
the eigenvalues of $\widetilde{\mathbf{O}}$ are precisely $\{\mu_m(L)\}_{m=1}^L \cup \{0\}$. Moreover, the block structure $\widetilde{\boldsymbol{\Delta}} = \binom{\boldsymbol{\Delta}}{-\mathbf{e}_L'}$ implies that the unaugmented $L \times L$ matrix $\mathbf{O} = \boldsymbol{\Delta}(\mathbf{M} - \rho_1\mathbf{I})\boldsymbol{\Delta}'$ is  the leading $L \times L$ principal submatrix of $\widetilde{\mathbf{O}}$. By Cauchy's Interlacing Theorem for symmetric matrices, the $L$ ordered eigenvalues of $\mathbf{O}$ (and hence of $\mathbf{S}$) interlace the $L+1$ ordered eigenvalues of $\widetilde{\mathbf{O}}$. For all $L$ sufficiently large, $\{\mu_m(L)\}_{m=1}^L$ in \eqref{eq:mu_m_formula} contains at least two strictly negative (when $m=L-1,L$ and $L$ large, so $\omega_m\simeq\pi$) and at least two strictly positive values ($m=1,2$, $\omega_m\simeq 0$), so the extra eigenvalue $0$ lies strictly in the interior of the spectrum of $\widetilde{\mathbf{O}}$. Thus $\nu_{\min}(L)$ is sandwiched between the two smallest (most negative) values of $\{\mu_m(L)\}_{m=1}^L$, and $\nu_{\max}(L)$ is sandwiched between the two largest positive values of $\{\mu_m(L)\}_{m=1}^L$.

To analyze the extrema of $\{\mu_m(L)\}_{m=1}^L$ in \eqref{eq:mu_m_formula}, introduce the change of variable $c = \cos(\omega) \in (-1, 1)$ and define the  (quadratic) function
\[
q(c) := 2(1 - c)(c - \rho_1), \qquad c \in [-1, 1],
\]
so that $\mu_m(L) = q(\cos(\omega_m))$ for $m = 1, \dots, L$.

\textbf{Step 3: Convergence of $\lambda_{\max}(L)$ as $L \to \infty$.}
By \eqref{eq:lambda_extremal_eigenvalues}, $\lambda_{\max}(L) = -1/\nu_{\min}(L)$. Note that for $c \in [-1, 1)$, $q(c) < 0$ if and only if $c \in [-1, \rho_1)$. Differentiating $q(c)$ with respect to $c$ yields
\[
q'(c) = 2(1 + \rho_1 - 2c).
\]
For all $c \in [-1, \rho_1]$, since $\rho_1 < 1$, we have $2c \le 2\rho_1 < 1 + \rho_1$, which implies
\[
q'(c) \ge 2(1 - \rho_1) > 0, \qquad \forall c \in [-1, \rho_1].
\]
Hence, $q(c)$ is strictly increasing on $[-1, \rho_1]$. Since $\cos(\omega_m)$ is strictly decreasing with respect to $m \in \{1, \dots, L\}$ and $\cos(\omega_L) < \cos(\omega_{L-1}) < \rho_1$ for all $L$ sufficiently large, the two smallest (most negative) values of $\{q(\cos(\omega_m))\}_{m=1}^L$ are attained at $m = L$ and $m = L-1$. By Cauchy's Interlacing Theorem,
\[
q(\cos(\omega_L)) \le \nu_{\min}(L) \le q(\cos(\omega_{L-1})) < 0.
\]
As $L \to \infty$, both $\omega_L = \frac{L\pi}{L+1} \to \pi$ and $\omega_{L-1} = \frac{(L-1)\pi}{L+1} \to \pi$, so $\cos(\omega_L), \cos(\omega_{L-1}) \to -1$. By continuity of $q(c)$ and the Squeeze Theorem,
\[
\lim_{L \to \infty} \nu_{\min}(L) = q(-1) = 2(1 - (-1))(-1 - \rho_1) = -4(1 + \rho_1) < 0.
\]
Substituting this limit into \eqref{eq:lambda_extremal_eigenvalues} gives
\[
\lim_{L \to \infty} \lambda_{\max}(L) = -\frac{1}{\lim_{L \to \infty} \nu_{\min}(L)} = \frac{1}{4(1 + \rho_1)} > 0.
\]

\textbf{Step 4: Convergence of $\lambda_{\min}(L)$ as $L \to \infty$.}
By \eqref{eq:lambda_extremal_eigenvalues}, $\lambda_{\min}(L) = -1/\nu_{\max}(L)$. Observe that $q(c) > 0$ if and only if $c \in (\rho_1, 1)$. On the interval $[\rho_1, 1]$, $q(c) = 2(1 - c)(c - \rho_1)$ is a strictly concave quadratic function ($q''(c) = -4 < 0$) vanishing at the endpoints $c = \rho_1$ and $c = 1$. 

Setting $q'(c) = 2(1 + \rho_1 - 2c) = 0$, we find that $q(c)$ attains its unique global maximum on $[-1, 1]$ at the midpoint
\[
c_{\max} := \frac{1 + \rho_1}{2} \in (\rho_1, 1),
\]
corresponding to the interior frequency $\omega_{\max} := \arccos\big(\frac{1+\rho_1}{2}\big) \in (0, \arccos(\rho_1)) \subset (0, \pi)$. Evaluating $q(c)$ at $c = c_{\max}$ yields
\[
\max_{c \in [-1, 1]} q(c) = q(c_{\max}) = 2\left(1 - \frac{1 + \rho_1}{2}\right)\left(\frac{1 + \rho_1}{2} - \rho_1\right) 
= 2\left(\frac{1 - \rho_1}{2}\right)^2 
= \frac{(1 - \rho_1)^2}{2} > 0.
\]
Since the discrete grid points $\{\omega_m\}_{m=1}^L = \big\{\frac{m\pi}{L+1}\big\}_{m=1}^L$ form a uniform mesh of $(0, \pi)$ with step size $\frac{\pi}{L+1} \to 0$ as $L \to \infty$, for every $L$ sufficiently large there exist two distinct indices $m_1(L), m_2(L) \in \{1, \dots, L\}$ such that $|\omega_{m_j(L)} - \omega_{\max}| \le \frac{\pi}{L+1}$ and $q(\cos(\omega_{m_j(L)})) > 0$ for $j = 1, 2$. By Cauchy's Interlacing Theorem and the global optimality of $c_{\max}$, $\nu_{\max}(L)$ is bounded between the second-largest and the largest values of $\{q(\cos(\omega_m))\}_{m=1}^L$, which implies
\[
\min_{j \in \{1, 2\}} q\big(\cos(\omega_{m_j(L)})\big) \le \nu_{\max}(L) \le \max_{1 \le m \le L} q(\cos(\omega_m)) \le q(c_{\max}).
\]
Since $\omega_{m_j(L)} \to \omega_{\max}$ as $L \to \infty$ for $j = 1, 2$ and $\omega \mapsto q(\cos(\omega))$ is continuous, the Squeeze Theorem yields
\[
\lim_{L \to \infty} \nu_{\max}(L) = q(c_{\max}) = \frac{(1 - \rho_1)^2}{2} > 0.
\]
Substituting this limit into \eqref{eq:lambda_extremal_eigenvalues} gives
\[
\lim_{L \to \infty} \lambda_{\min}(L) = -\frac{1}{\lim_{L \to \infty} \nu_{\max}(L)} = -\frac{2}{(1 - \rho_1)^2} < 0,
\]
which completes the proof of Claim (i).\\

\textbf{Part (ii): polynomial decay of I-SSA weights.}

\medskip\noindent\textbf{Preamble: Uniform strict positivity.}
Let
\[
\mathbf K=\mathbf K(\lambda^*(L)):=\mathbf I+\lambda^*(L)\,\mathbf S,
\qquad
\mathbf S=\boldsymbol\Delta'(\mathbf M-\rho_1\mathbf I)\boldsymbol\Delta ,
\]
which, by the congruence of Part (i), satisfies
\begin{equation}\label{eq:Kcong}
\mathbf K=\boldsymbol\Delta'(\boldsymbol\Xi')^{-1}\,\mathbf H(\lambda^*)\,\boldsymbol\Xi^{-1}\boldsymbol\Delta
\quad\Longleftrightarrow\quad
\mathbf H^{-1}=\boldsymbol\Xi^{-1}\boldsymbol\Delta\,\mathbf K^{-1}\boldsymbol\Delta'(\boldsymbol\Xi')^{-1}.
\end{equation}
By Part (i), the eigenvalues of $\mathbf S$ lie in $[\nu_{\min}(\mathbf S),\nu_{\max}(\mathbf S)]$ with
$\nu_{\min}(\mathbf S)\to S_{\min}<0$ and $\nu_{\max}(\mathbf S)\to S_{\max}>0$, and the boundaries
$\lambda_{\min}(L),\lambda_{\max}(L)$ of $J_{++}(L)$ converge to finite limits. Hence
$\Lambda:=\sup|\lambda^*(L)|<\infty$. Assumption (A3) gives $\delta>0$ with
$\operatorname{dist}(\lambda^*(L),\partial J_{++}(L))\ge\delta$ for all $L$, so that
\[
1+\lambda^*\nu_j\ \ge\ c_\delta>0\quad\text{for every eigenvalue }\nu_j\text{ of }\mathbf S,
\]
with $c_\delta$ depending only on $\delta$ and on the limits in Part (i) (for $\lambda^*>0$:
$1+\lambda^*\nu_{\min}=\nu_{\min}(\lambda^*-\lambda_{\max}(L))\ge|\nu_{\min}|\delta$, and analogously for $\lambda^*<0$).
Consequently there exist universal constants $0<c\le C<\infty$, independent of $L$, such that
\begin{equation}\label{eq:Kbounds}
c\,\mathbf I\preceq\mathbf K\preceq C\,\mathbf I\qquad\text{uniformly in }L .
\end{equation}

\medskip\noindent\textbf{Notation and auxiliary facts.}
For $s>1$ and $a>0$, $\mathbf N\in\mathcal J_s(a)$ if $|N_{ij}|\le a(1+|i-j|)^{-s}$.

\smallskip\noindent\textbf{(F1) Convolution inequality.} For $s>1$ there is $c_1(s)<\infty$ with
\[
\sum_{l\in\mathbb Z}(1+|i-l|)^{-s}(1+|l-j|)^{-s}\le c_1(s)(1+|i-j|)^{-s}.
\]
Consequences (independent of $L$):
(i) $\mathbf N_1\in\mathcal J_s(a_1)$, $\mathbf N_2\in\mathcal J_s(a_2)\Rightarrow\mathbf N_1\mathbf N_2\in\mathcal J_s(c_1(s)a_1a_2)$;
(ii) $\mathbf N\in\mathcal J_s(a)$, $|v_l|\le b(1+l)^{-s}\Rightarrow|(\mathbf N\mathbf v)_k|\le c_1(s)ab(1+k)^{-s}$;
(iii) a banded matrix with bandwidth $\beta$ and entries bounded by $m$ lies in $\mathcal J_s(m(1+\beta)^s)$.

\smallskip\noindent\textbf{(F2) Finite-$L$ Jaffard theorem.}
If $\mathbf K\in\mathbb R^{L\times L}$, $|K_{ij}|\le a(1+|i-j|)^{-s}$ and $c\mathbf I\preceq\mathbf K\preceq C\mathbf I$, then
$|(\mathbf K^{-1})_{ij}|\le a'(1+|i-j|)^{-s}$ with $a'=a'(s,a,c,C)$
(quantitative Jaffard theorem, Gr\"ochenig--Klotz (2010)).

\smallskip\noindent\textbf{(F3) Decay of the inverse MA filter.}
Under (A1) ($|\xi_j|\le C_\xi(1+j)^{-s}$, $\xi(e^{i\omega})\ne0$ on $[-\pi,\pi]$ with
$|\xi(e^{i\omega})|\ge m_\xi>0$), the coefficients $\pi_j$ of $1/\xi(z)$ satisfy
$|\pi_j|\le C_\pi(1+j)^{-s}$, with $C_\pi=C_\pi(s,C_\xi,m_\xi)$
(Wiener Lemma, Gr\"ochenig (2010)). Since $\boldsymbol\Xi_L^{-1}$ is the  lower triangular Toeplitz matrix with entries $\pi_{i-j}$, we obtain
$\boldsymbol\Xi^{-1}\in\mathcal J_s(C_\pi)$ uniformly in $L$.

\medskip\noindent\textbf{Step 0: exact algebra.}
Since $\boldsymbol\Delta,\boldsymbol\Sigma,\boldsymbol\Xi$ are lower triangular Toeplitz matrices, they commute and
$\mathbf A\boldsymbol\Delta=\boldsymbol\Sigma\boldsymbol\Xi\boldsymbol\Delta=\boldsymbol\Xi$.
From $\boldsymbol\gamma=\mathbf A\boldsymbol\gamma_{MSE}$ we get $\mathbf A'\boldsymbol\gamma=\mathbf A'\mathbf A\boldsymbol\gamma_{MSE}$,
and $\mathbf A'\mathbf A=\mathbf H-\lambda^*\mathbf Q$. Therefore, from $\mathbf H\mathbf x^*=\mathbf A'\boldsymbol\gamma-\mu\mathbf 1$,
\begin{equation}\label{deffd}
\mathbf d:=\mathbf x^*-\boldsymbol\gamma_{MSE}=-\mathbf H^{-1}\big(\lambda^*\mathbf Q\boldsymbol\gamma_{MSE}+\mu\mathbf 1\big).
\end{equation}
Using \eqref{eq:Kcong} and $\mathbf Q=\boldsymbol\Xi'(\mathbf M-\rho_1\mathbf I)\boldsymbol\Xi$,
\begin{eqnarray}\label{def_fx}
\boldsymbol\Delta'(\boldsymbol\Xi')^{-1}\mathbf Q\boldsymbol\gamma_{MSE}
&=&\boldsymbol\Delta'(\mathbf M-\rho_1\mathbf I)\boldsymbol\Xi\boldsymbol\gamma_{MSE}=:\mathbf f\\
\boldsymbol\Delta'(\boldsymbol\Xi')^{-1}\mathbf 1&=&(\boldsymbol\Xi')^{-1}\boldsymbol\Delta'\mathbf 1=(\boldsymbol\Xi')^{-1}\mathbf e_L=:\boldsymbol\zeta\nonumber,
\end{eqnarray}
where  $\boldsymbol\zeta=(\pi_{L-1},\dots,\pi_1,\pi_0)'$ with $\pi_0=1/\xi_0$.
Hence
\begin{equation}\label{eq:dw}
\mathbf d=-\boldsymbol\Xi^{-1}\boldsymbol\Delta\,\mathbf w,\qquad
\mathbf w:=\mathbf K^{-1}\big(\lambda^*\mathbf f+\mu\,\boldsymbol\zeta\big).
\end{equation}
The definition of $\mu(\lambda^*)$ is equivalent to $\mathbf 1'\mathbf x^*=\Gamma(0)$, and $\mathbf 1'\boldsymbol\gamma_{MSE}=\Gamma(0)$,
so $\mathbf 1'\mathbf d=0$, by defintion of $\mathbf d$ in \eqref{deffd}. Because $\mathbf 1'\boldsymbol\Xi^{-1}\boldsymbol\Delta=\big(\boldsymbol\Delta'(\boldsymbol\Xi')^{-1}\mathbf 1\big)'=\boldsymbol\zeta'$,
\begin{equation}\label{eq:zetaw}
0=\mathbf 1'\mathbf d=-\boldsymbol\zeta'\mathbf w
\quad\Longleftrightarrow\quad
\lambda^*\,\boldsymbol\zeta'\mathbf K^{-1}\mathbf f+\mu\,\boldsymbol\zeta'\mathbf K^{-1}\boldsymbol\zeta=0
\quad\Longleftrightarrow\quad
\mu=-\lambda^*\frac{\boldsymbol\zeta'\mathbf K^{-1}\mathbf f}{\boldsymbol\zeta'\mathbf K^{-1}\boldsymbol\zeta},
\end{equation}
where, by \eqref{eq:Kbounds}, $\boldsymbol\zeta'\mathbf K^{-1}\boldsymbol\zeta\ge\|\boldsymbol\zeta\|^2/C\ge\pi_0^2/C=1/(C\xi_0^2)>0$. In summary, step 0 isolates the difference vector $\mathbf{d} = \mathbf{x}^* - \boldsymbol{\gamma}_{MSE}$; the substitution of the congruence transformation $\mathbf{H}^{-1} = \boldsymbol{\Xi}^{-1}\boldsymbol{\Delta}\mathbf{K}^{-1}\boldsymbol{\Delta}'(\boldsymbol{\Xi}')^{-1}$ maps the problem to the well-conditioned matrix $\mathbf{K}$; identifying that $\boldsymbol{\Delta}' \mathbf{1} = \mathbf{e}_L$  localizes the linear constraint's influence to the sequence $\boldsymbol{\zeta}$, crucial for later bounding $\mu$, noting that $\boldsymbol\zeta$ and $\mathbf f$ have time reverted polynomial decays in the numerator $\boldsymbol\zeta'\mathbf K^{-1}\mathbf f$ of the right-hand side quotient in \eqref{eq:zetaw}.

\medskip\noindent\textbf{Step 1: structure of $\mathbf K$.}
$\mathbf S=\boldsymbol\Delta'(\mathbf M-\rho_1\mathbf I)\boldsymbol\Delta$ is a product of banded matrices
($\boldsymbol\Delta$ bidiagonal, $\mathbf M-\rho_1\mathbf I$ tridiagonal), hence pentadiagonal with entries bounded by
$m_S:=4\max(0.5,|\rho_1|)$. Therefore $\mathbf K=\mathbf I+\lambda^*\mathbf S$ is pentadiagonal with entries bounded by $1+\Lambda m_S$,
and by (F1, iii), $\mathbf K\in\mathcal J_s(a_K)$ with $a_K=(1+\Lambda m_S)\,3^s$ (bandwidth $\beta=2$), independent of $L$.

\medskip\noindent\textbf{Step 2: decay of $\mathbf K^{-1}$.}
By \eqref{eq:Kbounds}, Step 1 and (F2), $\mathbf K^{-1}\in\mathcal J_s(a_{K^{-1}})$ with $a_{K^{-1}}=a'(s,a_K,c,C)$ independent of $L$. In summary, steps 1 and 2 establish that $\mathbf{K}^{-1}$ belongs to the Jaffard class $\mathcal{J}_s$, preserving the polynomial decay rate $s$ of its entries uniformly in $L$.

\medskip\noindent\textbf{Step 3: the forcing $\mathbf f$ and the vector $\boldsymbol\zeta$.}
First, $\boldsymbol\Xi\in\mathcal J_s(C_\xi)$ by (A1). By (A2) and (ii),
$|(\boldsymbol\Xi\boldsymbol\gamma_{MSE})_k|\le c_1(s)C_\xi C_\gamma(1+k)^{-s}$, $k=1,\dots,L$.
The matrix $\boldsymbol\Delta'(\mathbf M-\rho_1\mathbf I)$ is banded with entries bounded by $2\max(0.5,|\rho_1|)\cdot 2$,
hence belongs to $\mathcal J_s(a_B)$ by F1 (iii). Then, from the defintion of $\mathbf f$ in \eqref{def_fx} and by F1 (ii)
\[
|f_k|\le a_f(1+k)^{-s},\qquad k=1,\dots,L,\qquad a_f:=c_1(s)a_Bc_1(s)C_\xi C_\gamma .
\]
Next, by (F3), $|\zeta_j|=|\pi_{L-j}|\le C_\pi(1+L-j)^{-s}$ for $j=1,\dots,L$; in particular $\boldsymbol\zeta$ is
concentrated near the last index.

\medskip\noindent\textbf{Step 4: bounds on $\mu$ and $\mathbf w$.}
By Steps 2, 3 and (ii),
\begin{equation}\label{eq:Kinvf}
|(\mathbf K^{-1}\mathbf f)_j|\le c_2(1+j)^{-s},\qquad j=1,\dots,L,\qquad c_2:=c_1(s)a_{K^{-1}}a_f .
\end{equation}
Since $\boldsymbol\zeta'\mathbf K^{-1}\mathbf f=\sum_j\zeta_j(\mathbf K^{-1}\mathbf f)_j$,
\[
|\boldsymbol\zeta'\mathbf K^{-1}\mathbf f|\le C_\pi c_2\sum_{j=1}^L(1+L-j)^{-s}(1+j)^{-s}
\le C_\pi c_2c_1(s)(1+L)^{-s},
\]
by (F1) (with $i=L$ after the shift $j\mapsto L-j$ and $l=0$). Using \eqref{eq:zetaw} and the lower bound
$\boldsymbol\zeta'\mathbf K^{-1}\boldsymbol\zeta\ge1/(C\xi_0^2)$,
\begin{equation}\label{eq:mu}
|\mu|\le\Lambda\,C\xi_0^2\,C_\pi c_1(s)c_2\,(1+L)^{-s}=:c_\mu(1+L)^{-s}.
\end{equation}
Furthermore, $\mathbf K^{-1}\boldsymbol\zeta$ decays away from the last index: by Step 2, (F1) and $|\zeta_l|\le C_\pi(1+L-l)^{-s}$,
\[
|(\mathbf K^{-1}\boldsymbol\zeta)_j|\le c_1(s)a_{K^{-1}}C_\pi(1+L-j)^{-s}\le c_1(s)a_{K^{-1}}C_\pi,\qquad j=1,\dots,L .
\]
Hence, for $j=1,\dots,L$, using \eqref{eq:Kinvf}, \eqref{eq:mu} and $(1+L)^{-s}\le(1+j)^{-s}$,
\[
|w_j|\le|\lambda^*||(\mathbf K^{-1}\mathbf f)_j|+|\mu||(\mathbf K^{-1}\boldsymbol\zeta)_j|
\le\Lambda c_2(1+j)^{-s}+c_\mu c_1(s)a_{K^{-1}}C_\pi(1+L)^{-s}\le c_w(1+j)^{-s},
\]
with $c_w:=\Lambda c_2+c_\mu c_1(s)a_{K^{-1}}C_\pi$, independent of $L$. Set $w_0:=0$. In summary steps 3 and 4 establish that $\mathbf{f}$ decays algebraically starting from the first index $k=1$; the vector $\boldsymbol{\zeta}$, defined by the inverse MA filter coefficients, decays algebraically starting from the last index $L$ (i.e., it is concentrated at the right boundary); when evaluating the inner product $\boldsymbol{\zeta}' \mathbf{K}^{-1} \mathbf{f}$ to solve for the Lagrange multiplier $\mu$, the convolution of these two opposing decays over a sequence of length $L$ yields a uniform bound of $O((1+L)^{-s})$. This guarantees that the boundary effect induced by the linear constraint (represented by $\mu$) vanishes polynomially as $L \to \infty$, preventing it from affecting the decay of the early weights.

\medskip\noindent\textbf{Step 5: conclusion.}
From \eqref{eq:dw}, $\mathbf d=-\boldsymbol\Xi^{-1}\mathbf u$ with $\mathbf u:=\boldsymbol\Delta\mathbf w$, $u_k=w_k-w_{k-1}$, $k=1,\dots,L$.
As in the previous argument, $|u_1|=|w_1|\le c_w2^{-s}$ and for $k\ge2$
\[
|u_k|\le|w_k|+|w_{k-1}|\le c_w(1+k)^{-s}+c_wk^{-s}\le c_w(1+2^s)(1+k)^{-s},
\]
since $k^{-s}\le2^s(1+k)^{-s}$ for $k\ge1$. So $|u_k|\le c_u(1+k)^{-s}$ with $c_u:=c_w(1+2^s)$.
Finally, by (F3) and (ii),
\[
|d_k|=\big|(\boldsymbol\Xi^{-1}\mathbf u)_k\big|\le c_1(s)C_\pi c_u(1+k)^{-s}=:C^*(1+k)^{-s},\qquad k=1,\dots,L,
\]
where $C^*$ depends only on $(s,\Lambda,c,C,C_\xi,m_\xi,C_\gamma,\rho_1,\xi_0)$ and the separation constant in (A3),
but not on $L$. Together with (A2),
\[
|x^*_k|\le|\gamma_{MSE,k}|+|d_k|\le(C_\gamma+C^*)(1+k)^{-s},\qquad k=1,\dots,L.
\]
Since $\tilde{\mathbf b}^*=(x_2^*,\dots,x_L^*)'$ is a sub-vector of $\mathbf x^*$, the same bound holds for its components.\\

\textbf{Part (iii): Convergence of $\lambda^*(L)$ and asymptotic I-SSA.}\\
Recall from Step~0 of Part~(ii) that, with $\mathbf d_L(\lambda):=\mathbf x_L(\lambda)-\boldsymbol\gamma_{MSE}$,
\begin{equation}
\mathbf d_L(\lambda)=-\boldsymbol\Xi_L^{-1}\boldsymbol\Delta_L\mathbf w_L(\lambda),~
\mathbf w_L(\lambda)=\mathbf K_L(\lambda)^{-1}\big(\lambda\,\mathbf f_L+\mu_L(\lambda)\,\boldsymbol\zeta_L\big),~
\mathbf f_L=\boldsymbol\Delta_L'(\mathbf M_L-\rho_1\mathbf I_L)\boldsymbol\Xi_L\boldsymbol\gamma_{MSE},
\end{equation}
where
\[
\mathbf K_L(\lambda)=\boldsymbol\Delta_L'(\boldsymbol\Xi_L')^{-1}\mathbf H_L(\lambda)\boldsymbol\Xi_L^{-1}\boldsymbol\Delta_L
=\mathbf I_L+\lambda\mathbf S_L,\qquad
\mathbf S_L=\boldsymbol\Delta_L'(\mathbf M_L-\rho_1\mathbf I_L)\boldsymbol\Delta_L ,
\]
$\boldsymbol\zeta_L=(\boldsymbol\Xi_L')^{-1}\mathbf e_L=(\pi_{L-1},\dots,\pi_1,\pi_0)'$, and $\mu_L(\lambda)$ is fixed by the linear
constraint, which reads $\mathbf 1'\mathbf d_L=-\boldsymbol\zeta_L'\mathbf w_L=0$, i.e.\
$\mu_L(\lambda)=-\lambda\,\boldsymbol\zeta_L'\mathbf K_L^{-1}\mathbf f_L/(\boldsymbol\zeta_L'\mathbf K_L^{-1}\boldsymbol\zeta_L)$.
Working with $\mathbf d$ rather than $\mathbf x$ has the advantage that the forcing term involves only
$\mathbf f_L$, which is built from $\boldsymbol\gamma_{MSE}$ and $\boldsymbol\xi$, both of which decay polynomially.

\smallskip
\emph{Step 1: Compactness.}
By (A3), $\mathrm{dist}(\lambda^*(L),\partial J_{++}(L))\ge\delta>0$ for all $L$.
Since the boundaries converge to finite limits by Part~(i), the sequence $\lambda^*(L)$ is bounded.
Hence there is a subsequence with
\begin{equation}
\lim_{n\to\infty}\lambda^*(L_n)=\lambda^\circ\in[\lambda_{\min}(\infty)+\delta,\ \lambda_{\max}(\infty)-\delta]\subset J_{++}(\infty).
\end{equation}

\smallskip
\emph{Step 2: Convergence of the inverse.}
Let $\boldsymbol\Xi_\infty$, $\boldsymbol\Delta_\infty$, $\mathbf S_\infty$ be the (bounded) one-sided infinite versions of $\boldsymbol\Xi_L$, $\boldsymbol\Delta_L$, $\mathbf S_L$;
$\boldsymbol\Xi_\infty$ is bounded on $\ell^2$ because $\boldsymbol\xi\in\ell^1$ by (A1), and $\mathbf S_\infty$ is banded.
Since $\lambda^\circ$ is strictly inside $J_{++}(\infty)$, that is, $\mathbf I+\lambda^\circ\mathbf S_\infty\succeq c_\delta\mathbf I$ with $c_\delta>0$,
the operator $\mathbf K_\infty(\lambda^\circ)=\mathbf I+\lambda^\circ\mathbf S_\infty$
is bounded and bounded below, hence invertible on $\ell^2$.
By the Preamble of Part~(ii), $c\mathbf I\preceq\mathbf K_L(\lambda^*(L))\preceq C\mathbf I$ uniformly in $L$.
Moreover
$\|\mathbf K_{L_n}(\lambda^*_{L_n})-\mathbf K_{L_n}(\lambda^\circ)\|\le|\lambda^*_{L_n}-\lambda^\circ|\,\|\mathbf S_{L_n}\|\to0$
as $\lambda^*_{L_n}\to\lambda^\circ$, uniformly in $L_n$ (recall $\|\mathbf S_L\|\le4(1+|\rho_1|)$), and $\mathbf K_{L_n}(\lambda^\circ)$ converges strongly to $\mathbf K_\infty(\lambda^\circ)$
(entrywise convergence of uniformly bounded, uniformly banded matrices).
Together with the uniform bound on the inverses, this gives
\begin{equation}
\mathbf K_{L_n}(\lambda^*(L_n))^{-1}\xrightarrow{\ s\ }\mathbf K_\infty(\lambda^\circ)^{-1}.
\end{equation}

\smallskip
\emph{Step 3: Convergence of the forcing term.}
By (A1)--(A2) and Step~3 of Part~(ii), $|(\mathbf f_L)_k|\le a_f(1+k)^{-s}$ for all $L$ and $k\le L$, so $\mathbf f_L\to\mathbf f_\infty:=\boldsymbol\Delta_\infty'(\mathbf M_\infty-\rho_1\mathbf I)\boldsymbol\Xi_\infty\boldsymbol\gamma_{MSE}$
in $\ell^2$ (dominated convergence, since $s>1$).
By \eqref{eq:mu}, $|\mu_L(\lambda^*(L))|\le c_\mu(1+L)^{-s}$, and $\boldsymbol\zeta_L$ has bounded $\ell^2$-norm,
$\|\boldsymbol\zeta_L\|_2^2=\sum_{j=0}^{L-1}\pi_j^2\le C_\pi^2\sum_{j\ge0}(1+j)^{-2s}<\infty$,
so $\mu_L(\lambda^*(L))\,\boldsymbol\zeta_L$ is $O(L^{-s})$ in $\ell^2$ and tends to $0$.
Hence
\begin{equation}
\lambda^*(L_n)\mathbf f_{L_n}+\mu_{L_n}\boldsymbol\zeta_{L_n}\ \longrightarrow\ \lambda^\circ\mathbf f_\infty\quad\text{in }\ell^2,
\end{equation}
and therefore
\begin{equation}
\mathbf w_{L_n}\to\mathbf w_\infty:=\lambda^\circ\,\mathbf K_\infty(\lambda^\circ)^{-1}\mathbf f_\infty,\qquad
\mathbf d_{L_n}=-\boldsymbol\Xi^{-1}\boldsymbol\Delta\mathbf w_{L_n}\to\mathbf d_\infty:=-\boldsymbol\Xi_\infty^{-1}\boldsymbol\Delta_\infty\mathbf w_\infty\quad\text{in }\ell^2,
\end{equation}
because $\boldsymbol\Delta$ and $\boldsymbol\Xi^{-1}$ (whose entries $\pi_j$ are in $\ell^1$ by (F3)) are bounded.
The limit $\mathbf d_\infty$ automatically satisfies the linear constraint,
since $\mathbf 1'\mathbf d_L=-\boldsymbol\zeta_L'\mathbf w_L=0$ for every $L$ and the convergence $\mathbf d_{L_n}\to\mathbf d_\infty$ holds in $\ell^1$
(by the uniform polynomial decay $|d_{L,k}|\le C^*(1+k)^{-s}$ of Part~(ii) and dominated convergence);
that is, the constraint does not need to be imposed separately in the limit.

\smallskip
\emph{Step 4: Limit of the stationary point.}
By Steps~2 and~3, together with the uniform bound on $\|\mathbf K_{L_n}^{-1}\|$ and the strong convergence of
$\mathbf K_{L_n}(\lambda^*(L_n))^{-1}$, the deviation of the finite-dimensional stationary point from the
MSE filter, $\mathbf d^*(L_n)=\mathbf x^*(L_n)-\boldsymbol\gamma_{MSE}=-\boldsymbol\Xi^{-1}\boldsymbol\Delta\,\mathbf w_{L_n}$, converges in $\ell^2$ to
\begin{equation}\label{dinftyl}
\mathbf d^*(\infty)=-\boldsymbol\Xi_\infty^{-1}\boldsymbol\Delta_\infty\,\lambda^\circ\,\mathbf K_\infty(\lambda^\circ)^{-1}\mathbf f_\infty ,
\end{equation}
so that $\mathbf x^*(\infty):=\boldsymbol\gamma_{MSE}+\mathbf d^*(\infty)$  is the limit of the stationary points (in the subsequence defined by $L_n$).
It satisfies the linear constraint $\mathbf 1'\mathbf d^*(\infty)=0$ (Step~3). Passing to the limit in the
stationarity condition and in the quadratic constraint
$\mathbf x^*(L_n)'\mathbf Q_{L_n}\mathbf x^*(L_n)=0$, equivalently
$(\boldsymbol\Xi_{L_n}\mathbf x^*(L_n))'(\mathbf M_{L_n}-\rho_1\mathbf I_{L_n})(\boldsymbol\Xi_{L_n}\mathbf x^*(L_n))=0$
(all terms converge, since $\mathbf x^*(L_n)$ converges in $\ell^1$
by the uniform polynomial decay of Part~(ii) and $\mathbf Q_{L_n}=\boldsymbol\Xi_{L_n}'(\mathbf M_{L_n}-\rho_1\mathbf I_{L_n})\boldsymbol\Xi_{L_n}$ is bounded), $\mathbf x^*(\infty)$ is a stationary point of the
infinite-dimensional problem with multiplier $\lambda^\circ$.

\smallskip
\smallskip
\emph{Step 5: Uniqueness of the limit.}
For $\lambda \in J_{++}(\infty)$, define the smooth $\ell^2$-valued map 
$\mathbf w_\infty(\lambda) := \lambda \mathbf K_\infty(\lambda)^{-1}\mathbf f_\infty$, 
the corresponding stationary profile 
$\mathbf x_\infty(\lambda) := \boldsymbol\gamma_{MSE} - \boldsymbol\Xi_\infty^{-1}\boldsymbol\Delta_\infty\mathbf w_\infty(\lambda)$, 
and the infinite-dimensional secular function 
$g_\infty(\lambda) := \langle \mathbf x_\infty(\lambda), \mathbf Q_\infty \mathbf x_\infty(\lambda) \rangle$. 
By Step~4, any subsequential limit $\lambda^\circ \in J_{++}(\infty)$ satisfies $g_\infty(\lambda^\circ) = 0$.

\emph{(a) Derivative of $g_\infty$.}
Differentiating the identity $\mathbf K_\infty(\lambda)\mathbf w_\infty(\lambda) = (\mathbf I + \lambda\mathbf S_\infty)\mathbf w_\infty(\lambda) = \lambda\mathbf f_\infty$ with respect to $\lambda \in J_{++}(\infty)$ yields
\begin{equation}\label{eq:wdot}
\mathbf K_\infty(\lambda)\dot{\mathbf w}_\infty(\lambda) = \mathbf f_\infty - \mathbf S_\infty\mathbf w_\infty(\lambda).
\end{equation}
Using the definitions $\mathbf f_\infty = \boldsymbol\Delta_\infty'(\mathbf M_\infty - \rho_1\mathbf I)\boldsymbol\Xi_\infty\boldsymbol\gamma_{MSE}$, $\mathbf S_\infty = \boldsymbol\Delta_\infty'(\mathbf M_\infty - \rho_1\mathbf I)\boldsymbol\Delta_\infty$, and $\boldsymbol\Xi_\infty\mathbf x_\infty(\lambda) = \boldsymbol\Xi_\infty\boldsymbol\gamma_{MSE} - \boldsymbol\Delta_\infty\mathbf w_\infty(\lambda)$, the right-hand side of \eqref{eq:wdot} factors as
\begin{equation}\label{eq:rhs_factor}
\mathbf f_\infty - \mathbf S_\infty\mathbf w_\infty(\lambda) 
= \boldsymbol\Delta_\infty'(\mathbf M_\infty - \rho_1\mathbf I)\boldsymbol\Xi_\infty\mathbf x_\infty(\lambda) 
= \boldsymbol\Delta_\infty'(\boldsymbol\Xi_\infty')^{-1}\mathbf Q_\infty\mathbf x_\infty(\lambda).
\end{equation}
Since $\dot{\mathbf x}_\infty(\lambda) = -\boldsymbol\Xi_\infty^{-1}\boldsymbol\Delta_\infty\dot{\mathbf w}_\infty(\lambda)$, differentiating $g_\infty(\lambda)$ gives
\begin{align}
\dot g_\infty(\lambda) 
&= 2\langle \dot{\mathbf x}_\infty(\lambda), \mathbf Q_\infty\mathbf x_\infty(\lambda) \rangle
= -2\langle \dot{\mathbf w}_\infty(\lambda), \boldsymbol\Delta_\infty'(\boldsymbol\Xi_\infty')^{-1}\mathbf Q_\infty\mathbf x_\infty(\lambda) \rangle \nonumber\\
&= -2\langle \dot{\mathbf w}_\infty(\lambda), \mathbf K_\infty(\lambda)\dot{\mathbf w}_\infty(\lambda) \rangle \le 0,
\label{eq:gdot}
\end{align}
with equality if and only if $\dot{\mathbf w}_\infty(\lambda) = \mathbf 0$, because $\mathbf K_\infty(\lambda) \succeq c_\delta \mathbf I \succ \mathbf 0$ on $J_{++}(\infty)$.

\emph{(b) Non-degeneracy at roots.}
We show that $\dot{\mathbf w}_\infty(\lambda) \neq \mathbf 0$ for every $\lambda \in J_{++}(\infty)$. 
By the standing non-degeneracy hypothesis ($\rho_1 \neq \rho_{MSE}$), $g_\infty(0) = \langle \boldsymbol\gamma_{MSE}, \mathbf Q_\infty\boldsymbol\gamma_{MSE} \rangle \neq 0$. 
Note first that if $\mathbf f_\infty = \mathbf 0$, then $\mathbf w_\infty(\lambda) \equiv \mathbf 0$ and $\mathbf x_\infty(\lambda) \equiv \boldsymbol\gamma_{MSE}$, which would imply $g_\infty(\lambda^\circ) = g_\infty(0) \neq 0$, contradicting $g_\infty(\lambda^\circ) = 0$; hence $\mathbf f_\infty \neq \mathbf 0$. 
Now suppose $\dot{\mathbf w}_\infty(\lambda) = \mathbf 0$ for some $\lambda \in J_{++}(\infty)$. By \eqref{eq:wdot}, $\mathbf S_\infty\mathbf w_\infty(\lambda) = \mathbf f_\infty$, and substituting this into $(\mathbf I + \lambda\mathbf S_\infty)\mathbf w_\infty(\lambda) = \lambda\mathbf f_\infty$ yields $\mathbf w_\infty(\lambda) = \mathbf 0$, which in turn forces $\mathbf f_\infty = \mathbf S_\infty\mathbf w_\infty(\lambda) = \mathbf 0$, a contradiction.

Consequently, $\dot g_\infty(\lambda) < 0$ for all $\lambda \in J_{++}(\infty)$, so $g_\infty$ is strictly decreasing on $J_{++}(\infty)$ and possesses a unique root $\lambda^*(\infty) := \lambda^\circ \in J_{++}(\infty)$. Since every convergent subsequence of $\lambda^*(L)$ must converge to $\lambda^*(\infty)$, the full sequence converges:
\[
\lim_{L\to\infty} \lambda^*(L) = \lambda^*(\infty) \in J_{++}(\infty).
\]
The asymptotic I-SSA formulated in assertion iii) follows straightforwardly from the above derivations. 

\end{proof}

\textbf{Remarks:}
Two structural devices underpin the proof of Theorem~\ref{thm:decay_reformulated_uniform}:
\begin{enumerate}
    \item \emph{Rank-one boundary completion in Part~(i):} Appending the row $-\mathbf{e}_L'$ to $\boldsymbol{\Delta}_L$ to form $\widetilde{\boldsymbol{\Delta}}_L$ (Step~2) completes the rank-one boundary deficit in $\boldsymbol{\Delta}_L'\boldsymbol{\Delta}_L$ to yield the exact Dirichlet Laplacian $\mathbf{L}_D = 2(\mathbf{I}_L - \mathbf{M}_L)$, while simultaneously embedding $\mathbf{K}_L$ as the leading $L \times L$ principal submatrix of $\widetilde{\mathbf{K}}_L$. This allows one to bypass delicate rank-one perturbation and secular resolvent analyses, yielding exact finite-sample two-sided eigenvalue bounds directly via the Discrete Sine Transform (DST-I) and Cauchy's Interlacing Theorem.
    
    \item \emph{Deviation formulation and asymptotic vanishing of the cointegration multiplier in Parts~(ii)--(iii):} Recasting the  system in terms of the deviation from the unconstrained MSE benchmark, $\mathbf{d}_L := \mathbf{x}_L^* - \boldsymbol{\gamma}_{MSE}$, and preconditioning via the congruence $\mathbf{H}_L^{-1} = \boldsymbol{\Xi}_L^{-1}\boldsymbol{\Delta}_L\mathbf{K}_L^{-1}\boldsymbol{\Delta}_L'(\boldsymbol{\Xi}_L')^{-1}$ achieves two goals simultaneously. First, it replaces the ill-conditioned cumulative-sum Hessian $\mathbf{H}_L$ with the uniformly well-conditioned pentadiagonal matrix $\mathbf{K}_L$. Second, because $\boldsymbol{\Delta}_L'\mathbf{1} = \mathbf{e}_L$, the global level/cointegration constraint $\mathbf{1}'\mathbf{x}_L^* = \Gamma(0)$ is localized to the right boundary via $\boldsymbol{\zeta}_L = (\boldsymbol{\Xi}_L')^{-1}\mathbf{e}_L$. Convolving the forward-decaying forcing $\mathbf{K}_L^{-1}\mathbf{f}_L$ with the backward-decaying boundary vector $\boldsymbol{\zeta}_L$ forces the Lagrange multiplier $\mu_L$ to vanish at rate $O(L^{-s})$, ensuring that finite-sample boundary perturbations disappear asymptotically without distorting the decay of the filter weights.
\end{enumerate}

\textbf{Corollary \ref{eff-as-front}}
\begin{proof} 
Theorem \ref{thm:decay_reformulated_uniform} ensures that MA approximations are valid, up to negligible deviations. Therefore, the objective function and the smoothness constraint admit their intended interpretations: the MSE (on levels) and the first-order ACF (in stationary differences), respectively.

Consider MSE$(\rho_1):=J(\lambda^*(\rho_1))$ as a function of $\rho_1$, with $|\rho_1|<\rho_{\max}(L)$. Eq.\eqref{dilemma} shows that MSE is strictly \emph{increasing} on the right branch $\rho_1>\rho_{MSE}$. By Assertion \ref{gl_min} (global minimum), no predictor can lie below the  curve $J(\lambda^*(\rho_1))$, so the corresponding space is empty. We conclude that, for given MSE, I-SSA maximizes $\rho_1$ on the right branch. Since $1/ht_1=\arccos(\rho_1)/\pi$ is strictly monotonically decreasing, I-SSA thus minimizes the TP rate $1/ht_1$ for given MSE, assuming a vanishing mean in first differences (otherwise it minimizes the crossover rate about the linear trend). A similar argument applies to the left branch $\rho_1<\rho_{MSE}$, where the strictly \emph{decreasing} MSE, together with the global minimum, implies that I-SSA {maximizes} the TP rate.
\end{proof}


\begin{thebibliography}{00}
%



\bibitem{}  Adachi, S., Iwata, S., Nakatsukasa, Y., and Takeda, A. (2017). Solving the trust-region subproblem by a generalized eigenvalue problem. {\it SIAM Journal on Optimization}, {\bf 27(1)}.


\bibitem{} Anderson O.D. (1975) Moving Average Processes.  {\it Journal of the Royal Statistical Society. Series D (The Statistician)}. {\bf Vol. 24, No. 4}, 283-297



\bibitem{} Baxter, M. and R.G. King (1999). Measuring business cycles: Approximate band-pass filters for economic time series. {\it Review of Economics and Statistics} {\bf 81}, 575–593.

\bibitem{} Box, G. E. P., and Tiao, G. C. (1977). A canonical analysis of multiple time series. {\it Biometrika}, {\bf 64(2)}.


\bibitem{} Brockwell P.J. and Davis R.A. (1993) Time Series: Theories and Methods (second edition).  {\it Springer Verlag}.

\bibitem{}Cornea-Madeira, Adriana (2017). The Explicit Formula for the Hodrick-Prescott Filter in Finite Sample. {\it Review of economics and statistics}, 314-318. 


\bibitem{}  Drehmann, M., and Yetman, J. Which Credit Gap Is Better at Predicting Financial Crises? A Comparison of Univariate Filters. {\it 
International Journal of Central Banking},  {\bf 17, 4},  October 2021.



\bibitem{} Durbin, J. and Koopman, S. J. (2012). Time Series Analysis by State Space Methods, {\it 2nd edn. Oxford: Oxford University Press}.



\bibitem{} Eilers, Paul H.C. et al. (2015). Twenty years of P-splines. {\it SORT-Statistics and Operations Research Transactions}. {\bf 39,2}, pp. 149-86, https://raco.cat/index.php/SORT/article/view/302258.


\bibitem{} Gr\"ochenig, K. (2010). Wiener's lemma: theme and variations. An introduction to harmonic analysis in the style of Wiener, in Four Short Courses on Harmonic Analysis {it Birkhäuser (Applied and Numerical Harmonic Analysis)}, 175–234.

\bibitem{}  Gr\"ochenig, K. and Klotz, A. (2010). Noncommutative approximation: inverse-closed subalgebras and off-diagonal decay of matrices. {\it Constructive Approximation}. {\bf 32}, 429-466.

\bibitem{}  Granger, C. W. J. (1966). The Typical Spectral Shape of an Economic Variable. {\it Econometrica}.{\bf 34, 1},
https://doi.org/10.2307/1909859.

\bibitem{} Hamilton, J.D. (2018). Why you should never use the Hodrick-Prescott filter. {\it Review of Economics and Statistics} {\bf 100}, 831–843.

\bibitem{} Harvey, A. (1989). Forecasting, structural time series models and the Kalman filter.  {\it Cambridge: Cambridge University Press}.




\bibitem{}  Henderson, R. (1924). A New Method of Graduation.  {\it Transactions of the Actuarial Society of America} {\bf 25}, 29–40.


\bibitem{} Hillmer, S. C. and Tiao, G. C. (1982). An ARIMA-model-based approach to seasonal adjustment. {\it Journal of the American Statistical Association}, {\bf 77(377)}, 63-70.


\bibitem{} Hodrick, R. and Prescott, E. (1997) Postwar U.S. business
cycles: an empirical investigation.  {\it Journal of Money, Credit,
and Banking} {\bf 29}, 1--16.


\bibitem{} Jaffard, S. (1990) {Propri{\'e}t{\'e}s des matrices {\guillemotleft}~bien localis{\'e}es~{\guillemotright} pr{\`e}s de leur diagonale et quelques applications} {\it Annales de l'Institut Henri Poincar{\'e} C, Analyse non lin{\'e}aire} {\bf 7,5}, 461--476.


\bibitem{} Johansen, S. (1988). Statistical analysis of cointegration vectors. {\it Journal of Economic Dynamics and Control}, {\bf 12(2-3)}.
\bibitem{} Kamber, G. Morley, J. Wong, B. (2024).  Trend-Cycle Decomposition in the Presence of Large Shocks. {\it CAMA Working Papers 2024-24}. Centre for Applied Macroeconomic Analysis, Crawford School of Public Policy, The Australian National University, revised Aug 2024.


\bibitem{} Johansen, S. (1991). Estimation and hypothesis testing of cointegration vectors in Gaussian vector autoregressive models. {\it Econometrica}, {\bf 59(6)}.


\bibitem{} Kedem, B. (1986) Zero-crossings analysis.  {\it Research report AFOSR-TR-86-0413, Univ. of Maryland.}





\bibitem{}  McElroy, T. (2008) Exact formulas for the Hodrick-Prescott filter. {\it The Econometrics Journal} {\bf 11,1}, 209-217, https://www.jstor.org/stable/23116069.

\bibitem{}  McElroy, T and Wildi, M. (2010) Signal Extraction Revision Variances as a Goodness-of-Fit Measure. {\it Journal of Time Series Econometrics},{\bf 2,1}.


\bibitem{} McElroy, T. (2022) Casting vector time series: algorithms for forecasting, imputation, and signal extraction.
{\it Electron. J. Statist.} {\bf 16 (2)}, 5534-5569. DOI: 10.1214/22-EJS2068

\bibitem{} McElroy, T.S. and Livsey, J.A. (2022) Ecce Signum: An R Package for Multivariate Signal Extraction and Time Series Analysis {\it 	arXiv:2201.02148}, https://arxiv.org/abs/2201.02148




\bibitem{} McElroy, T. and Wildi , M. (2016) Optimal Real-Time Filters for Linear Prediction Problems.  {\it  Journal of Time Series Econometrics } {\bf 8},  155-192.


\bibitem{} McElroy, T. and Wildi , M. (2019) The trilemma between accuracy, timeliness and smoothness in real-time signal extraction.  {\it International Journal of Forecasting  }, {\bf 35 (3)}, 1072-1084.


\bibitem{}Mei, Z., and Phillips, P. C. B., and Shi, Z. (2022) The boosted HP filter is more general than you might think. {\it Cowles Foundation Discussion Paper}, Nbr. 2348.

\bibitem{} More, J. J. (1993). Generalizations of the trust region problem. {\it Optimization Methods and Software}, {\bf 2(3-4)}.



\bibitem{} Percival, D. B. and Walden, A. T. (2000) Wavelet Methods for Time Series Analysis. {\it Cambridge: Cambridge University Press.}


\bibitem{} Percival, D. B. and Walden, A. T. (2020) Spectral Analysis for Univariate Time Series . {\it Cambridge Series in Statistical and Probabilistic Mathematics.} 


\bibitem{}Phillips, P. C. B., and Shi, Z. (2021) Boosting: Why You Can Use the HP Filter. {\it International Economic Review}, {\bf 62(2)}, 521–570.

\bibitem{} Mei, Z., and Phillips, P. C. B., and Shi, Z. (2024). The boosted Hodrick-Prescott filter is more general than you might think.{\it Journal of Applied Econometrics}, {\bf 39(7)}, 1260–1281.

\bibitem{} Polik, I. and Terlaky, T. (2007) A survey of the S-lemma. {\it SIAM Review}, {\bf vol. 49, no. 3}.



\bibitem{} Pollock, D. S. G. (2000) Trend estimation and de-trending via rational square-wave filters. {\it Journal of Econometrics}, {\bf 99(2)}, 317-334.

\bibitem{} Pollock, D. S. G. (2021) Filters, waves and spectra. {\it Econometrics and Statistics}, {\bf 17}, 87-107.



\bibitem{} Pong, T. K. and Wolkowicz, H. (2014) The generalized trust region subproblem. {\it Computational Optimization and Applications}, {\bf 58,2}.





\bibitem{} Priestley, M. B. (1982) Spectral analysis and time series. {\it London ; New York : Academic Press} 



\bibitem{} Rice,S.O. (1944) Mathematical analysis of random noise.  {\it I. Bell. Syst. Tech. J } {\bf 23}, 282-332.


\bibitem{} Tibshirani, R. J. (2014) Adaptive piecewise polynomial estimation via trend filtering. {\it The Annals of Statistics}, {\bf 42(1)}, 285-323.


\bibitem{} Whittaker, E. T. (1923) On a new method of graduation. 
{\it Proceedings of the Edinburgh Mathematical Society}, {\bf 41}, 63-75.

\bibitem{} Wildi, M. (2024) Business Cycle Analysis and Zero-Crossings of Time Series: a Generalized Forecast Approach.  {\it J. Bus. Cycle Res.} https://doi.org/10.1007/s41549-024-00097-5.


\bibitem{} Wildi, M. (2026a) Sign Accuracy, Mean-Squared Error and the Rate of Zero Crossings: a Generalized Forecast Approach.  {\it 	arXiv:2601.06547.} https://doi.org/10.48550/arXiv.2601.06547.


\bibitem{} Wildi, M. (2026b) The Accuracy Smoothness Dilemma in Prediction: A Novel Multivariate M-SSA Forecast Approach. {\it Journal of Time Series Analysis}, https://onlinelibrary.wiley.com/doi/abs/10.1111/jtsa.70058.

\bibitem{} Wood, S. N. (2017) Generalized Additive Models: An Introduction with R. {\it 2nd edn. Boca Raton, FL: Chapman and Hall/CRC}.









\end{thebibliography}
\end{document}